\documentclass[12pt]{article}
\usepackage{graphicx}
\usepackage{hep}
\usepackage{amsmath}
\usepackage{amssymb}
\usepackage{cite}
\usepackage{a4wide}
\usepackage[a4paper]{hyperref}
\usepackage{rotating}

\def\Title#1{\begin{center} {\Large\bf #1 } \end{center}}
\def\Author#1{\begin{center}{ \sc #1} \end{center}}
\def\Address#1{\begin{center}{ \it #1} \end{center}}

\newcommand\pubblock{\rightline{\begin{tabular}{l} \pubnumber\\
      \hepnumber \\
      \pubdate\end{tabular}}}
\newcommand\pubnumber{} 
\newcommand\pubdate{September 2012}
\newcommand\hepnumber{arXiv:1209.5214 [hep-ph]}

\newcommand{\SUSY}{Nilles:1984ex,Haber:1985rc,Lahanas:1987uc,Ferrara87}
\newcommand{\TESLA}{Aguilar-Saavedra:2001rg}
\newcommand{\LHC}{ATLAS,Ball:2007zza}

\newcommand{\vacuum}{Frere:1983ag,Claudson:1983et,Kounnas:1983td,Gunion:1987qv}

\begin{document}
\pubblock

\vfill
\def\thefootnote{\fnsymbol{footnote}}
\Title{Effective squark/chargino/neutralino couplings:
MadGraph implementation}
\Author{Arian Abrahantes$^{a}$, Jaume Guasch$^{b,c}$, Siannah Pe{\~n}aranda$^{a,b,c}$,\\ Ra{\"u}l S{\'a}nchez-Florit$^{d,c}$}
\Address{\textsl{
$^a$ Departamento de F{\'\i}sica Te{\'o}rica,
Facultad de Ciencias,\\ Universidad de Zaragoza,
            E-50009 Zaragoza, Spain\\
$^b$ Departament de F{\'\i}sica Fonamental,\\ Universitat de Barcelona,
    Diagonal 645, E-08028 Barcelona, Catalonia, Spain\\
$^c$ Institut de Ci{\`e}ncies del Cosmos (ICC), \\ 
Universitat de Barcelona,
    Diagonal 645, E-08028 Barcelona, Catalonia, Spain\\
$^d$ Departament d'Estructura i Constituents de la Mat{\`e}ria,\\ Universitat de Barcelona,
    Diagonal 645, E-08028 Barcelona, Catalonia, Spain\\
E-mails: arian@unizar.es, jaume.guasch@ub.edu, siannah@unizar.es, florit@ffn.ub.es}
} \vspace{1cm}

\begin{abstract}
We have included the effective description of squark 
interactions with char\-gi\-nos/neu\-tra\-li\-nos in the MadGraph MSSM model. 
This effective description includes the effective Yukawa couplings, and another
logarithmic term which encodes the supersymmetry-breaking.
We have performed an extensive test of our implementation analyzing the 
results of the 
partial decay widths of squarks into charginos and neutralinos obtained by 
using FeynArts/FormCalc programs and the new model file in MadGraph. 
We present results for 
the cross-section of top-squark production decaying into
charginos and neutralinos. 
\end{abstract}
\vfill

\def\thefootnote{\arabic{footnote}}
\newpage

\section{Introduction}

The Standard Model (SM) of the strong and electroweak interactions is the
present paradigm of particle physics. Its validity has been tested to a
level better than one per mile at particle
accelerators~\cite{Beringer:2012}. 
Nevertheless, there are
arguments against the SM being the fundamental model of
particle interactions~\cite{Haber:1993kz}, giving rise to the
investigation of competing alternative or extended models, which
can be tested at high-energy colliders, such as the Large Hadron Collider
(LHC)~\cite{\LHC}, or a $500-1000\GeV$, $e^+e^-$ International Linear
Collider (ILC)~\cite{\TESLA,Weiglein:2004hn}. 
One of the most promising possibilities for physics beyond the Standard 
Model is Supersymmetry (SUSY)~\cite{\SUSY}, which leads to a renormalizable field 
theory with precisely calculable predictions to be tested in present and 
future experiments. The simplest model of this kind is the Minimal 
Supersymmetric Standard Model (MSSM). Among the most important 
phenomenological consequences of SUSY models, is the prediction of 
new particles, the SUSY partners of SM particles, sometimes called sparticles. There is much excitement 
for the possibility of discovering these new particles at 
LHC~\cite{Aad:2009wy,Ball:2007zza}. 
The LHC collaborations are already performing searches on these
particles, and excluding portions of the SUSY parameter space, see e.g.\cite{ATLASCMS:ICHEP2012,Chatrchyan:2011zy,Aad:2011ib,ATLAS:2011ad,CMS:SUS12009,Aad:2012hm,Chatrchyan:2012mf,Collaboration:2012si,Collaboration:2012ar,Collaboration:2012as,Aad:2012ms,CMS:2012rg,ATLAS:2012st,ATLAS:2012st2,ATLASweb,CMSweb}.
Recently, the CMS and ATLAS collaborations have reported a 5
  standard deviations signal on a new boson particle, at a mass $m\simeq125\GeV$, which is compatible with the interpretation of a SM Higgs
boson\cite{CMSHiggs, ATLASHiggs,ATLASHiggs:2012a,CMSHiggs:2012a}. 
The CDF and D0 collaborations at the Tevatron also found (less significant)
signals compatible with this interpretation~\cite{:2012zzl,Aaltonen:2012qt}. This
signal is also compatible with the lightest neutral Higgs boson of the
MSSM~(see e.g.~\cite{MahmoudiICHEP}).
Precision measurements and precision computations are both mandatory 
nowadays. Under the huge avalanche of new data at present, an accurate 
prediction of sparticles couplings to other particles and their production 
cross-section is needed. In this work we focus on the properties of the 
squarks -- the SUSY partners 
of SM quarks. In particular, we concentrate on the squark
 decay channels involving charginos and neutralinos --
the fermionic SUSY partners of the electroweak
gauge and Higgs bosons.

Once produced,  squarks will decay in a way dependent on the
model parameters  (see e.g.~\cite{Bartl:1994bu}). If gluinos
(the fermionic SUSY partners of the SM 
gluons) are light enough, squarks will mainly decay into gluinos
and quarks ($\squark\to q\tilde{g}$)~\cite{Beenakker:1996dw,Beenakker:1996de}, which proceeds through a coupling
constant of strong strength. If the mass difference among different
squarks is large enough, some squarks can decay via a bosonic channel
into an electroweak gauge boson and another squark 
($\squark_a \to \squark_b'(Z,\,W^\pm)$), and if Higgs bosons are
light enough, also the scalar decay channels are available ($\squark_a
\to \squark_b'(h^0,\,H^0,\,A^0,\,H^\pm)$)~\cite{Bartl:1997pb,Bartl:1998xk,Bartl:1998xp,Bartl:1999bg}, which can be dominant for
third generation squarks due to the large Yukawa couplings. Otherwise,
the main decay channels of squarks are the fermionic ones:
chargino/neutra\-lino and a quark ($\squark \to q'\chi$)\cite{Guasch:2002ez,Guasch:2002qa,Guasch:2008fs}. Some of those
channels are expected to be always open, given the large mass
difference between quarks and squarks, and that the
charginos/neutra\-linos are expected to be lighter than most of squarks in
the majority of SUSY-breaking models. In the few cases in which these
channels are closed, the squarks will decay through flavour changing
neutral channels~\cite{Hikasa:1987db,Han:2003qe,delAguila:2008iz,Muhlleitner:2011ww}, or
through three- or four-body  decay channels involving a non-resonant
SUSY particle~\cite{Porod:1996at,Porod:1998yp,Boehm:1999tr,Djouadi:2000aq,Das:2001kd,Djouadi:2000bx}.

Here we will concentrate on the squark decay channels involving
  charginos and neutralinos.
Their partial decay widths were computed some time ago,
including the radiative corrections due to the strong
(QCD)~\cite{Hikasa:1995bw,Kraml:1996kz,Djouadi:1996wt}, and the electroweak
(EW)\cite{Guasch:1998as,Guasch:2001kz,Guasch:2002ez,Guasch:2002qa} sectors of the
theory. These radiative corrections are large in certain regions of the
parameter space\cite{Guasch:2002ez}, and their complicated expressions
are not suitable for their introduction in the Monte-Carlo programs used
in experimental analysis. 
Recently Ref.~\cite{Guasch:2008fs} presented 
an improved description of squark/chargino/neutralino couplings, 
more simple to write and to introduce in computer codes. 
This computation combines the effective description (which includes
higher order terms) with the complete one-loop description (which
includes all kinetic and mass-effects factors) and defines a new
effective coupling. It includes a non-decoupling logarithmic gluino
mass term, which implies a deviation of the higgsino/gaugino and
Higgs/gauge couplings equality predicted by exact SUSY. This deviation
is important and has to be taken into account in the experimental
measurement of SUSY relations. Ref.\cite{Guasch:2008fs} showed that
the effective description approximates the improved description within a
$10\%$ precision, except in special uninteresting corners of the
parameter space,
where the corresponding branching ratios are practically zero.
Ref.\cite{Guasch:2008fs} applies the description only
to squark decays. The present work expands the results of
Ref.\cite{Guasch:2008fs} by applying those results to the production
cross-section of squarks at the LHC. To that end we have implemented
this effective description in MadGraph's~\cite{Stelzer:1994ta,Alwall:2007st,Alwall:2011uj}
MSSM framework~\cite{Cho:2006sx}, we have applied it to the partial decay widths of
squarks into charginos and neutralinos and we have computed the
corresponding cross-section.

In section~\ref{sec:secLsquarks} we present the theoretical framework, 
by introducing
the notation for particles and
couplings. Section~\ref{sec:qcd-corrections} summarizes the results for the
one-loop QCD-corrections to the 
squark partial decay widths into charginos and neutralinos. 
A few details about the effective description approach described in~\cite{Guasch:2008fs} are given in this section. This will allow readers to know what we have coded into our effective MSSM model for squarks within MadGraph. 
In section~\ref{sec:MC} we describe our Monte-Carlo tool
  implementation of the MSSM effective couplings.
Section~\ref{sec:resultsnumerical} presents the numerical
  analysis: numerical setup and parameter choices
  (section~\ref{sec:numericalsetup}), partial decay widths analysis
  (section~\ref{resultsGamma}), and cross-section results
  (section~\ref{sec:crosssection}).
Finally section~\ref{sec:conclusions} shows our conclusions.

\section{Tree-level relations and parameter definitions}
\label{sec:secLsquarks}

Here we introduce our notation for SUSY particles and couplings.
Throughout this work we will use a
third-generation notation to describe quarks and squarks, but the analytic
results and conclusions are completely general, and can be used for
quarks-squarks of any generation. We will show numerical results only
for top-squarks ($\stopp$), since their signals are the most
 phenomenologically interesting.

To describe the computation of the partial decay widths, we will follow
the conventions of Ref.~\cite{Hahn:2001rv}. We will study the partial 
decay widths of sfermions into fermions and
charginos/neutra\-linos,
\begin{equation}
\Gamma(\sfr \to f' \chi)\,\,.
\label{eq:gammadef}
\end{equation}

We denote {the two sfermion-mass eigenvalues
by $m_{\sfr_a}\,(a=1,2)$, with
$m_{\sfr_1}<m_{\sfr_2}$}.
The sfermion-mixing angle $\osf$
is defined by the transformation relating the weak-interaction
($\sfr^\prime_a=\sfr_L, \sfr_R$) and the mass eigenstate
 ($\sfr_a=\sfr_1, \sfr_2$) sfermion bases:
\begin{equation}
\label{eq:defsq}
  \sfr_a=R_{ab}^{(f)}\, \sfr^\prime_b\,\,; \,\,\,\,
  R^{(f)}=\left(\begin{array}{cc}
      \cos\osf&-\sin\osf \\
      \sin\osf &\cos\osf
    \end{array}\right)\,.
\end{equation}
By this  basis transformation, the sfermion mass matrix,
\begin{equation}
{\cal M}_{\sfr}^2 =\left(\begin{array}{cc}
\MsfL^2+m_f^2+c_{2\beta}(T_3-Q\,s_W^2)\,M_Z^2 
 &  m_f\, M^{LR}_f\\
 m_f\, M^{LR}_f &
 \MsfR^2+m_f^2+Q\,c_{2\beta} \,s_W^2\,M_Z^2  
\end{array} \right)\,,
\label{eq:sbottommatrix}
\end{equation}
becomes diagonal: 
$R^{(f)}\,{\cal M}_{\sfr}^2\,R^{(f)\dagger}=
{\rm diag}\left\{m_{\sfr_1}^2,
  m_{\sfr_2}^2\right\}$. $\MsfL^2$ is the
soft-SUSY-breaking mass parameter of the $SU(2)_L$
doublet\footnote{{With $M_{\stopp_L}=M_{\sbottom_L}$ due to $SU(2)_L$ gauge invariance.}}, whereas
$\MsfR^2$ is the soft-SUSY-breaking mass parameter of the
singlet. $T_3$ and 
$Q$ are the usual third component of the isospin and the
electric 
charge respectively, $m_f$ is the corresponding fermion mass, 
$\mz$ is the electroweak $Z$ boson mass, and $s_W$
is the sinus of the weak mixing angle.\footnote{We abbreviate trigonometric functions by their initials,
 like $s_W\equiv\sin \theta_W$, $c_{2\beta}\equiv\cos (2\beta)$, $t_W\equiv s_W/c_W$, etc.} 
The mixing parameters in the non-diagonal entries read
$$
M^{LR}_b=A_b-\mu\tb\ \ \ ,\ \ \ M^{LR}_t=A_t-\mu/\tb\,\,.
$$
$A_{b,t}$ are the trilinear soft-SUSY-breaking couplings, $\mu$ is the
higgsino mass parameter, and $\tb$ is the ratio between the vacuum
expectation values of the two Higgs doublets, $\tb=v_2/v_1$. The input
parameters in the sfermion sector are then:
\begin{equation}
(\MsfL,M_{\sbottom_R},M_{\stopp_R},A_b,A_t,\mu,\tb)\ \ ,
\label{eq:inputsf}
\end{equation}
for each sfermion doublet. From them, we can derive the masses and
mixing angles:
\begin{equation}
(\msbo,\msbt,\osb)\ ,\ (\msto,\mstt,\ost)\ \ .
\label{eq:outputsf}
\end{equation}
For the trilinear couplings, we require the approximate (necessary) condition
\begin{equation}
A_q^2<3\,(m_{\stopp}^2+m_{\sbottom}^2+M_H^2+\mu^2)\,,
\label{eq:necessary}
\end{equation}
to avoid colour-breaking minima 
in the MSSM scalar potential\,\cite{\vacuum}. Here 
$m_{\squark}$ is of the order of the average squark masses
for $\squark=\stopp,\sbottom$, and $M_H$ is the
  Soft-SUSY-breaking mass parameter of the Higgs fields,
  see~\cite{\vacuum} for details.

Although the tree-level chargino ($\cplus$)-neutralino ($\neut$) sector is well known, we
give here a short description, in order to set our conventions. 
We start by constructing the following set of Weyl spinors:
\begin{equation}
\begin{array}{lcl}
\Gamma^+&\equiv&(-i \tilde W^+,\tilde H_2^+) \,\,,\\
\Gamma^-&\equiv&(-i \tilde W^-,\tilde H_1^-) \,\,,\\
\Gamma^0&\equiv&(-i \tilde B^0,-i \tilde W_3^0,\tilde H_1^0,\tilde H_2^0) \,\,.
\end{array}
\label{eq:inosweak}
\end{equation}
The mass Lagrangian in this basis reads
\begin{equation}
{\cal L}_M=-\frac{1}{2}\begin{pmatrix}\Gamma^+,\Gamma^-\end{pmatrix}
\begin{pmatrix}0&{\cal M}^T\\
{\cal M}&0\end{pmatrix}
\begin{pmatrix} \Gamma^+\\ \Gamma^-\end{pmatrix}
-\frac{1}{2} \begin{pmatrix}\Gamma_1,\Gamma_2,\Gamma_3,\Gamma_4\end{pmatrix}
{\cal M}^0 \begin{pmatrix}\Gamma_1\\ \Gamma_2 \\ \Gamma_3 \\ \Gamma_4\end{pmatrix}
+\mbox{ h.c.}\,\,,
\end{equation}
where we have defined
\begin{eqnarray}
{\cal M}&=&\begin{pmatrix}
M&\sqrt{2} M_W \sbtt \cr
\sqrt{2} M_W \cbtt&\mu
\end{pmatrix}\,\,,\nonumber\\
{\cal M}^0&=&
\begin{pmatrix}
M^\prime&0&M_Z\cbtt\swp&-M_Z\sbtt\swp \cr
0&M&-M_Z\cbtt\cwp&M_Z\sbtt\cwp\cr
M_Z\cbtt\swp&-M_Z\cbtt\cwp&0&-\mu \cr
-M_Z\sbtt\swp&M_Z\sbtt\cwp&-\mu&0
\end{pmatrix}\,\,\,,
\label{eq:massacplusneut}\end{eqnarray}
{with $M$ and $M'$} the $SU(2)_L$ and $U(1)_Y$
soft-SUSY-breaking gaugino 
masses. 
The four-component mass-eigenstate fields are related to the ones
in~(\ref{eq:inosweak}) by 
\begin{equation}
  \chi_i^{+}= \begin{pmatrix}
    V_{ij}\Gamma_j^{+} \cr U_{ij}^{*}\bar{\Gamma}_j^{-}
  \end{pmatrix}
  \; \;\; \;\;,\;\;\;\;\;
  \chi_i^{-}= {\cal C}\bar{\chi_i}^{+T} =\begin{pmatrix}
    U_{ij}\Gamma^{-}_j \cr V_{ij}^{*}\bar{\Gamma}_j^{+} 
  \end{pmatrix}
\ \ , \ \  \chi_{\alpha}^0= \begin{pmatrix}N_{\alpha\beta}\Gamma_{\beta}^0 \cr 
    N_{\alpha\beta}^{*}\bar{\Gamma}_{\beta}^0
  \end{pmatrix}=  
  {\cal C}\bar{\chi}_{\alpha}^{0T}\ ,
  \label{eq:ninos} 
  \nonumber
\end{equation}
where $U$, $V$ and $N$ are in general complex matrices that 
diagonalize the mass-matrices~(\ref{eq:massacplusneut}):
\begin{equation}
\begin{array}{lcccl}
U^* {\cal M} V^\dagger&=&{\cal M}_D&=&{\rm diag}\left(M_1,M_2\right)\,\,(0<M_1<M_2)\,\,,\\
 N^*{\cal M}^0 N^\dagger &=&{\cal M}^0_D&=&
{\rm diag}\left(M_1^0,M_2^0,M_3^0,M_4^0\right)\,\,(0<M_1^0<M_2^0<M_3^0<M_4^0)\,\,.
\end{array}\label{eq:defUVN}
\end{equation}

Using this notation, the tree-level
interaction Lagrangian between fermion-sfermion-(chargino or neutralino)
reads~\cite{Guasch:2002ez}
    \begin{eqnarray}
      \label{eq:Lqsqcn}
      {\cal L}_{\chi \sfr f'}&=&\sum_{a=1,2}\sum_{r} {\cal L}_{\chi_r
      \sfr_a f'} + \mbox{ h.c.}\,\,,\nonumber\\
     {\cal L}_{\chi_r \sfr_a f'}&=& -g\,\sfr_a^* \bar{\chi}_r
      \left(A_{+ar}^{(f)}\pl +  A_{-ar}^{(f)}\pr\right) f'\,\,.
    \end{eqnarray}
Here we have adopted a compact notation, where
$f'$ is either $f$ or its 
$SU(2)_L$ partner for $\chi_r$ being a neutralino or a chargino,
respectively. Roman characters 
   $a,b\ldots$ are reserved for sfermion indices and 
   $i,j,\ldots$ for chargino indices, 
   Greek indices $\alpha,\beta,\ldots$ denote neutralinos,
   Roman indices $r,s\ldots$ indicate either a chargino or a
   neutralino. For example, the top-squark interactions with charginos
   are obtained by replacing $f\to t$, $f'\to b$, $\chi_r\to \cmin_r$,
   $r=1,2$. The coupling matrices that encode the dynamics are given by
    \begin{eqnarray}
        \label{eq:V1Apm}
        \Apit &=& \Rot\Vo^*-\lt\Rtt\Vt^*\, ,\nonumber\\
        \Amit &=& -\lb\Rot\Ut\, ,\nonumber\\
        \Apat &=&\frac{1}{\sqrt{2}} \left(
            \Rot\left(\Nt^*+\YL\tw\No^*\right)
            +\sqrt{2}\lt\Rtt\Nf^*
          \right)\, ,\nonumber\\
        \Amat &=& \frac{1}{\sqrt{2}} \left(
            \sqrt{2}\lt\Rot\Nf
            -\YRt\tw\Rtt\No
            \right)\, ,\nonumber\\
        \Apib &=& \Rob\Uo^*-\lb\Rtb\Ut^*\, ,\nonumber\\
        \Amib &=& -\lt\Rob\Vt\, ,\nonumber\\
        \Apab &=& -\frac{1}{\sqrt{2}} \left(
          \Rob\left(\Nt^*-\YL\tw\No^*\right)
          -\sqrt{2}\lb\Rtb\Nth^*
        \right)\, ,\nonumber\\
        \Amab &=& -\frac{1}{\sqrt{2}} \left(
          -\sqrt{2}\lb\Rob\Nth
          +\YRb\tw\Rtb\No
          \right) \, ,
     \end{eqnarray}
with $Y_{L}$ and $Y_{R}^{t,b}$ the weak hypercharges of the left-handed $SU(2)_{L}$ doublet and right-handed singlet fermion, and $\lambda_{t}=m_{t}/(\sqrt{2}M_{W}\sin\beta)$ and $\lambda_{b}=m_{b}/(\sqrt{2}M_{W}\cos\beta)$ are the Yukawa couplings normalized to the $SU(2)_{L}$ gauge coupling constant $g$. Note the following, each coupling is formed  by two parts: the gaugino parts ($G$), formed exclusively by gauge couplings, and the higgsino part ($H_{\pm}$), which contains factors of the quark masses, each of these parts will receive different kind of corrections. First line from equations (\ref{eq:V1Apm}) can be rewritten as:
\begin{equation}
A^{(t)}_{+ai}=H_{+} + G\,.
\label{twoterms}
\end{equation}

The tree-level partial partial decay width reads
\begin{eqnarray}
\Gamma_{ar}^{Tree}=\Gamma^{Tree}(\sfr_{a}\rightarrow f^{\prime}\chi_{r})=\frac{g^{2}}{16\pi m^{3}_{\sfr_{a}}}\lambda(m^{2}_{\sfr_{a}},M^{2}_{r},m_{f^{\prime}}^{2}) \times \nonumber \\
\times \left[ (m^{2}_{\sfr_{a}}-M^{2}_{r}-m_{f^{\prime}}^{2}) \left( |A^{(f)}_{+ar}|^{2}+|A^{(f)}_{-ar}|^{2} \right) 
-4m_{f^{\prime}}M_{r}\Re\left( A^{(f)}_{+ar}A^{(f)\ast}_{-ar} \right)
\right]
\end{eqnarray}
with $\lambda(x^{2},y^{2},z^{2})=\sqrt{[ x^{2} - (y-z)^{2}][ x^{2} - (y+z)^{2}]}$.

\section{QCD Corrections}
\label{sec:qcd-corrections}

It is known that QCD corrections
to the squark partial decay widths into charginos and
neutralinos can be numerically large, specially in
certain regions of the parameter space~\cite{Guasch:2002ez}. 
An effective description of squark/chargino/neutralino couplings, 
simple to write and to introduce in computer codes, was given in~\cite{Guasch:2008fs}.
The complete one-loop corrections to squark partial decay
  widths are already available~\cite{Guasch:2002ez,SFdecay}\footnote{Ref.~\cite{Hlucha:2011yk}
  provides a computer program to compute the sfermion two-body partial
  decay widths at the one-loop level in the $\overline{DR}$
  renormalization scheme. The comparison with Ref.~\cite{SFdecay} is not
  straightforward due to the use of different renormalization schemes.}, 
but their
complicated expressions are not suitable for the introduction in
Monte-Carlo programs used for experimental analysis.  
In this article we present the results of the implementation of the effective description in MadGraph's MSSM framework. 
While the EW corrections can be important in some regions of the
parameter space, they do not admit a simple effective description as
the one described here, and therefore one can not compute an
approximation to the cross-section as the one performed in the present work.

In the following we present few details of the effective description approach
as given in \cite{Guasch:2008fs}. In this approach, following hints from Higgs-boson physics, an effective Yukawa coupling is defined as:
\begin{equation}
\lambda_{b}^{eff}\equiv \frac{m_{b}^{eff}}{v_{1}}\equiv \frac{m_{b}(Q)}{v_{1}(1+\Delta m_{b})}, \hspace{0.5cm}
\lambda_{t}^{eff}\equiv \frac{m_{t}^{eff}}{v_{2}}\equiv \frac{m_{t}(Q)}{v_{2}(1+\Delta m_{t})}
\label{eq:hbhbteff1}
\end{equation}
where $m_{q}(Q)$ is the running quark mass and $\Delta m_{q}$ is the finite threshold correction. The SUSY-QCD contributions to $\Delta m_{q}$ are
\begin{eqnarray}
\label{eq:deltamq}
\Delta m_{b}^{SQCD}&=&\frac{2\alpha_{s}}{3\pi}\mg\mu\tan\beta\, I(m_{\sbottom_{1}},m_{\sbottom_{2}},\mg), \nonumber \\
\Delta m_{t}^{SQCD}&=&\frac{2\alpha_{s}}{3\pi}\mg\frac{\mu}{\tan\beta}\, I(m_{\stopp_{1}},m_{\stopp_{2}},\mg)
\end{eqnarray}
where $I(a,b,c)$ is the scalar three-point function at zero momentum transfer:
\begin{equation}
I(a,b,c)=\frac{a^{2}b^{2}\ln{(a^{2}/b^{2})}+b^{2}c^{2}\ln{(b^{2}/c^{2})+a^{2}c^{2}\ln{(c^{2}/a^{2})}}}
{(a^{2}-b^{2})(b^{2}-c^{2})(a^{2}-c^{2})}
\end{equation} 
The effective description of the squark interaction consist in
replacing the
tree-level quark masses in the couplings~(\ref{eq:V1Apm}) by the effective
Yukawa couplings of eq.~(\ref{eq:hbhbteff1}), and use this lagrangian to
compute the partial decay width, schematically
$\Gamma^{Yuk-eff}=\Gamma^{Tree}(m_q^{eff})$ (see~\cite{Guasch:2008fs} for details). 
This expression contains the large one-loop corrections from the
finite threshold corrections~(\ref{eq:hbhbteff1}), but it also contains
higher order corrections. At this point one can 
make a computation that combines the higher order effects (which ignore
the effects of external momenta) and the fixed one-loop (which ignore
the higher order effects). At the same time, this will allow us to
quantify the degree of accuracy obtained by the effective
description. A \textit{Yukawa-improved decay width computation} have been
defined in ~\cite{Guasch:2008fs} and they showed that the effective description 
using just the Yukawa threshold corrections~(\ref{eq:hbhbteff1}) 
is not enough for the squark decay
widths description. The one-loop corrections develop a term which grows
as the gluino mass $\mg$~\cite{Hikasa:1995bw}, 
which is absent in the effective Yukawa couplings~(\ref{eq:hbhbteff1}).
Therefore, the QCD corrections to squark decay widths produce explicit
non-decoupling terms of the sort $\log \mg$.
To understand those terms a renormalization group analysis is in 
order~\cite{Guasch:2008fs}. They constructed an
effective theory below the gluino mass scale, which contains only
squarks, quarks, charginos, neutralinos and gluons in the light sector
of the theory, and integrate out the gluino contributions. Then, they
found out
the renormalization group equations (RGE) of the gaugino and higgsino
couplings, and performed the matching with the full MSSM couplings at the
gluino mass scale $\mg$. Only the logarithmic RGE effects have been
considered, neglecting the possible threshold effects at the gluino mass scale.
Since the effective theory does not contain
gluinos, only the contributions from the gluon have to be taken into
account. We do not present here details of the renormalization group analysis
and we restrict ourselves to give the results that are relevant for our purpose.

The squark-chargino  running coupling constant as a
function of the gauge and Higgs boson couplings at the
renormalization scale $Q$, up to $\mathcal{O}(\alpha_{s})$, is given 
by~\cite{Guasch:2008fs}:
\begin{equation}
A^{(t)}_{+ai}(Q) \simeq H_{+}(m_{q}(Q))\left( 1+ \frac{\alpha_{s}(Q)}{\pi}\log\frac{Q}{\mg} \right) + G\left( 1- \frac{\alpha_{s}(Q)}{\pi}\log\frac{Q}{\mg} \right)\,.
\label{effectivecoup}
\end{equation}
 The $\log \mg$ term arises as a factor to each higgsino 
($H_{\pm}$) or gaugino ($G$) term, eq.~(\ref{twoterms}). Note that the expressions for the 
higgsino and gaugino couplings are different due to the different 
running of the gauge and Higgs-boson couplings 
between the scales $\mg$ and $Q$. 
The same can be safely extended to the rest of the couplings in (\ref{eq:V1Apm}). 

Then, the renormalization group running of the coupling constant can be
summarized as follows: we can use effective gaugino and higgsino
couplings given by~\cite{Guasch:2008fs},
\begin{eqnarray}
g^{eff}(Q)&=&g\,\left(\frac{\al(Q)}{\al(\mg)}\right)^\frac{2}{\beta_{0}}\simeq g\left( 1-\frac{\al(Q)}{\pi}\log  \frac{Q}{\mg}\right)\ \ , \nonumber\\
\tilde{\lambda}_{b,t}^{eff}(Q)&=&\lambda_{b,t}^{eff}(Q)\,\left(\frac{\al(Q)}{\al(\mg)}\right)^\frac{-2}{\beta_{0}}
\simeq\lambda_{b,t}^{eff}(Q)\left( 1+\frac{\al(Q)}{\pi}\log \frac{Q}{\mg}\right)\ \ ,
\label{eq:logterms}
\end{eqnarray} 
where $\lambda^{eff}(Q)$ are the effective Yukawa couplings
  defined in~(\ref{eq:hbhbteff1}), and $\beta_0$ is the QCD $\beta$-function. At this point, a simple expression for 
the effective description of squark/chargino/  neu\-tra\-lino couplings is given
by eqs.~(\ref{eq:hbhbteff1}), (\ref{eq:deltamq}), (\ref{eq:logterms}).
Ref.\cite{Guasch:2008fs} showed that the effective
description~(\ref{eq:logterms}) approximates the \textit{improved}
computation to within $2-5\%$ for large enough gluino masses
($\mg\gtrsim 1\TeV$). The effects of the
log-terms are better visible in the gaugino-like channels, where the
Yukawa couplings play no role, and the bulk of the corrections
corresponds to the log-terms. In the
higgsino-like channels their importance is less apparent.
After introducing these expressions in computer codes a reasonable description for squark
decays into charginos and neutralinos is accomplished. They can be used in Monte-Carlo
generators and other computer programs that provide predictions for the
LHC and the ILC to improve their accuracy, requiring little computational costs.

\section{Monte-Carlo Tool}
\label{sec:MC}

The effective couplings of section~\ref{sec:qcd-corrections}
have been implemented in a standard Monte-Carlo tool to allow their use
in multiple processes. We have chosen to include them in the
MadGraph/MadEvent framework~\cite{Stelzer:1994ta,Alwall:2007st,Alwall:2011uj,Alwall:2008pm}.

The program MadGraph~\cite{Stelzer:1994ta} automatically generates
Fortran code to calculate arbitrary tree-level helicity amplitudes. 
Its later implementation, the MadGraph/MadEvent
package~\cite{Alwall:2007st,Alwall:2011uj}, allows the computation of physical
processes at the partonic or hadronic level, and includes interfaces to
hadronization routines. Recent additions allow also to compute the
partial decay widths of unstable particles~\cite{Alwall:2008pm}. A
number of physical models are supplied with the standard version of
MadGraph/MadEvent, aside from the SM, we will be using the
MSSM implementation~\cite{Cho:2006sx}. For our computation, we use
MadGraph/MadEvent 5.1.4.8~\cite{Alwall:2011uj}, and we make heavy use of its
possibility to extend/modify internal physical models. MadGraph accepts
inputs in the form of  SUSY Les Houches Accord
(SLHA)~\cite{Skands:2003cj} file format, which allows easy interfacing
with programs computing SUSY spectra and SUSY particles decay widths. 

We have modified MadGraph's standard MSSM file, changing the
squark-quark-neu\-tra\-li\-no/char\-gi\-no vertices by the ones of
expression~(\ref{eq:logterms}). Using our model file, we are able to
compute any physical processes involving these vertices,
including the leading radiative corrections.

In section~\ref{resultsGamma} we compute the partial decay widths of
squarks, using the recent addition to MadGraph implemented
in~\cite{Alwall:2008pm}. These results are checked against an
independent implementation of the partial decay widths in programs based
in the
Feyn\-Arts/Form\-Calc/Loop\-Tools~\cite{Hahn:2000kx,Hahn:2001rv,Hahn:1998yk}
(FAFCLT) chain, which were used in Ref.\cite{Guasch:2008fs}. Our set of
FAFCLT-based programs prepares a standard SLHA~\cite{Skands:2003cj}
input file, which contains the SUSY inputs, as well as the SUSY spectrum
and mixing matrices. This SLHA file is then used as input for both, the
MadGraph-based and the FAFCLT-based computation. In this way we can make
meaningful comparisons of the output of both computations. We find that
both computations agree among them (see section \ref{resultsGamma}) and
with Ref.\cite{Guasch:2008fs}.

In section~\ref{sec:crosssection} we use our MadGraph-based programs to
compute the production cross-section of squarks decaying into
chargino/neutralino. We generate an SLHA input file with our set of
FAFCLT-based programs, which we use as input for MadGraph. 
MadGraph needs the full decay width of on-shell particles for the
treatment of resonant propagators, while our decay programs (FAFCLT- or
MadGraph-based) can only compute the fermionic decay channels. We use SDECAY
1.1a~\cite{Muhlleitner:2003vg} for the computation of the bosonic decay
channels partial decay widths. All along the computation a single SLHA
file is used for all steps, ensuring consistency among the different
inputs/outputs. The computing flow is the following:
\begin{enumerate}
\item we use the FAFCLT-based programs to write a standard
    SLHA input file, which includes the SUSY inputs
    as well as the SUSY particle masses and mixing matrices;
\item we use this file as input for
    SDECAY to obtain the 
    squark decay tables at the tree-level;
\item then, we use our set of FAFCLT-based programs to replace,
  in the above file, the squark partial decay widths into
  charginos/neutralinos, and the full decay width, by using the
  effective coupling approach\footnote{The result would be the same if
    we used our MadGraph-based programs, since we have checked that we
    obtain the same results. It is a matter of practical convenience.}; 
\item this SLHA file is used as an input for the
    MadGraph/MadEvent cross-section computation, which uses the input
    section for its internal parameter setup, and the full decay width
    in the resonant particles propagators.  
\end{enumerate}

In this way a cross-section at the parton level in the final state is obtained. 
MadGraph/MadEvent interfaces may
process further this output through parton showers,
hadronization and detector simulation packages, however this analysis is beyond
the scope of the present work.

\section{Numerical analysis}
\label{sec:resultsnumerical}

\subsection{Numerical set up}
\label{sec:numericalsetup}
Our programs are able to perform computations for any MSSM
parameter space point. They admit SLHA\cite{Skands:2003cj} input for
easy interaction with other programs/routines.
As an example of the effects of the new included terms, we
  will show numerical analysis for fixed values of the
SUSY parameters, and make plots by changing one parameter at a time. 
On one side, we choose a set of SUSY parameters as described below
(see eqs.~(\ref{eq:SUSYparams}) and (\ref{eq:physicalmasses})) and, 
on the other side, SUSY parameters are chosen from the
  {\it{Snowmass Points and Slopes}} (SPS)~\cite{Allanach:2002nj} set.
It should be noted that some of the SPS scenarios are already
in conflict with LHC data~\cite{Dolan:2011ie,AbdusSalam:2011fc}, but we
have chosen them as reference points to perform our computation since
they have been studied at length in the literature.
The results for the decay widths serve as a test of correct
  implementation, they are checked against previous computer
  programs\cite{Guasch:2008fs}.

For the SM
parameters we use $\mt=172\GeV$, $\mb=4.7\GeV$, $\al(\mz)=0.1172$,
 $s_{W}^2=0.221$, $\mz=91.1875\GeV$, $1/\alpha=137.035989$. 
The renormalization scale $Q$ is taken to be the physical mass of the
decaying squark.

First, we take for the central values of the parameters:
\begin{equation}
\begin{array}{c}
\tb=5\ ,\ 
\mu=300\GeV\ ,\ 
M=200 \GeV\ ,\ 
\MsfL=800\GeV\ ,\ \mg=3000\GeV \ ,\\
\msusy\equiv \MsfR=1000\GeV\ ,\ 
A_{t}=A_{b}=2\MsfL+\mu/\tb=1660\GeV\ ,
\end{array}
\label{eq:SUSYparams}
\end{equation} 
where we have introduced a parameter $\msusy$ as a shortcut for all the
SUSY mass parameters which are not explicitly given. 
We use the grand unification relation $M'=5/3 \,\tw^2 \,M$ for the bino mass parameter. 
The value of the trilinear couplings $A_{b,t}$ is given by the algebraic
expression, the given numerical value corresponds to
the default values of the other parameters, this numerical value will
change in the plots, the chosen expression allows to show plots with a
significant parameter variation avoiding colour-breaking-vacuum
conditions~(\ref{eq:necessary}).
The gluino mass is chosen to be large to enhance the effects of the 
logarithmic terms. Note also that if the gluino decay channel is open,
it will be the dominant decay channel for squarks, rendering the
chargino/neutralino channels phenomenologically irrelevant. Therefore
our region of interest is:
\begin{equation}
\mg+m_q > \msq \,\,.
\label{eq:gluinocondition}
\end{equation}
With
these input parameters, the central values for the physical SUSY particle
masses are:
\begin{eqnarray}
  M_{\cplus}&=&(170.40, 337.50)\GeV\ \ ,\nonumber\\
  M_{\neut}&=&(89.52, 172.28, 305.46, 338.58)\GeV\ \ ,\nonumber\\
  m_{\sbottom}&=&(802.05, 1000.30)\GeV \ \ , \nonumber\\
  m_{\stopp}&=&(720.00, 1084.87)\GeV \ \ .
  \label{eq:physicalmasses}  
\end{eqnarray}
Of course, the parameters in eq.~(\ref{eq:SUSYparams}) are just an
example for illustrative purposes. 
We denote this parameter choice ``\textit{Def}'' in the
  numerical analysis below.
In addition we will be analyzing the following parameter space
points from the SPS\cite{Allanach:2002nj} collection: 1b, 3, 7 (see below). The
corresponding values of the parameters are reproduced in
Appendix~\ref{sec:SPS}.
We have checked that our conclusions
hold for a wide range of the parameter space.

At present the ATLAS collaboration has already put some limits
on the top-squark mass from pair
production\cite{Collaboration:2012si,Collaboration:2012ar,Collaboration:2012as,ATLAS:2012st,ATLAS:2012st2}, the largest
excluded top-squark mass at 95\% C.L. is $\sim
465\GeV$\cite{Collaboration:2012si} assuming 
$BR(\stopp_1\to t\neut_1)=1$. Our top-squark mass parameter choices are
larger than the excluded ones, moreover the exclusion limits would be
loosened by allowing the existence of several decay channels for the
top-squark.

\subsection{Partial Decay Widths }
\label{resultsGamma}

Our aim is the computation of the production cross-section, however,
some of the numerical features of the production cross-section are
traceable to the numerical behaviour of the partial and total decay
widths. For this reason we show in the present section a short
numerical analysis of the partial decay widths of top-squarks into
charginos and neutralinos, following the computation and setup of the
previous sections. We show detailed results only for the SPS
scenarios, since the \textit{Def} scenario~(\ref{eq:SUSYparams}) has been
largely explored in~\cite{Guasch:2008fs}.

First of all, we have made a complete check of our FAFCLT-based
and MadGraph-based computations, both at the tree-level and effective
approximations, and we have found perfect agreement between both methods.
We have checked explicitly that for all possible squarks
decay channels into charginos and neutralinos the relative deviations
between both methods of calculations is within the precision of
MadGraph's Monte-Carlo numerical integrations (smaller than $0.08\%$, and
typically of the order of $0.02\%$).

The analysis of the accuracy of the effective approximation was
performed in Ref.\cite{Guasch:2008fs}, using the parameter set
\textit{Def}~(\ref{eq:SUSYparams}) as an example. We have
additionally checked that the same conclusions hold for the SPS
scenarios analyzed in the present work, namely that the effective
approximation provides a good description of the radiative-corrected
partial decay widths, if the gluino mass is heavier than the
top-squark mass ($\mg\gtrsim 800\GeV$).

For future reference, we comment briefly on the partial decay widths
in the \textit{Def}~(\ref{eq:SUSYparams}) scenario without showing any
plots, details are available in Ref.~\cite{Guasch:2008fs}. In this
scenario, in the explored range of gluino masses, the corrections shrink in more than a 
$10\%$ the full decay width of squarks compared to the tree-level result, 
and the effect is highly noticed for $\tan\beta>35$ growing up to a $40\%$ in some regions of the 
parameter space. As expected, the largest difference with respect to the
tree-level calculation is 
achieved at the largest gluino mass evaluated.

\begin{figure}
\centering
\begin{tabular}{cc}
\includegraphics[width=8.0cm]{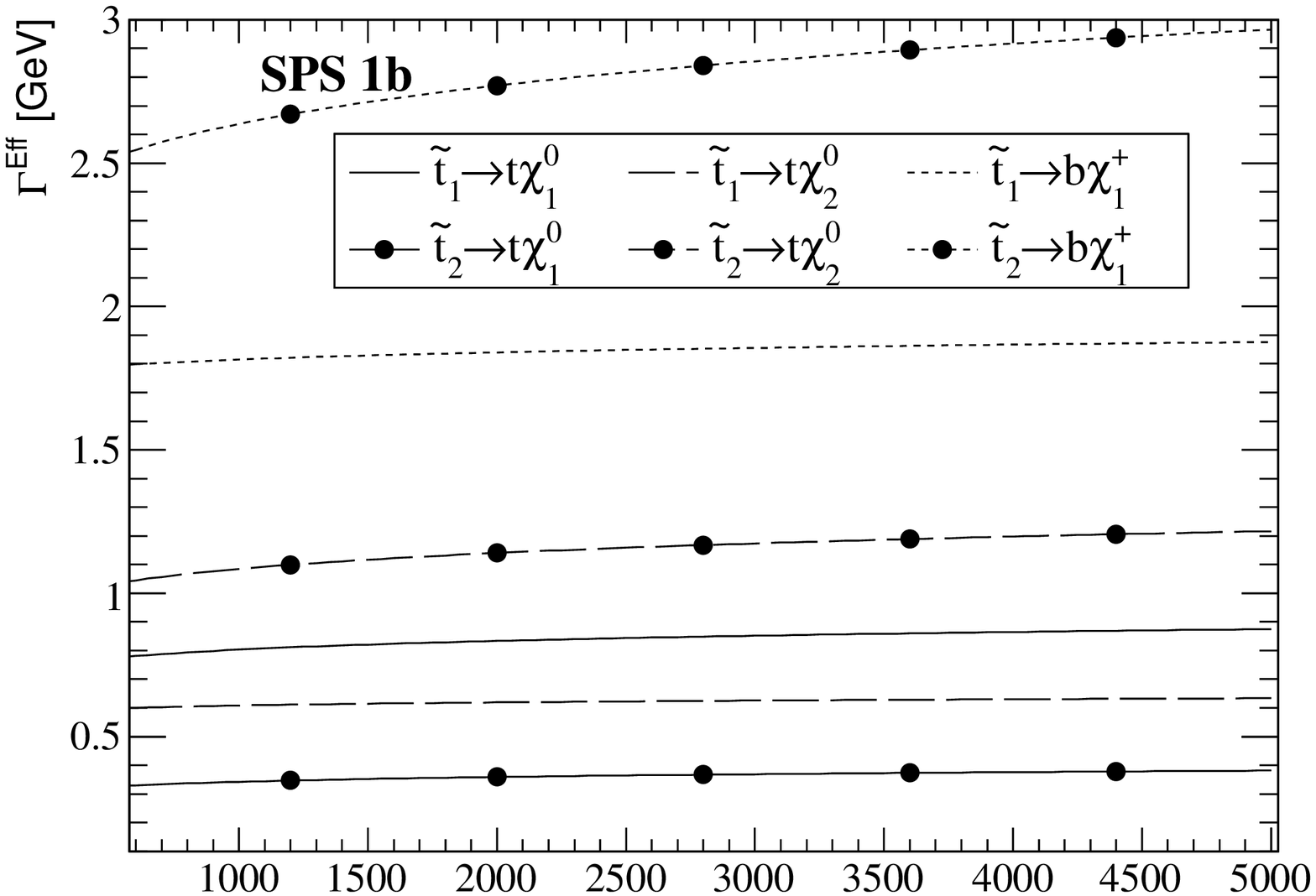}&
\includegraphics[width=8.0cm]{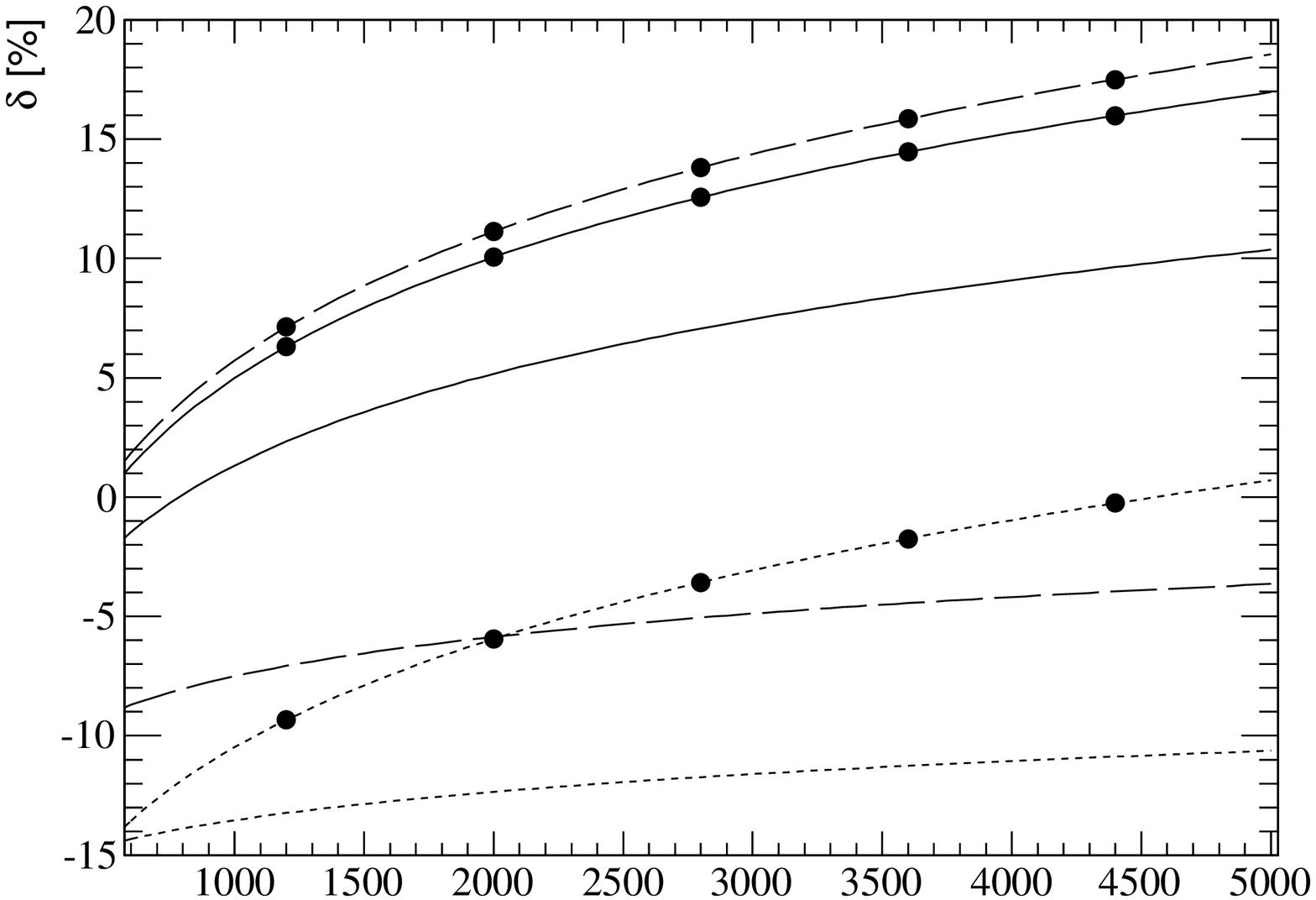}\\
\includegraphics[width=8.0cm]{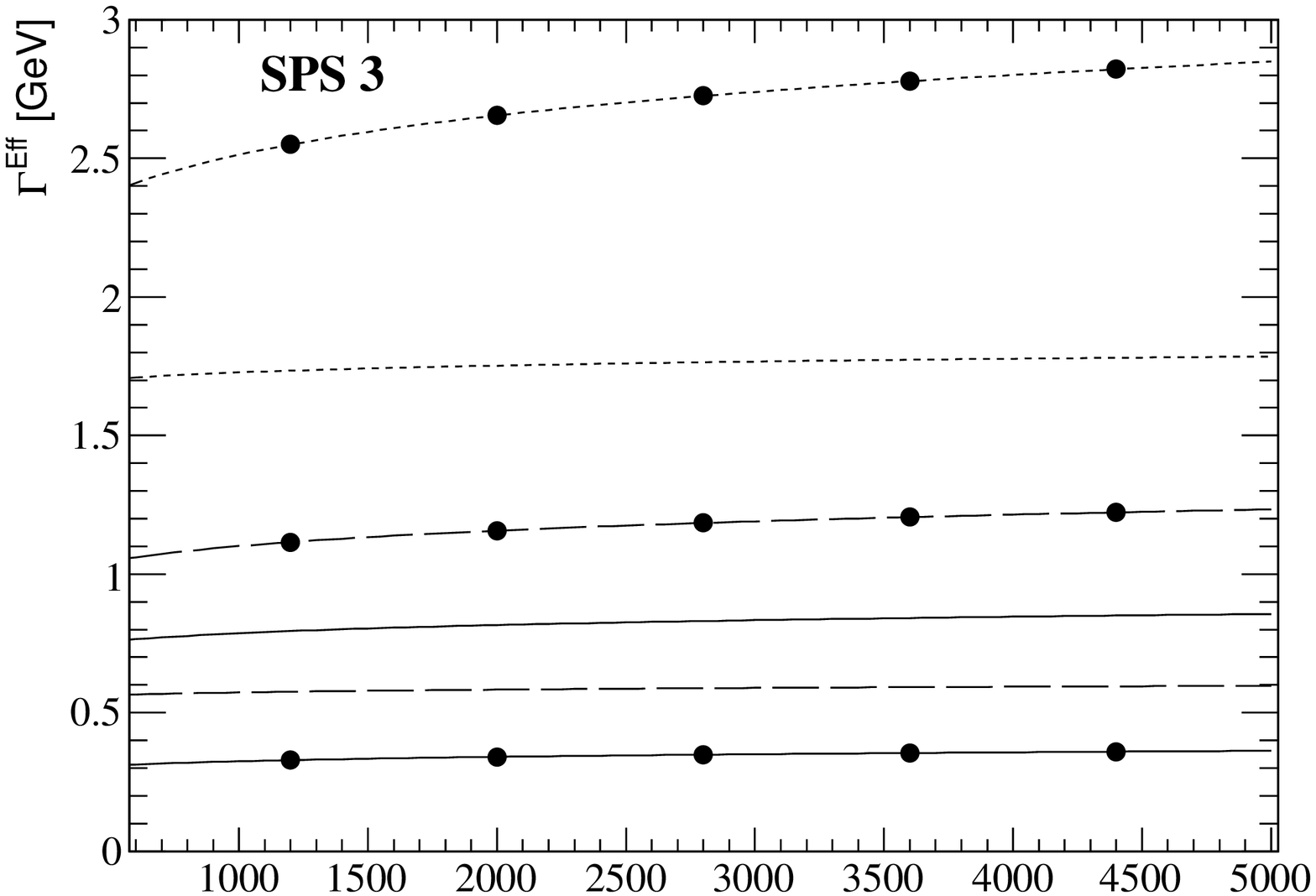}&
\includegraphics[width=8.0cm]{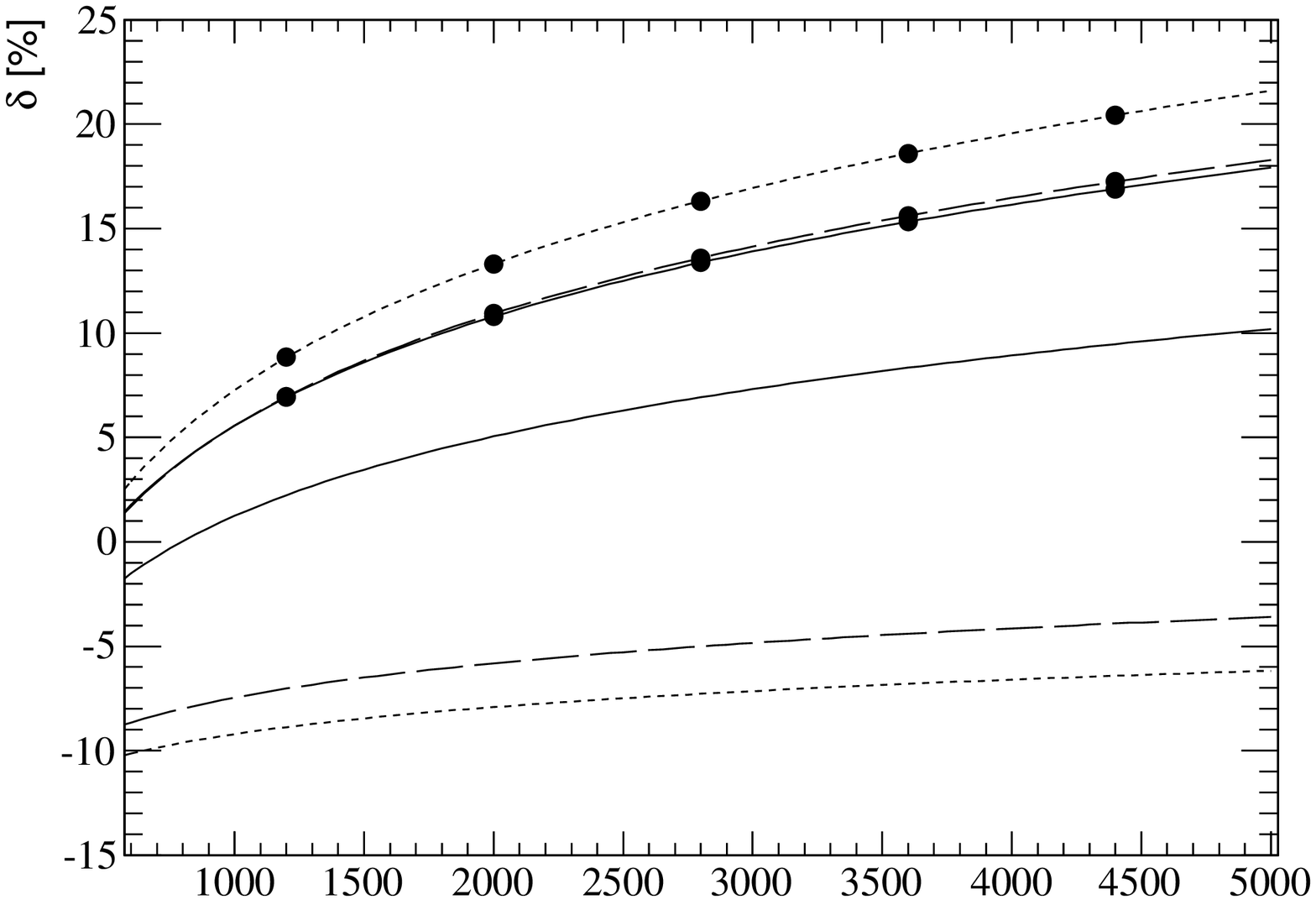}\\
\includegraphics[width=8.0cm]{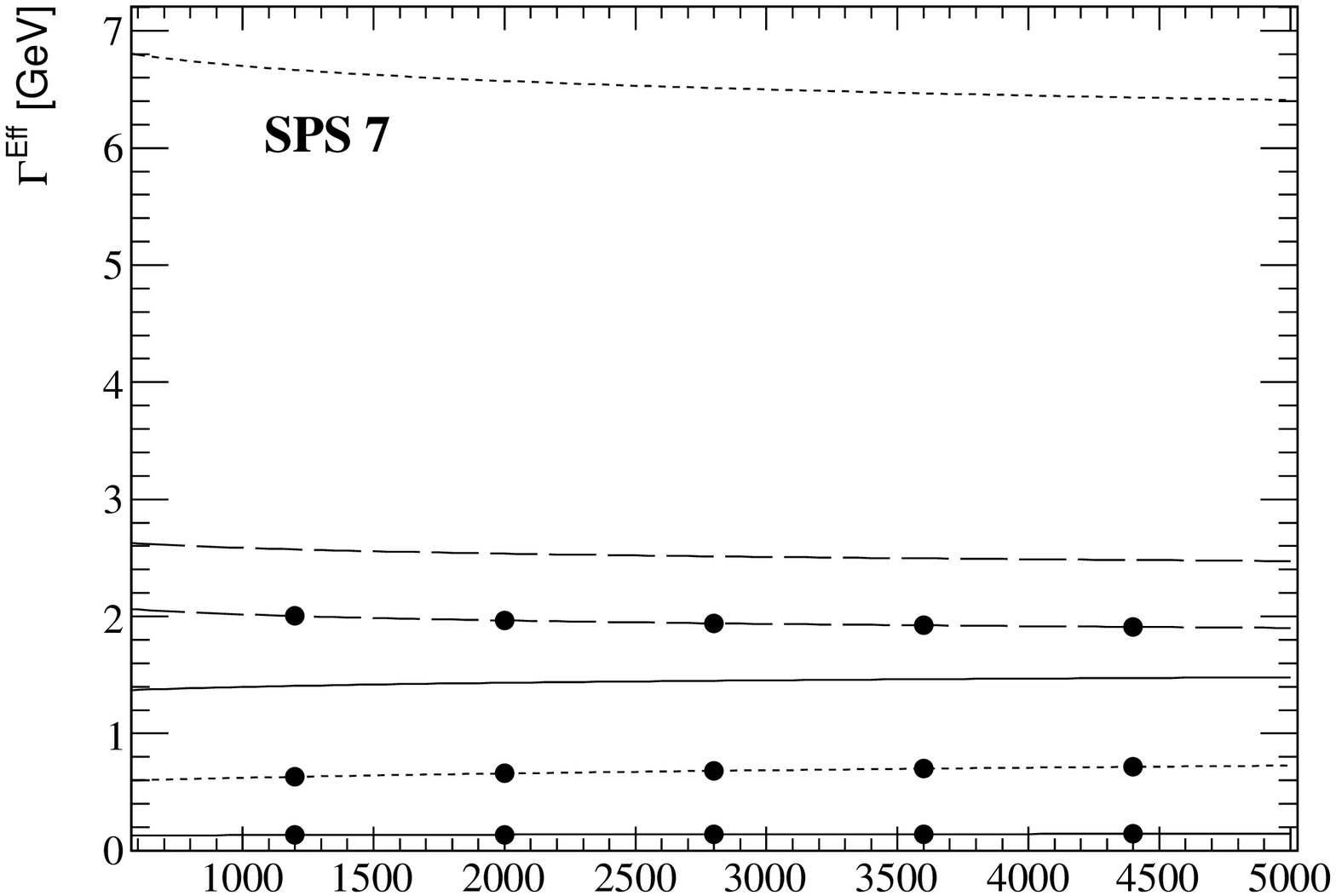}&
\includegraphics[width=8.0cm]{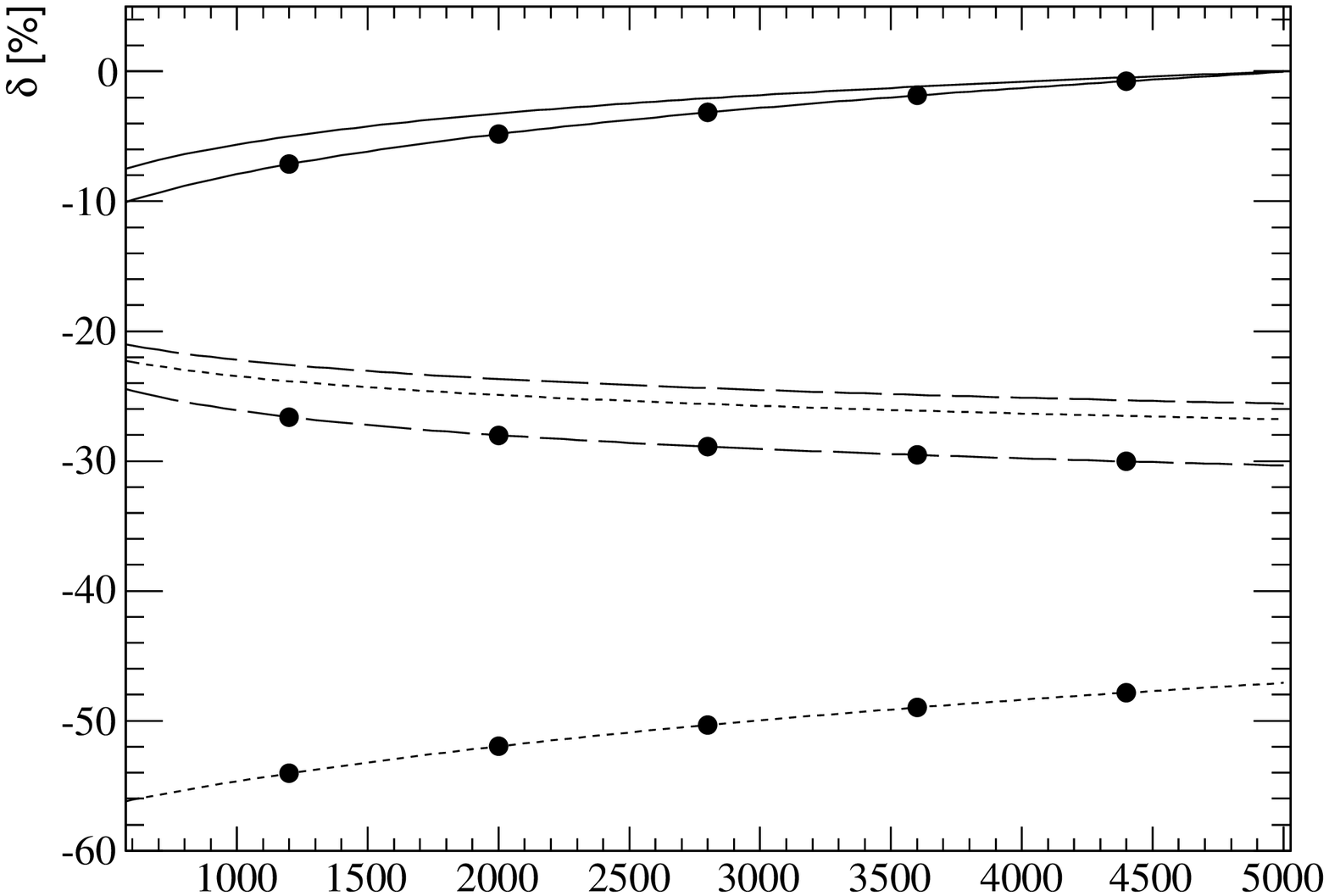}\\
\multicolumn{2}{c}{\small $\mg\ [{\rm GeV}]$}\\
(a)&(b)\\
\end{tabular}
\caption{Effective approximation prediction (\textit{Eff})
for top-squarks, $\tilde{t}_{1}$ and $\tilde{t}_{2}$, 
\textbf{(a)} partial decay widths into $\chi_{1}^{0}$ (solid lines),
$\chi_{2}^{0}$ (dashed lines) and $\chi_{1}^{+}$ (dotted lines);
  and \textbf{(b)} relative corrections, $\delta$ eq.~(\ref{eq:devcorrec}),
  in \%, as a function of $\mg$, for SPS1b (first row), SPS 3 (second row) 
  and SPS 7 (third row) scenarios.}
\label{partialsps1b37-mgl}
\end{figure}

\begin{figure}
\centering
\begin{tabular}{cc}
\includegraphics[width=8.0cm]{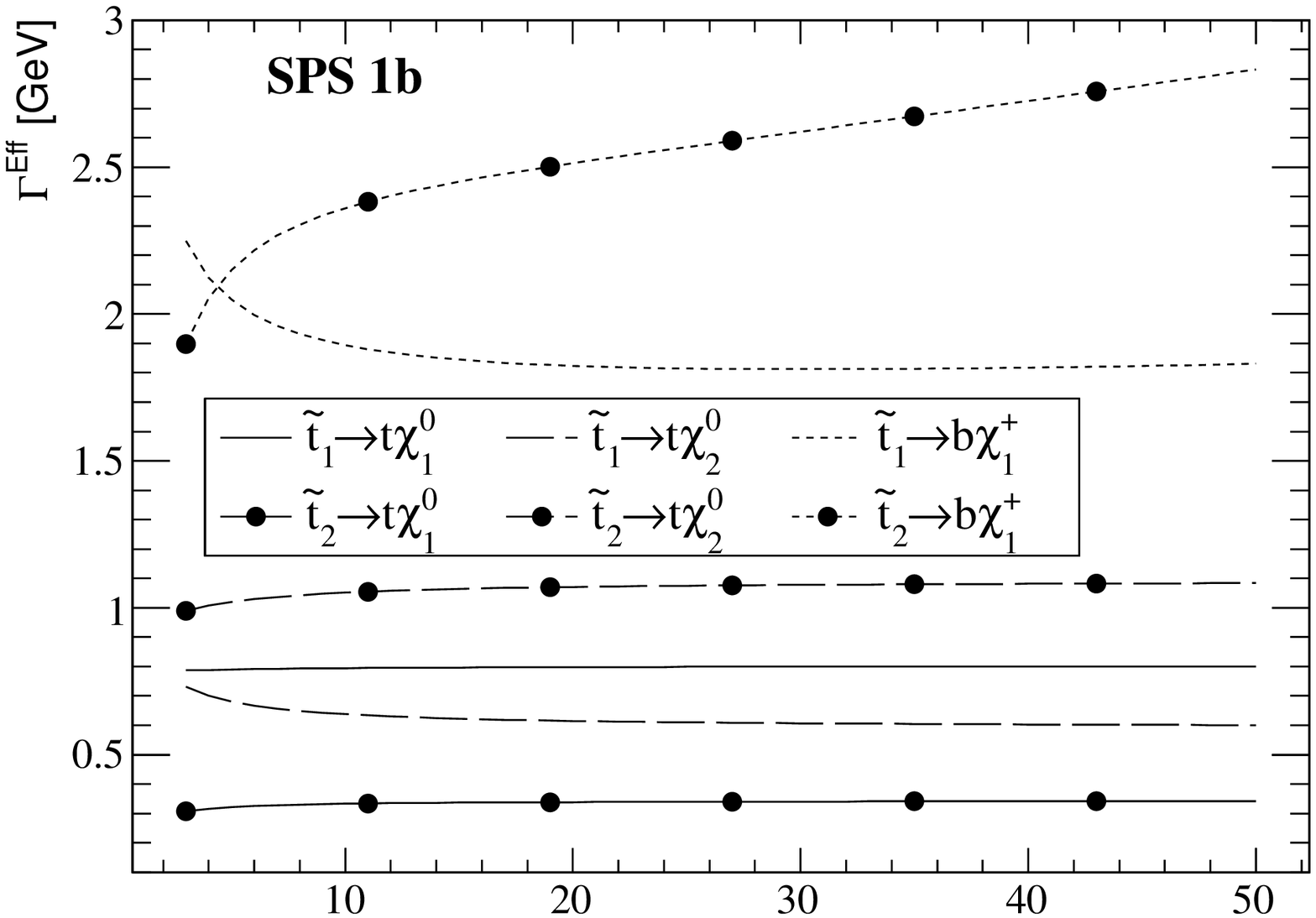}&
\includegraphics[width=8.0cm]{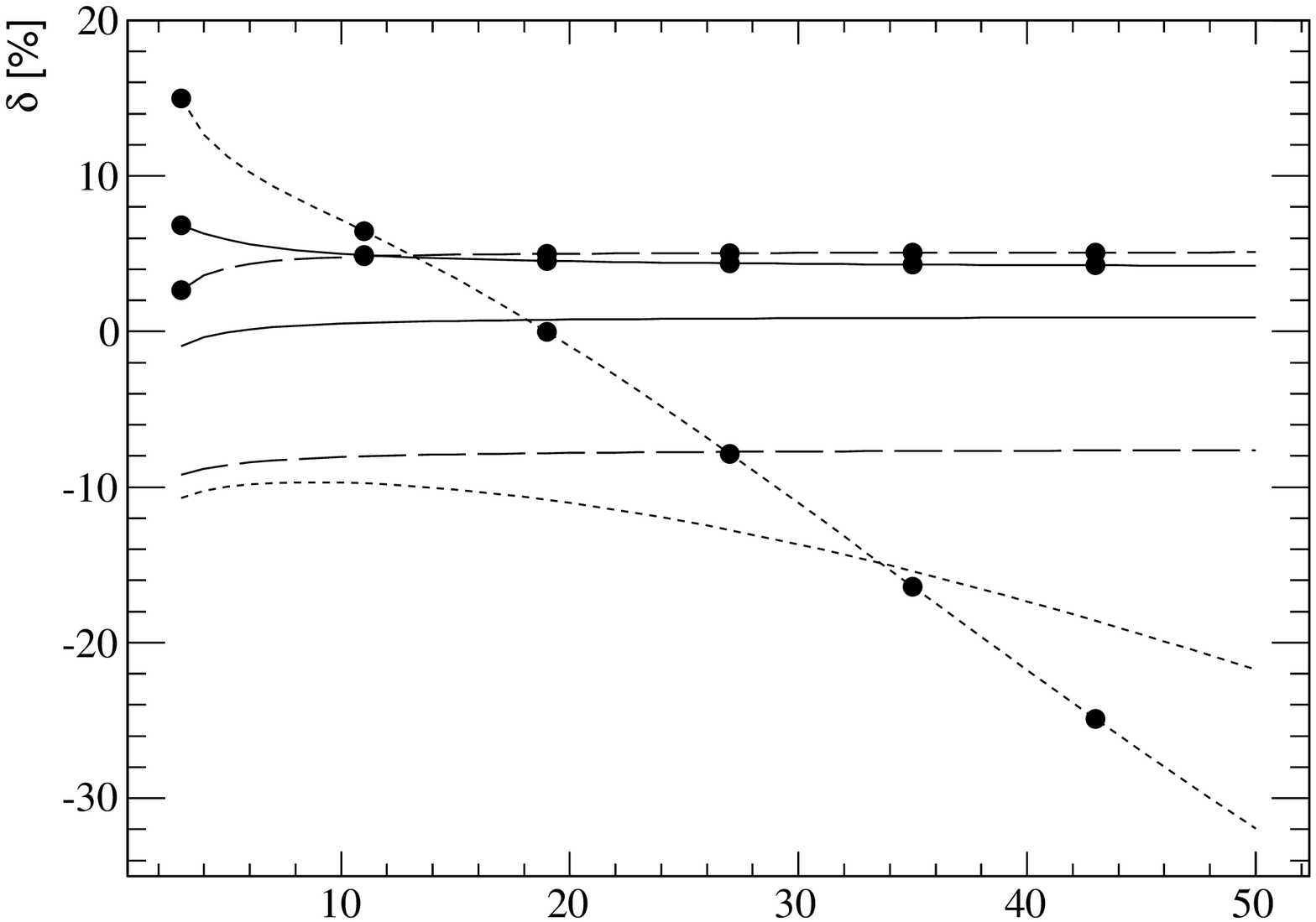}\\
\includegraphics[width=8.0cm]{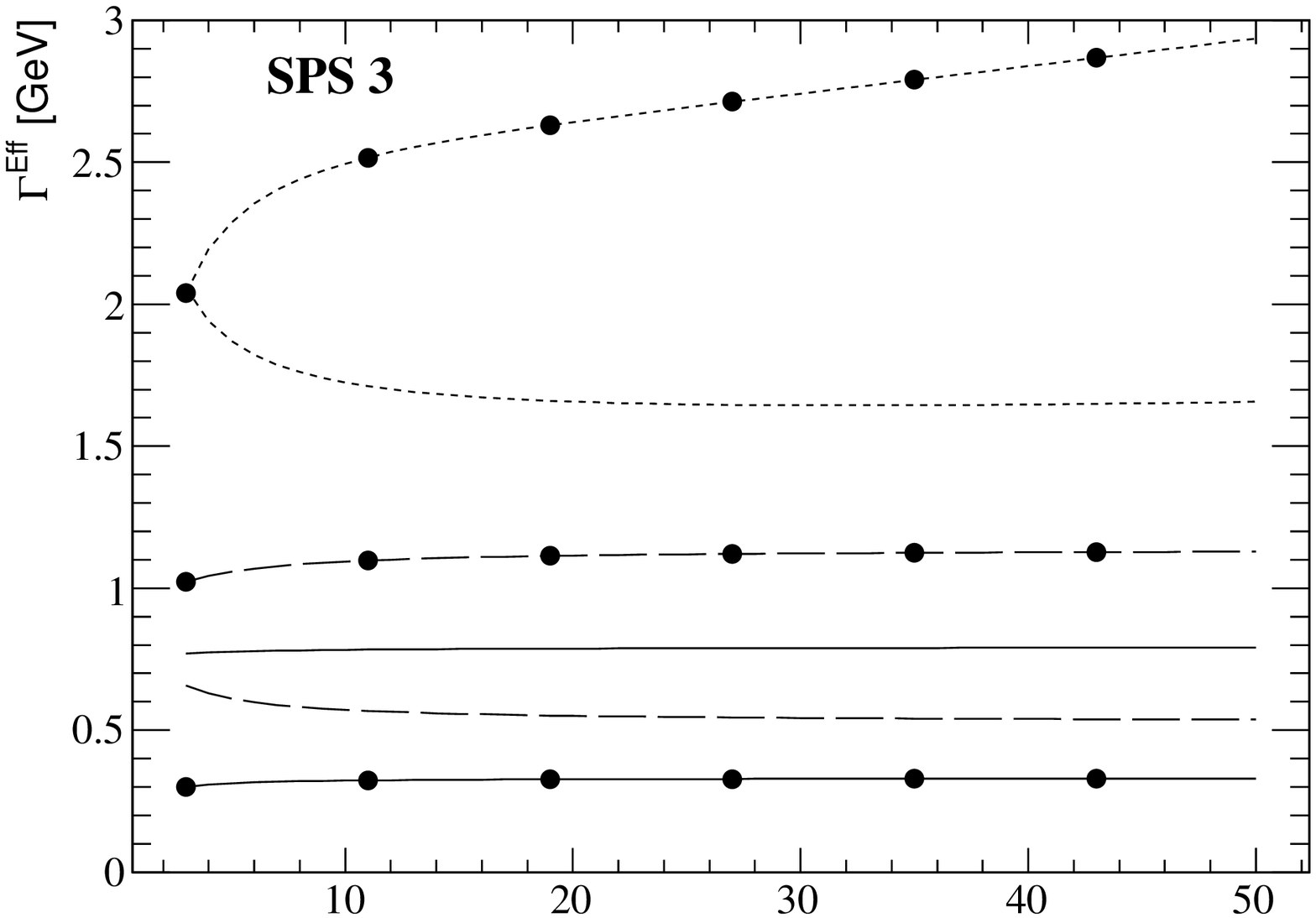}&
\includegraphics[width=8.0cm]{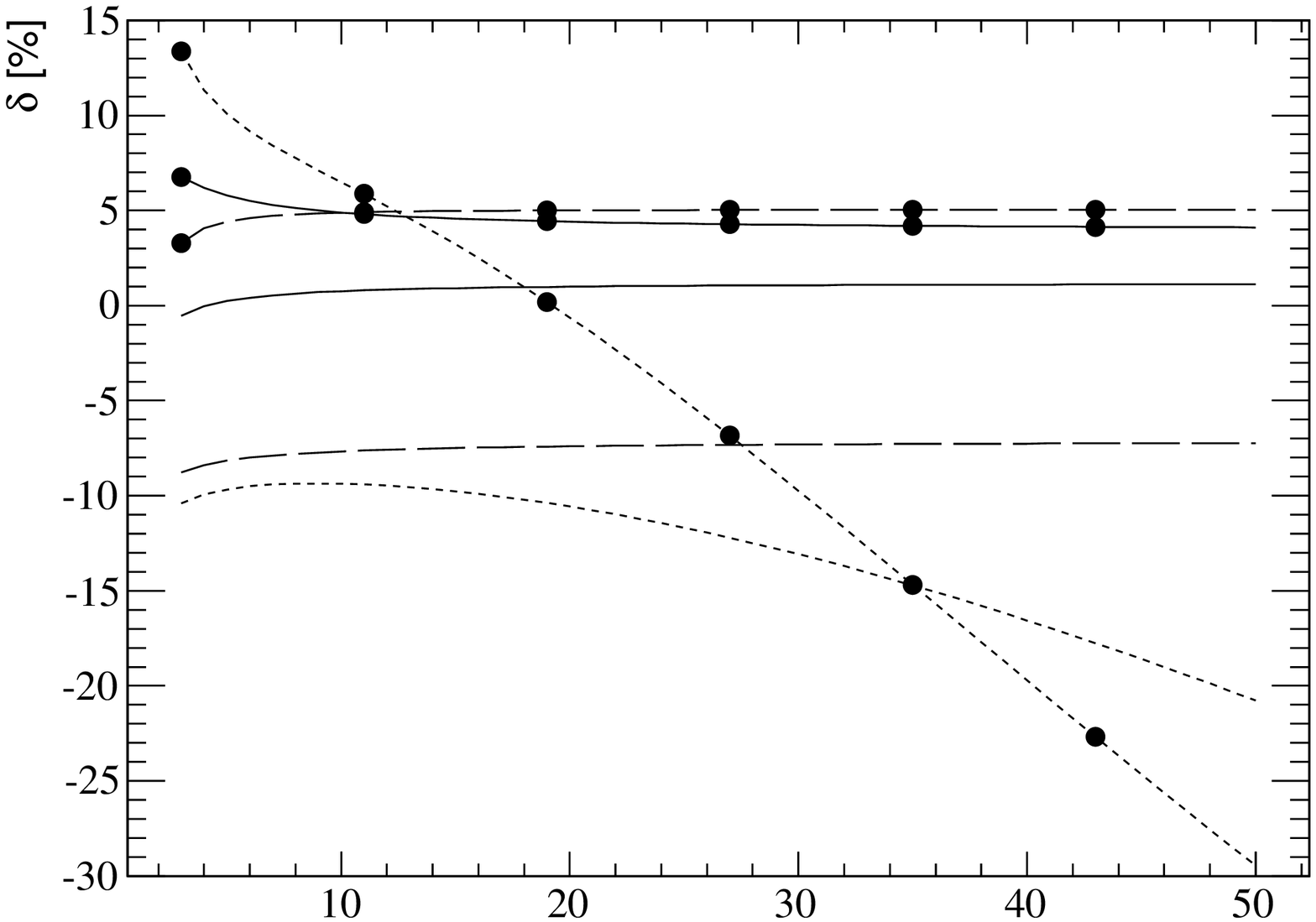}\\
\includegraphics[width=8.0cm]{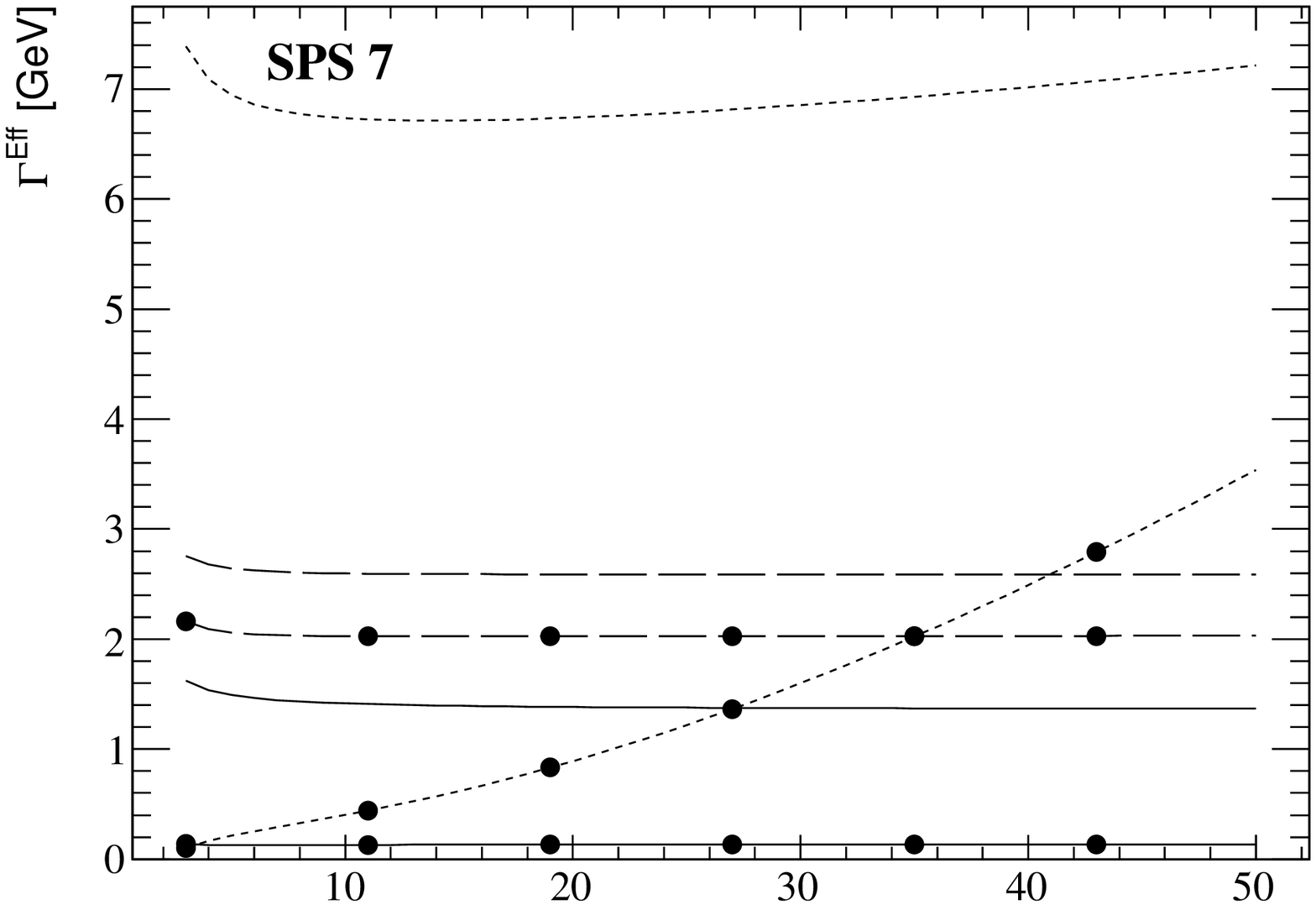}&
\includegraphics[width=8.0cm]{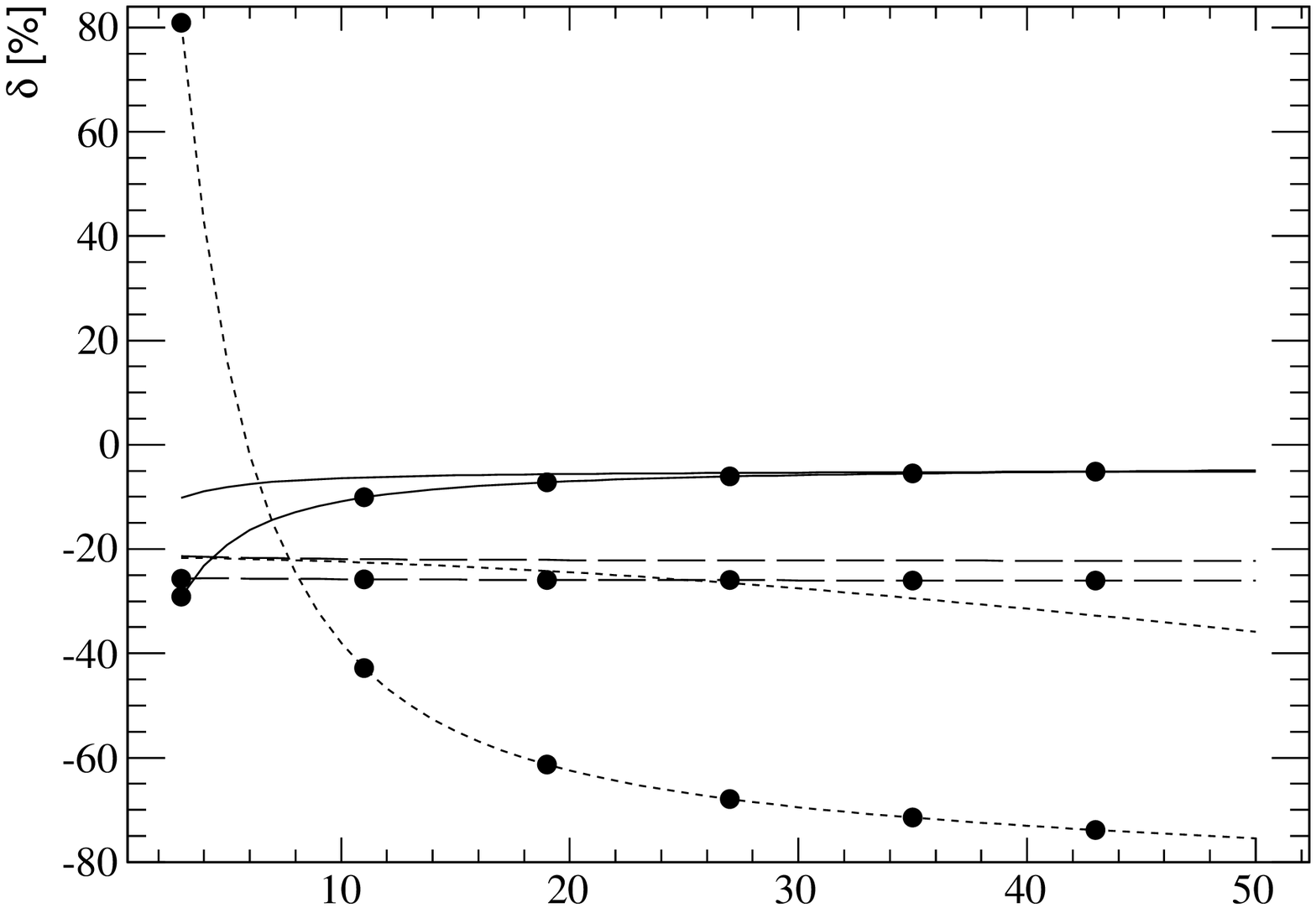}\\
\multicolumn{2}{c}{\small {$\tan\beta$}}\\
(a)&(b)\\
\end{tabular}
\caption{Same as Figure~\ref{partialsps1b37-mgl}, but as a function of $\tb$.}
\label{partialsps1b37-tb}
\end{figure}

We have made a complete numerical analysis for the Snowmass
Points and Slopes parameters: SPS1b, SPS3, SPS5 and SPS7.
SPS5 provides a very narrow range of useful SUSY parameters for our investigation 
due to a relatively light scalar top quark, therefore we are not
showing these results in the present paper.
The other SPS parameters are away of the 
range of SUSY parameters interesting for our analysis.  
Notice that the gluino mass is around $1 \TeV$ for 
the three points mentioned above,
and therefore the effects of the logarithmic terms is not as enhanced as in 
the \textit{Def}~(\ref{eq:SUSYparams}) scenario, where $\mg\sim 3 \TeV$. 
We explore the SUSY parameter space for SPS1b, 3 and 7 
first by scanning $\mg$ in the interval $[400, 5000] \GeV$ while keeping 
$\tan\beta$ fixed, then by scanning $\tan\beta$  between $3$ and $50$
with $\mg$ fixed. We remind that the best accuracy of the effective
description of squarks interactions for these points is obtained when
$\mg\gtrsim 800 \GeV$~\cite{Guasch:2008fs}. 
Figures~\ref{partialsps1b37-mgl} and~\ref{partialsps1b37-tb} show a
  summary of the analysis. We analyzed all possible decay channels, 
however, in these figures we present results
for the most interesting channels 
in the analysis of the cross-section in the next section, i.e.
the decays of the squarks into the lightest chargino and the 
two lightest neutralinos. 
Figures~\ref{partialsps1b37-mgl}, \ref{partialsps1b37-tb} show the
partial decay widths in the effective approximation (labelled
\textit{Eff}), results for the tree-level prediction (\textit{Tree})
and fixed-order one-loop prediction (\textit{1-loop}) are not shown in
the figures, but will be commented in the text if necessary. 
The effective description
follows the logarithmic behaviour of the full one-loop corrections. 
In addition to the partial
decay widths ($\Gamma$) we also show the relative corrections defined as:
\begin{equation}
\delta^{correc}=\frac{\Gamma^{correc}-\Gamma^{Tree}(m_{q})}{\Gamma^{Tree}(m_{q})}\ \,
\label{eq:devcorrec}
\end{equation} 
where the label $correc$ is, in general, \textit{Eff} for the effective description
computation and \textit{1-loop} for the full one-loop corrections. 
Results for the tree-level computation, $\Gamma^{Tree}$, and the relative
corrections, $\delta^{\textit{Eff}}$ and  $\delta^{\textit{1-loop}}$, for the
scenarios analyzed in this section are shown in Table \ref{tablepartial}. 

Figures~\ref{partialsps1b37-mgl} and~\ref{partialsps1b37-tb}
show the effective approximation prediction 
for the partial decay widths, $\Gamma$, and relative
corrections, $\delta$~(\ref{eq:devcorrec}), of the 
top-squarks $\stopp_1$ and $\stopp_2$ 
decaying into the lightest chargino $(\stopp_{1,2} \rightarrow
b\,\cplus_1)$ and the two lightest neutralinos 
$(\stopp_{1,2}\rightarrow t\,\neut_{1,2})$, 
as a function of $\mg$ and $\tb$, respectively.
The results are presented for the SPS1b, SPS3 and SPS7 scenarios. 
In general, one expects the effective approximation to be valid when the gluino mass
is much larger than the squark mass scale. We found that the shape of the effective
and one-loop approximation can deviate significantly for light gluino
masses  $400\GeV<\mg<700\GeV$, and in this case the effective
approximation is not appropriate. Since the study presented in this work is
only relevant when the gluino is heavy, we can use the effective
approximation since it is valid for $\mg\gtrsim 900\GeV$. 

Figure~\ref{partialsps1b37-mgl} shows that, for the SPS1b 
and SPS3 scenarios, 
$\Gamma(\stopp_{1}\rightarrow t\neut_1)$ has positive corrections
 for $\mg\gtrsim850 \GeV$. The largest difference with respect  
to tree-level calculation is obtained at the largest value of the gluino
mass $\mg=5\TeV$ ($10\%$). 
For $\stopp_{1}\rightarrow t\neut_2$ and 
$\stopp_{1}\rightarrow b\cplus_1$ decay channels the
corrections are negative and slightly decreasing 
(in absolute value) with $\mg$.  
Opposite to the SPS1b and SPS3 situation, in the SPS7 scenario, 
$\neut_{1,2}$ and $\cplus_1$ are 
mostly of higgsino-type and $\neut_{3,4}$ and 
$\cplus_2$ are mostly of gaugino-type.
Then, the SPS7 scenario has larger partial decay width values into the
lightest chargino and neutralinos, and the radiative corrections show a
different behaviour. 
In this scenario all decay channels receive negative corrections 
in the whole analyzed range of gluino mass.
While the (absolute value) corrections in the 
$\stopp_{1}\rightarrow t\neut_1$ channel decrease with
$\mg$, the $\stopp_{1}\rightarrow t\neut_2$ and 
$\stopp_{1}\rightarrow b\cplus_1$ channels have an
opposite behaviour, becoming more negative as $\mg$ increases. 
For all the SUSY parameter choices, the largest partial decay width values 
correspond to the decay channel $\stopp_{1}\rightarrow b\cplus_1$.

Figure~\ref{partialsps1b37-tb} shows the same partial decay widths and
relative corrections as before, 
for SPS1b, SPS3 and for SPS7, as a function of $\tb$.
In SPS1b, SPS3 $\stopp_{1}\rightarrow t\neut_1$ receives positive 
corrections in almost the whole analyzed range of $\tan\beta$, 
with the exception of the region of small $\tb$ ($\lesssim 6,13$ for
SPS1b, SPS3 respectively).
For $\stopp_{1}\rightarrow t\neut_2,b\cplus_1$ decay channels, 
the radiative corrections are always negative.
We found that, for all SPS at $\tb\approx10$, 
$\Gamma^{Tree}(\stopp_{1}\rightarrow b\cplus_1)$ has a
minimum and a steep increase for larger $\tb$. The radiative corrections
are large and negative, compensating this increase. The one-loop
corrections are proportional to $\tb$, and grow faster than the
tree-level value, the effective computation, on the other hand, resums
the large corrections obtained at large $\tb$, providing a smoother behaviour.
For example, in the SPS1b scenario, while
$\Gamma^{Tree}(\stopp_{1}\rightarrow b\cplus_1)$
increases a $25\%$ in the range $\tb=20-50$, the effective prediction
$\Gamma^{Eff}(\stopp_{1}\rightarrow b\cplus_1)$ increases
only a $0.4\%$.

For a gluino mass around $\sim 900 \GeV$, corresponding to the nominal
value of the chosen SPS, the corrections  to 
$\Gamma(\stopp_{1}\rightarrow t\neut_1)$ 
in SPS1b and SPS3 scenarios are small,
below $1\%$ in absolute value
in the whole $\tb$ range (see Fig~\ref{partialsps1b37-tb}b). For SPS7 $\delta$  is larger (in absolute value), 
evolving from $-9\%$ close to $-5\%$.
For the $\neut_2$ decay channel, the deviations $\delta$ are
 negative: $-10\%<\delta<-7\%$ for SPS1b and SPS3 and 
$\delta\approx -22\%$ for SPS7.
For $\cmin_1$ the situation changes, $\delta^{Eff}$
reaches $-36\%$ for SPS7 and $-21\%$ for SPS1b and 3, 
in absolute values they always increase 
at a faster rate with $\tan\beta$ if compared to the neutralino channels. 
We want to stress that neutralino channels show a nearly flat behaviour
for $\delta$ because the couplings
only involve terms of $\lambda_{t}$ strength ($\propto 1/\tb$), 
a suppression which wipes out new dependence with $\tb$
introduced by $\Delta m_{t}^{SQCD}$ and log-terms beyond certain $\tb$ ($\approx 10$) 
for the effective approximation.
Meanwhile chargino channels contain terms of $\lambda_{b}$ 
strength ($\propto \tb$) which enhance higher order corrections.
As pointed out above, the SPS7 scenario has the largest radiative corrections.
In this scenario, $\neut_1$, $\neut_2$ and 
$\cplus_1$ are of higgsino-type, 
enhancing higgsino terms ($H_{+}$) in the generic couplings 
of expression~(\ref{effectivecoup}). It is clear
 that large part of the dependence in $\tan\beta$ is encoded in 
$H_{+}$, and those are also the terms most affected by radiative corrections.
SPS1b and SPS3 scenarios have the opposite situation, 
$\neut_1$, $\neut_2$ and $\cplus_1 $ 
are of gaugino-type, therefore gaugino terms ($G$) 
--depending much less on $\tb$ in the effective approximation--
are amplified over $H_{+}$.

For the heavy
top-squark ($\stopp_2$) decaying into the lightest neutralinos and
chargino ($\neut_{1,2},\cplus_1$) as a function of the gluino mass
($\mg$) and $\tb$, Fig.~\ref{partialsps1b37-mgl} shows that the largest partial decay
widths correspond to the chargino channel ($\cplus_1$) for SPS1b/SPS3
and the second neutralino channel ($\neut_2$) for SPS7. The corrections
show the characteristic logarithmic behaviour. For SPS1b/SPS3 the
corrections are moderate to large (up to $20\%$)  and they are positive,
except for the chargino channel in SPS1b. For SPS7 the corrections are
negative, they are around $-10\%$ for $\neut_1$, $-30\%$ for $\neut_2$
and $-60\%$ for $\cplus_1$.
Fig.~\ref{partialsps1b37-tb} shows that $\tb$ has its largest
impact in the chargino $\cplus_1$ channel, both in the partial decay
width and in the radiative corrections. We found that the effective
approximation moderates the impact of the radiative corrections (see Table~\ref{tablepartial}). 
While the one-loop corrections may surpass the $-100\%$ value in some
cases (e.g. the $\cplus_1$ channel in SPS7 for $\tb>41$), 
meaning that the one-loop
correction is not reliable, the effective computation predicts always 
values that, although quite large, are compatible with
perturbation theory (e.g. $-75\%$ in that case), and we can rely on this computation.

\begin{table}
\begin{tabular}{|c|c|c|c||c|c|c|c|c|}
\hline
Decay
&\multicolumn{3}{|c||}{SPS1b}&\multicolumn{3}{|c|}{SPS3} &\multicolumn{2}{c|}{}\\
\cline{1-7}
&$\Gamma^{Tree}$&$\delta^{Eff}$&$\delta^{1-loop}$&$\Gamma^{Tree}$&$\delta^{Eff}$&$\delta^{1-loop}$&\multicolumn{2}{c|}{}\\
\cline{1-7}
$\stopp_{1}\rightarrow t\neut_1$ &0.7927&0.8&0.2&0.7769&0.8&0.2&\multicolumn{2}{c|}{}\\
$\stopp_{1}\rightarrow t\neut_2$ &0.6577&-7.7&-7.3&0.6193&-7.7&-7.2&\multicolumn{2}{c|}{}\\
$\stopp_{1}\rightarrow t\neut_3$ &-&-&-&-&-&-&\multicolumn{2}{c|}{}\\
$\stopp_{1}\rightarrow t\neut_4$ &-&-&-&-&-&-&\multicolumn{2}{c|}{}\\
$\stopp_{1}\rightarrow b\cplus_1$ &2.099&-13.7&-18.0&1.903&-9.4&-11.5&\multicolumn{2}{c|}{}\\
$\stopp_{1}\rightarrow b\cplus_2$ &0.9378&-39.9&-47.8&0.4941&-30.9&-31.9&\multicolumn{2}{c|}{}\\
\cline{1-7}
$\stopp_{2}\rightarrow t\neut_1$&0.3262&4.4&3.7&0.3075&4.9&4.2&\multicolumn{2}{c|}{}\\
$\stopp_{2}\rightarrow t\neut_2$&1.026&5.1&4.4&1.043&4.9&4.3&\multicolumn{2}{c|}{}\\
$\stopp_{2}\rightarrow t\neut_3$&1.489&-29.0&-28.7&1.362&-29.0&-28.5&\multicolumn{2}{c|}{}\\
$\stopp_{2}\rightarrow t\neut_4$&3.764&-25.6&-27.6&3.509&-25.7&-27.6&\multicolumn{2}{c|}{}\\
$\stopp_{2}\rightarrow b\cplus_1$&2.946&-11.0&-21.5&2.343&6.5&3.6&\multicolumn{2}{c|}{}\\
$\stopp_{2}\rightarrow b\cplus_2$&4.900&-51.7&-73.1&2.472&-28.8&-33.8&\multicolumn{2}{c|}{}\\
\hline\hline
&
\multicolumn{3}{|c||}{ SPS7}&\multicolumn{3}{|c|}{ Def $\,\,\mg=1\TeV$}&\multicolumn{2}{|c|}{ Def $\,\,\mg=3\TeV$}\\
\hline
&$\Gamma^{Tree}$&$\delta^{Eff}$&$\delta^{1-loop}$&$\Gamma^{Tree}$&$\delta^{Eff}$&$\delta^{1-loop}$
&$\delta^{Eff}$&$\delta^{1-loop}$\\
\hline
$\stopp_{1}\rightarrow t\neut_1$ &1.481&-5.9&-6.5&0.1264&11.5&7.8&22.9&15.2\\
$\stopp_{1}\rightarrow t\neut_2$ &3.323&-22.0&-22.1&1.174&-2.1&-3.2&3.6&1.9\\
$\stopp_{1}\rightarrow t\neut_3$ &5.776&-27.5&-28.4&2.963&-29.1&-24.8&-33.9&-27.5\\
$\stopp_{1}\rightarrow t\neut_4$ &0.5845&-40.2&-40.2&5.298&-23.7&-20.9&-26.7&-22.4\\
$\stopp_{1}\rightarrow b\cplus_1$ &8.754&-23.3&-25.4&1.642&16.1&10.1&29.4&20.2\\
$\stopp_{1}\rightarrow b\cplus_2$ &1.493&-43.4&-47.3&3.737&-20.5&-19.3&-22.0&-19.7\\
\hline
$\stopp_{2}\rightarrow t\neut_1$&0.1435&-8.2&-13.1&2.103&-5.5&-8.7&-1.2&-1.4\\
$\stopp_{2}\rightarrow t\neut_2$&2.730&-25.9&-27.6&3.299&-16.9&-17.6&-16.7&-14.1\\
$\stopp_{2}\rightarrow t\neut_3$&4.723&-29.4&-30.2&9.944&-29.2&-28.2&-33.1&-29.2\\
$\stopp_{2}\rightarrow t\neut_4$&5.624&-19.0&-20.5&4.515&-34.7&-31.9&-40.4&-35.2\\
$\stopp_{2}\rightarrow b\cplus_1$&1.371&-54.9&-69.6&7.531&-16.9&-18.6&-16.8&-15.0\\
$\stopp_{2}\rightarrow b\cplus_2$&6.262&-16.7&-19.6&8.662&-34.5&-33.4&-40.0&-36.5\\
\hline\hline
\end{tabular}
  \caption{Tree-level decay width, $\Gamma^{Tree}$ [GeV], and 
relative radiative corrections, $\delta^{Eff}$ and  $\delta^{1-loop}$
$[\%]$~(\ref{eq:devcorrec}), for the various SPS scenarios 
and \textit{Def} (with two values of $\mg=1,\,3\TeV$).}
\label{tablepartial}
\end{table}
Finally, Table \ref{tablepartial} summarizes the results for the tree-level 
computation, $\Gamma^{Tree}$ [GeV], and the relative radiative corrections,
 $\delta^{Eff}$ $[\%]$ and $\delta^{1-loop}$~(\ref{eq:devcorrec}), of all surveyed partial decay widths.
Numbers not shown correspond to kinematically closed channels.
The results are shown for the three SPS scenarios analyzed in this paper
SPS1b, SPS3, SPS7, and \textit{Def}, eq.~(\ref{eq:SUSYparams}). We
include two sets 
of \textit{Def} parameters: one with $\mg=3\TeV$ as in
eq.~(\ref{eq:SUSYparams}), and one with $\mg=1\TeV$. The reason is a
fair comparison among the various SUSY scenarios: since the radiative
corrections grow as the logarithm of $\mg$, the corrections in the \textit{Def}
parameter set are enhanced over those of the SPS, which
have a $\mg\simeq 1\TeV$. 
A quick look at Table~\ref{tablepartial} shows that the radiative
corrections are quite different in each channel, which means that they
will have an impact on the decay branching ratios (and production
cross-sections). As an example, in SPS1b the tree-level prediction is
that the largest branching ratio of the heavy top-squark corresponds to
the second chargino ($\stopp_2\to b\cplus_2$), followed by $\neut_4$ and
$\cplus_1$, however, after computing the radiative corrections one
finds that the largest branching ratio corresponds to
$\Gamma(\stopp_2\to t\neut_4)\simeq 2.8\GeV$, 
followed by $\Gamma(\stopp_2\to b\cplus_1)\simeq 2.6\GeV$, 
and $\Gamma(\stopp_2\to b\cplus_2)\simeq 2.4\GeV$ in third place.

\subsection{Cross-Section Computation}
\label{sec:crosssection}

After successfully implementing and checking our effective MSSM couplings in 
MadGraph\footnote{Files are available on request.},
we have computed the cross-section of top-squark pair
production\footnote{Gauge invariance requires one to consider pair
  production together with single production and non-resonant diagrams,
  see below.}
for the \textit{Def}, SPS1b, SPS3 and SPS7 scenarios. We have focussed on reactions
where top-squarks decay into the two lightest
neutralinos ($\neut_{1,2}$) and the lightest chargino 
($\cpmin_1$), since these are the search channels used at the LHC~\cite{ATLAS,Ball:2007zza,Aad:2009wy}. 
The ATLAS collaboration has already used them to analyze LHC
  data and set limits on top-squark pair-production~\cite{Collaboration:2012si,Collaboration:2012ar,Collaboration:2012as,ATLAS:2012st,ATLAS:2012st2}.

We have simulated proton-proton ($pp$) collisions at $14 \TeV$ using the
CTEQ6L set of parton distribution functions. 
Our aim is to use the standard
MadGraph input files as far as possible. 
By default, MadGraph sets renormalization and
factorization scales dynamically.  For the production of a pair of heavy
particles, as is our case, these scales are set equal to the
geometric mean of $M^{2}+p_{T}^{2}$ of both particles, where $M$ is the
mass of the particle and $p_{T}$ is its transverse momentum.
The scale $Q$ of the effective couplings~(\ref{eq:logterms}) is
fixed at the mass of the internal top-squark for each process.

In this section we present the results for the top-squark pair production
cross-section in $pp$ collisions, followed by the decay into a quark and
charginos and neutralinos, including the effect of the radiative
corrections in the effective coupling approximation.

\subsubsection{$\sigma(pp
\rightarrow ({q}^{'}{\chi}_{r})({\bar{q}}^{''}\chi_{s}))$}

\begin{figure}[tb!]
\centering
\begin{tabular}{cccc}
\includegraphics[height=3cm,angle=0]{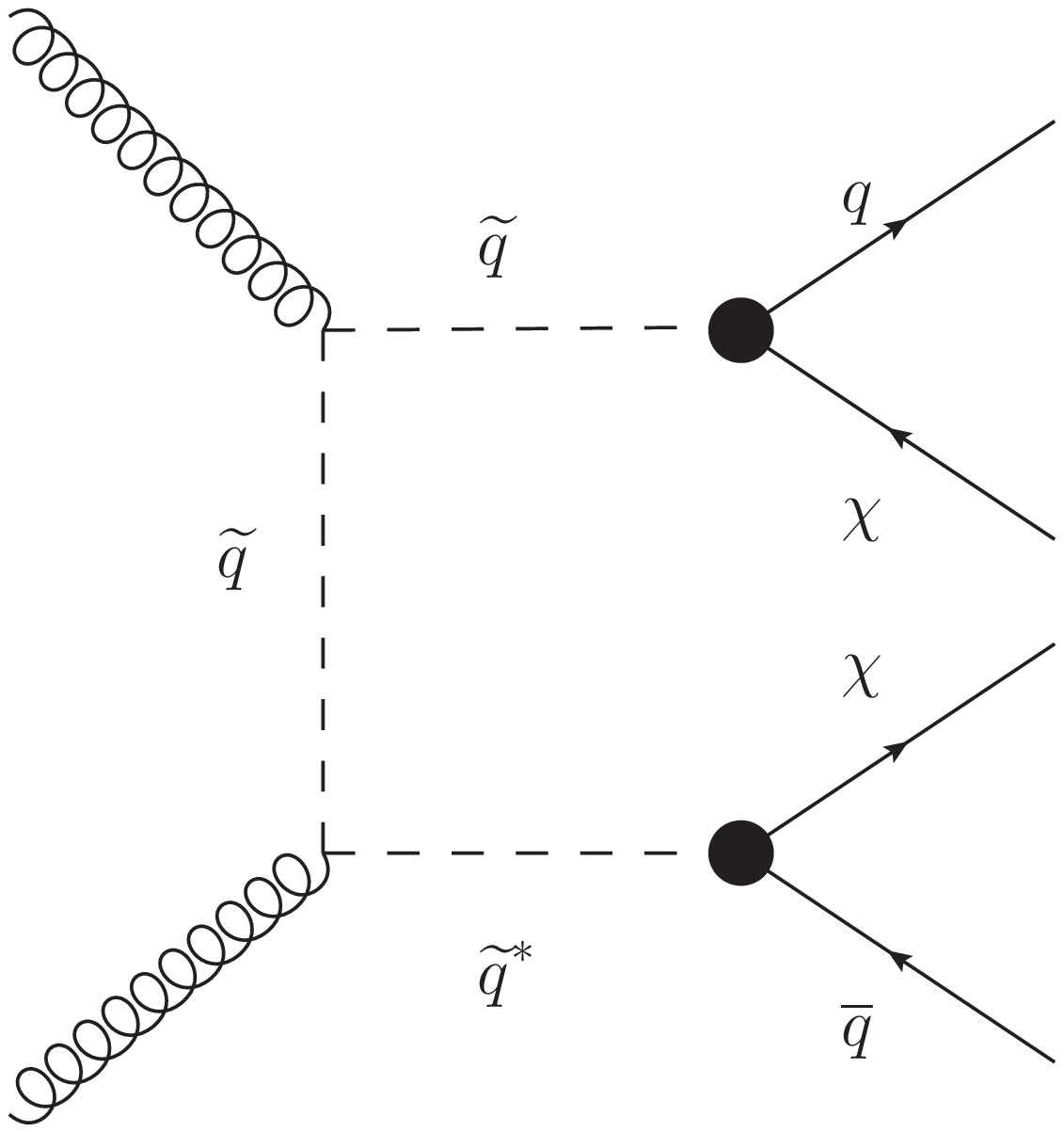}&
\includegraphics[height=3cm,angle=0]{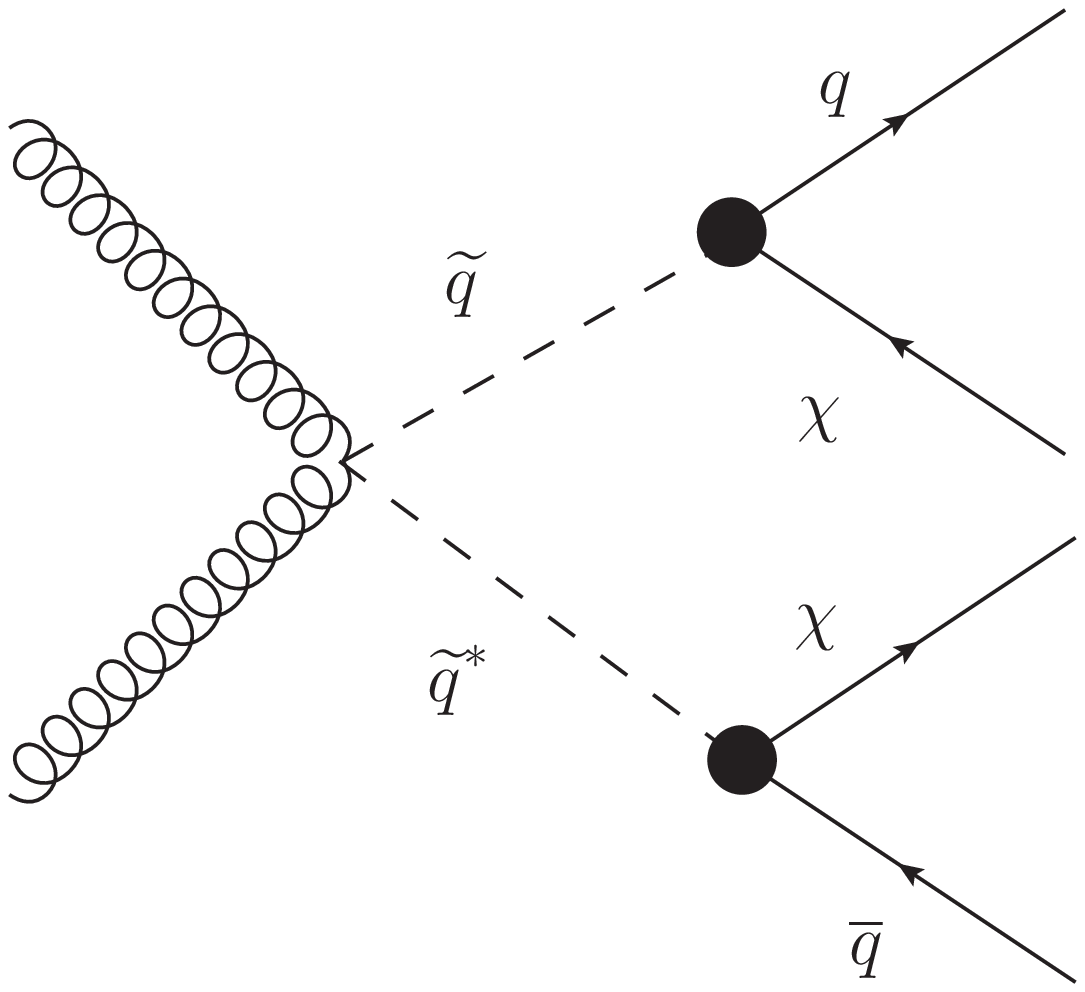}&
\includegraphics[height=3cm,angle=0]{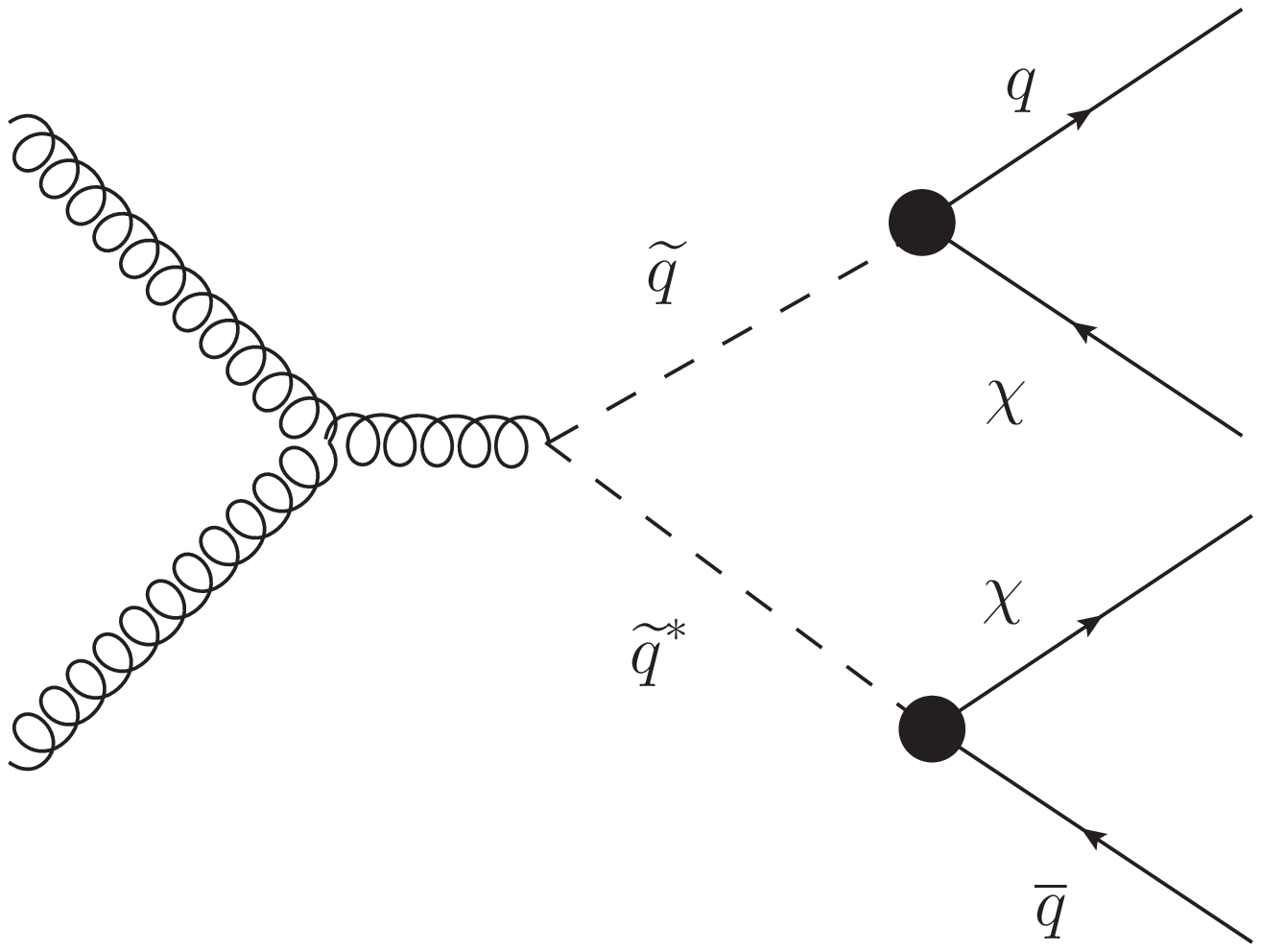}\\
(a)&(b)&(c)\\
\includegraphics[height=3cm,angle=0]{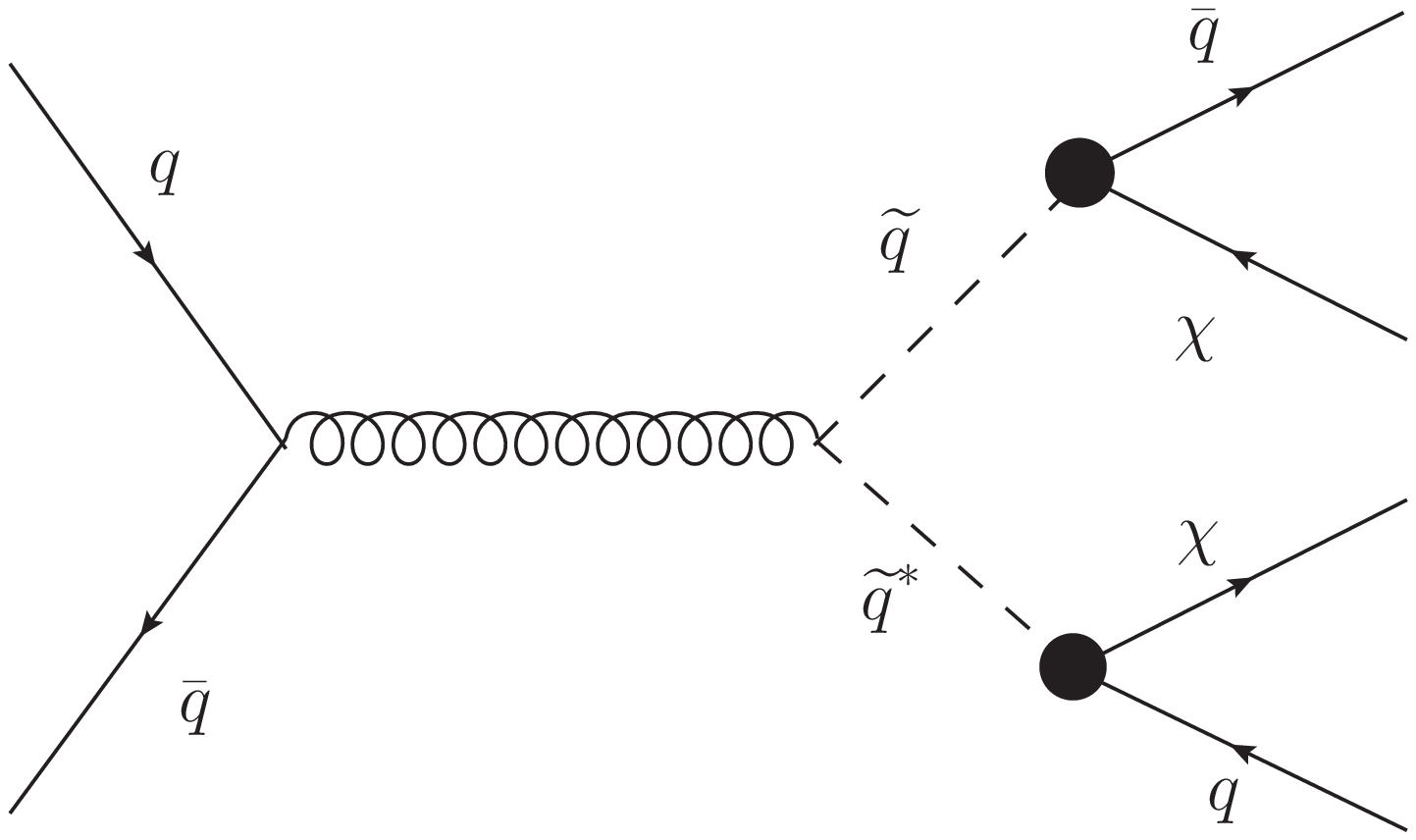}&
\includegraphics[height=3cm,angle=0]{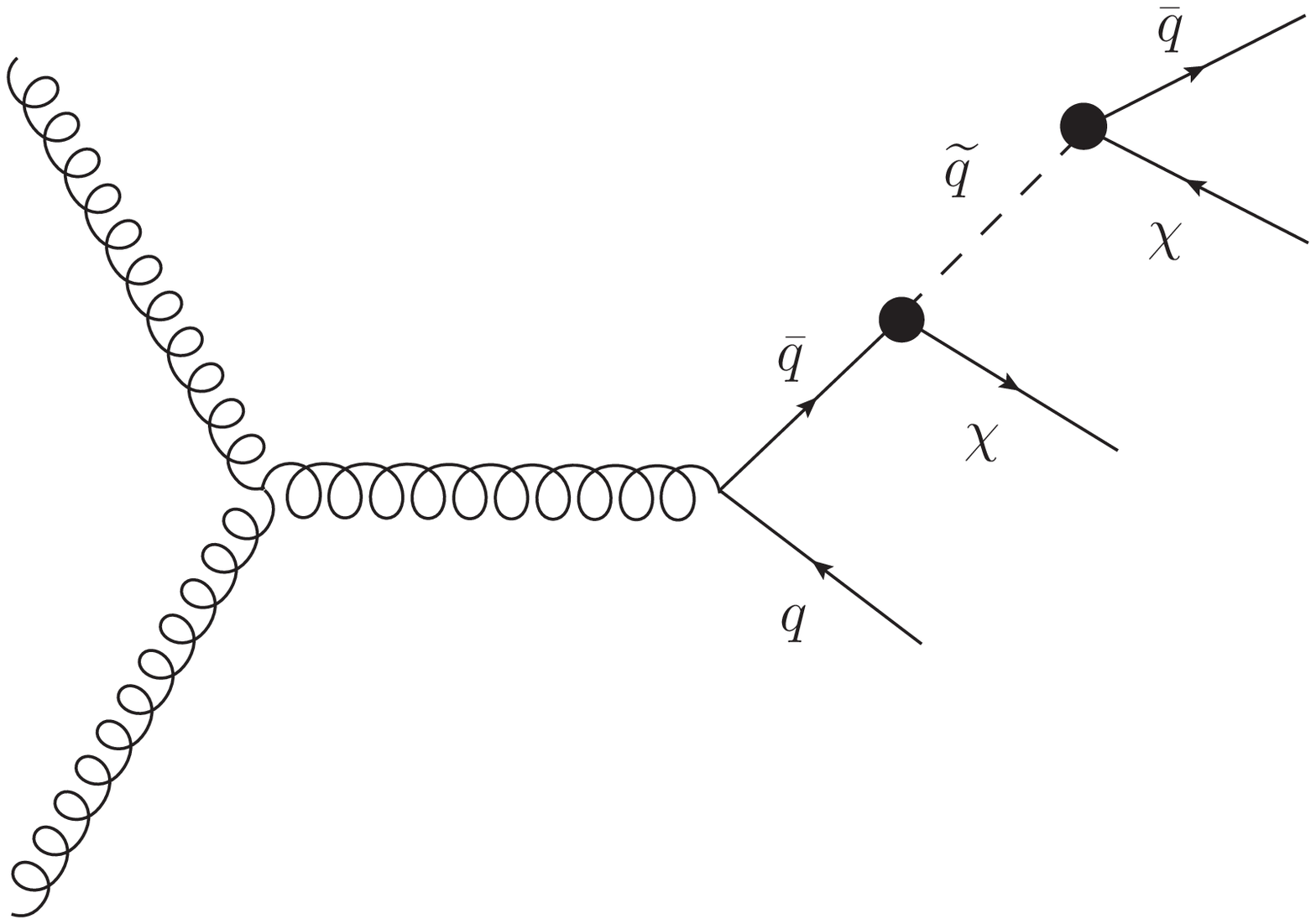}&
\includegraphics[height=3cm,angle=0]{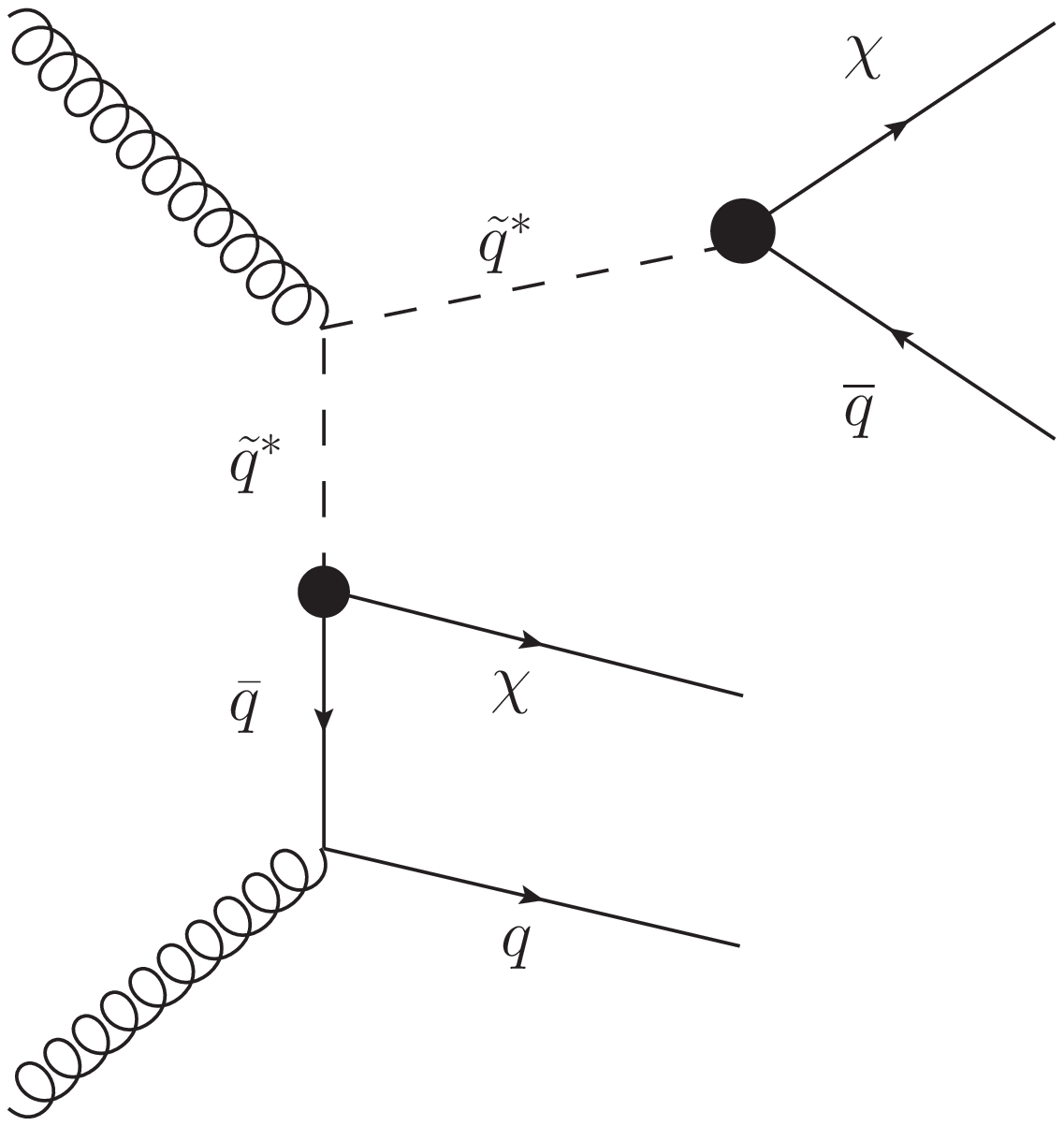}\\
(d)&(e)&(f)\\
\includegraphics[height=3cm,angle=0]{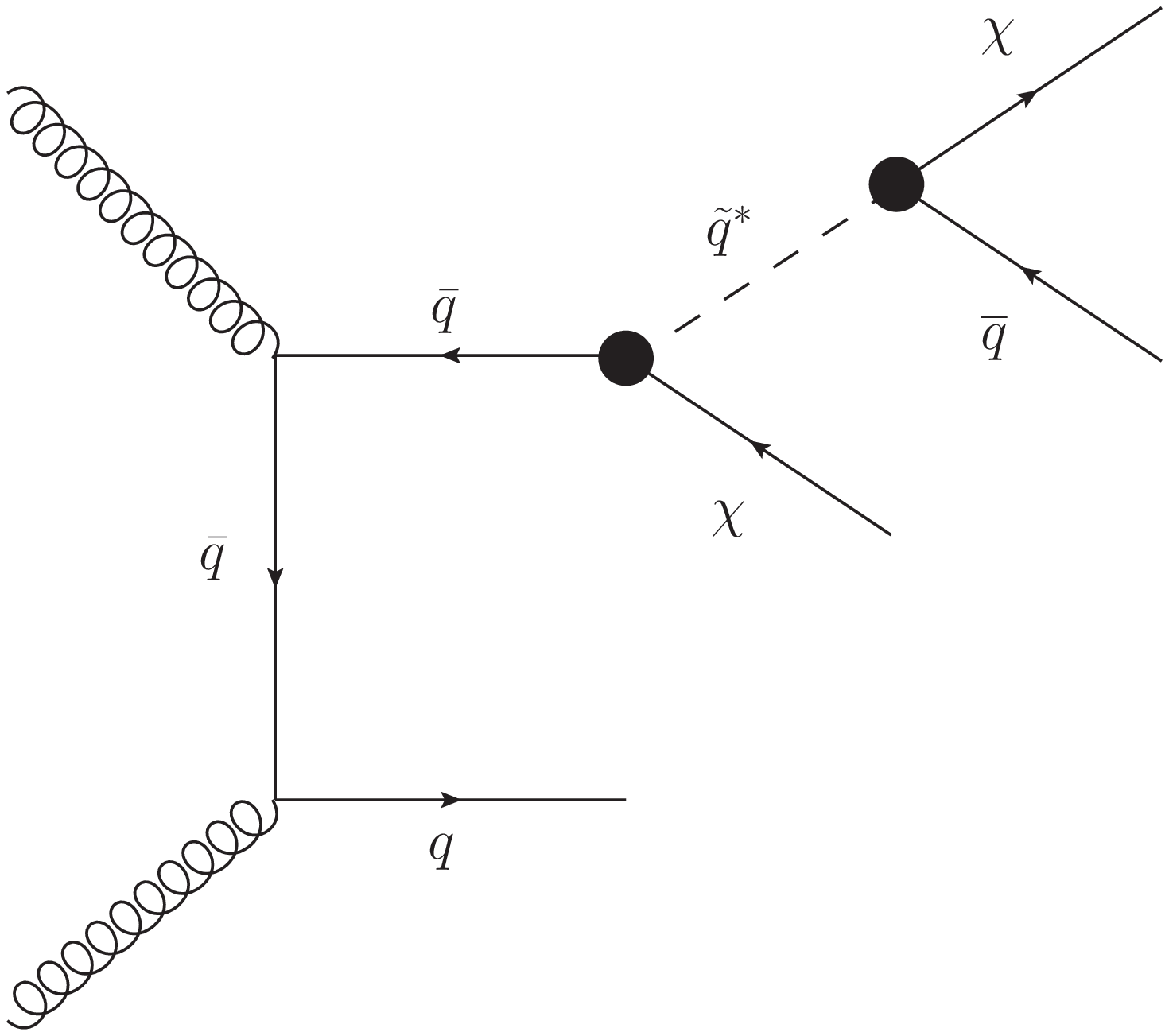}&
\includegraphics[height=3cm,angle=0]{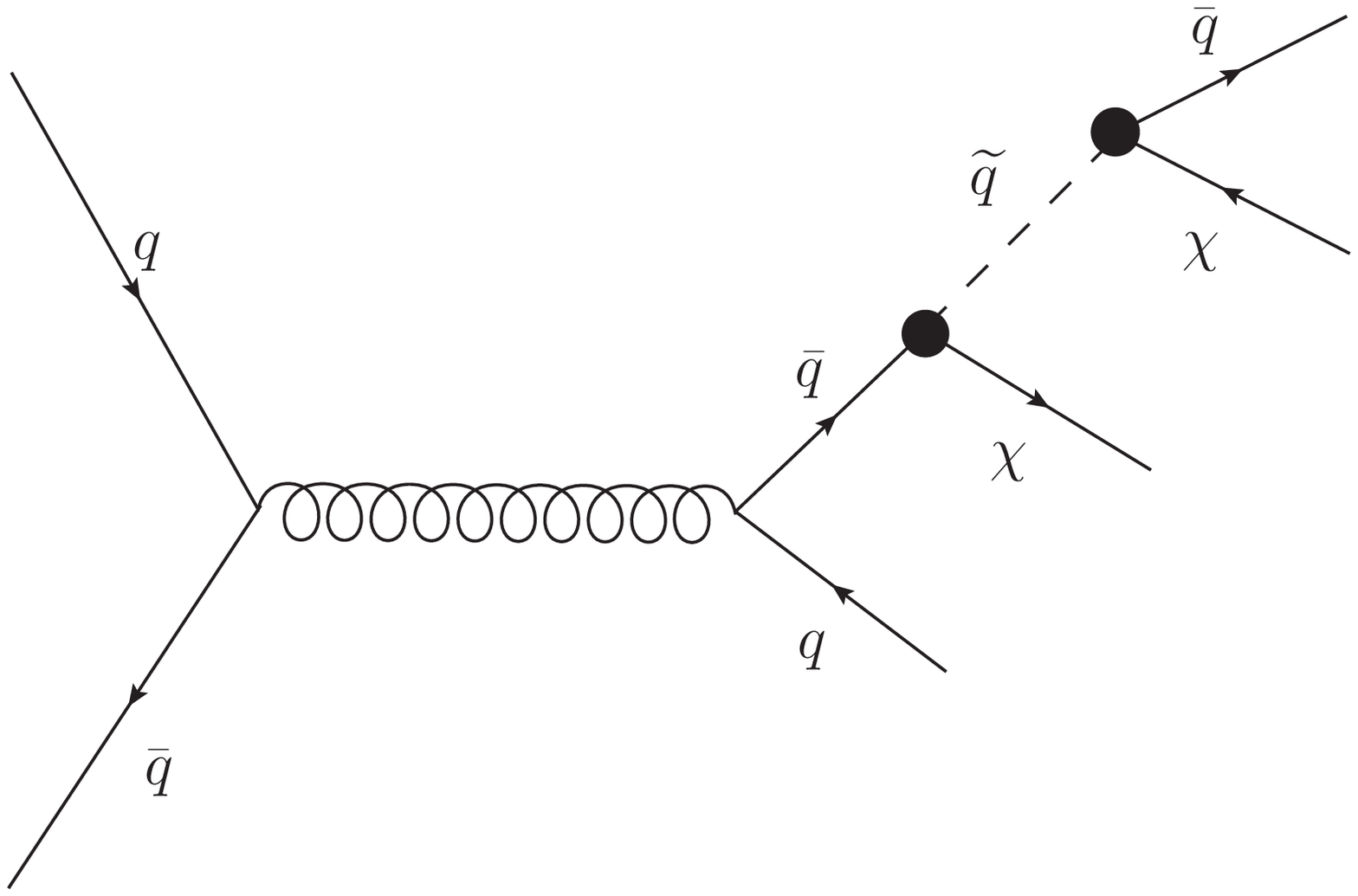}&
\includegraphics[height=3cm,angle=0]{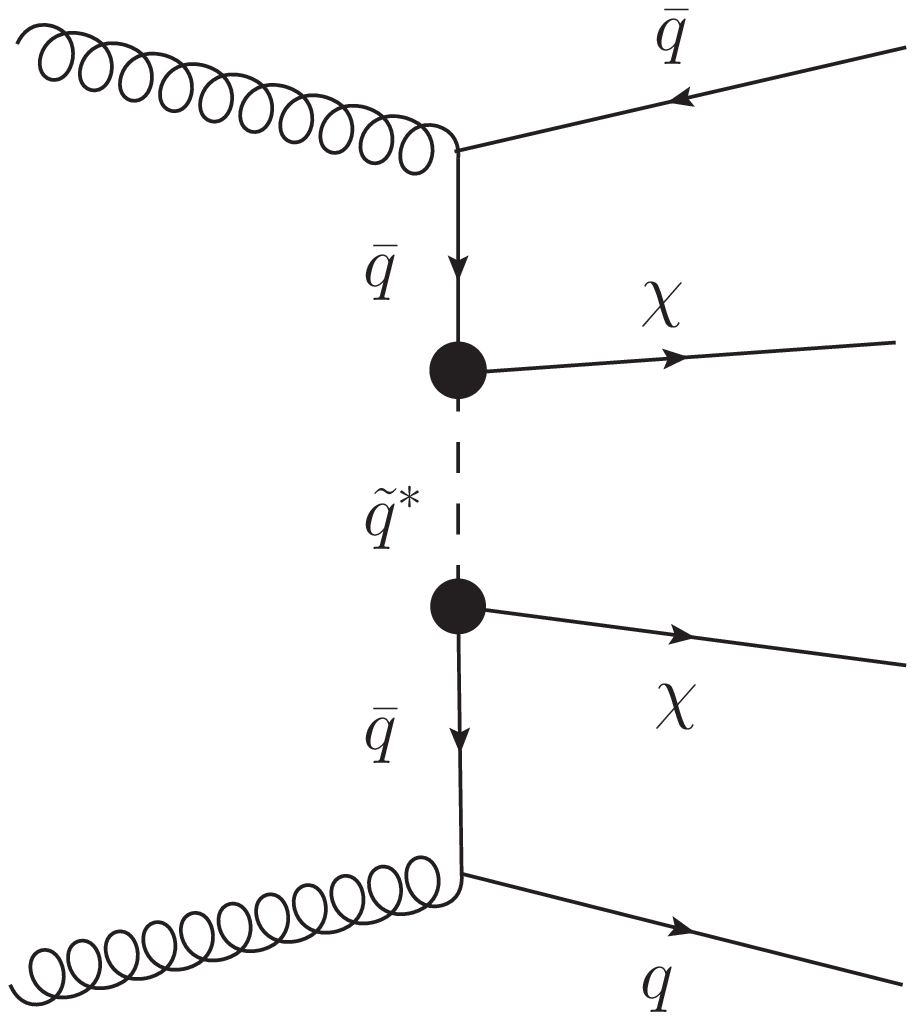}\\
(g)&(h)&(i)
\end{tabular}
\caption{Generic partonic Feynman diagrams contributing to 
$\sigma(pp\rightarrow ({q}\chi)({\bar{q}}\chi))$, \textbf{(a)-(d)}
double resonant diagrams 
($\sigma(pp\to\squark_a\squark_a^*\rightarrow
({q}\chi)({\bar{q}}\chi))$), \textbf{(e)-(h)} single resonant diagrams,
\textbf{(i)} non-resonant diagram.}
\label{pairdiag}
\end{figure}
The cross-section we consider in this work is\footnote{Our programs are
  prepared to compute any final state $\chi_{r,s}=\cpmin_{1,2}\,,\neut_{1,2,3,4}$.}
\begin{equation}
\sigma\left(pp(\to [\squark_a,\squark_a\squark_a^*])\rightarrow
({q}^{'}\chi_{r})({\bar{q}}^{''}\chi_{s})\right)\ \ , \ \ 
\squark_{a}=\stopp_{1,2}\ ,\ \ 
{q}^{'}, {\bar{q}}^{''}=t,b\ , \ \ 
\chi_{r,s}=\cpmin_1,
\neut_{1,2}\ ,
\label{eq:x-secfull}
\end{equation}
 that is: chargino or neutralino production,
associated with top and bottom quarks, which proceeds through a single
or double top-squark in the intermediate states.
Figure~\ref{pairdiag} shows the partonic Feynman
diagrams. Figs.~\ref{pairdiag}a-d show the 
double-resonant diagrams contributing to the process we are interested in:
$\sigma(pp\to\squark_a\squark_a^*\rightarrow({q}\chi)({\bar{q}}\chi))$,
however, this set of diagrams is not gauge invariant. The complete set
of gauge-invariant diagrams includes also single-resonant diagrams
(Fig.~\ref{pairdiag}e-h) and non-resonant diagrams
(Fig.~\ref{pairdiag}i). 
We have checked that quark-anti-quark channels ($q\bar{q}$, Fig.~\ref{pairdiag}d,h) are 
three times smaller than the gluon-gluon channel ($gg$), i.e. they
represent less than a $30\%$ of the total cross-section.
Both channels are included in our computation. Since all top-squarks
contribute to the same final state, one should add up all the amplitudes
in Fig.~\ref{pairdiag} for $\stopp_1$ and $\stopp_2$, however in squark
search studies it is costumary to separate the $\stopp_1$ and $\stopp_2$
channels, and look for each signal separately. For this reason we will
show results for the cross-section with only one of the top-squarks in
the intermediate states of Fig.~\ref{pairdiag}. This separation is
possible because we have checked that the interference effects between
$\stopp_1$ and $\stopp_2$ channels is less than $0.8\%$. We will denote this process as:
$$
pp \to [\squark_a] \to X
$$
where $\squark_a$ is the squark appearing in the intermediate state,
and $X$ is the final state, 
under the understanding that all diagrams (double-resonant,
single-resonant, and non-resonant) contribute.

Before performing the full simulation it is useful to analyze
approximations to the quantity under study. Under the narrow width
approximation the total cross-section~(\ref{eq:x-secfull}) can be
computed as the squark pair production cross-section followed by the
decay branching ratios,
\begin{equation}
\sigma\left(pp\rightarrow[\squark_{a}]
\rightarrow ({q}^{'}\chi_{r})({\bar{q}}^{''}\chi_{s})\right)\simeq
\sigma(pp\rightarrow\squark_{a}\squark_{a}^{\ast})\times
BR(\squark_{a}\rightarrow {q}^{'}\chi_{r})\times
BR(\squark_{a}^{\ast}\rightarrow {\bar{q}}^{''}\chi_{s})\,.
\label{eq:sigmaintoBRs}
\end{equation}
In this approximation only the double-pole diagrams
(Fig.~\ref{pairdiag}a-d) contribute.
In the present process the first part
$\sigma(pp\rightarrow\squark_{i}\squark_{i}^{\ast})$ proceeds purely
through standard QCD couplings, the only SUSY parameters being the
squark masses. All the information on SUSY couplings is in the second
part of the expression, that is, in the decay branching ratios or
partial decay widths, which
have been analyzed in section~\ref{resultsGamma}. 
The top-squark production cross-section has been computed to
NLO-SUSY-QCD~\cite{Beenakker:1997ut} and
NLL-SUSY-QCD~\cite{Beenakker:2010nq}, the radiative corrections increase
the production cross-section at the $14\TeV$ LHC by a $25\%-50\%$
depending on the top-squark mass, with a mild dependence on other SUSY
parameters~\cite{Beenakker:1997ut,Beenakker:2010nq}. 

In order to asses the effects of the radiative corrections, it is useful
to define a factor $K_{SUSY}$ as: 
\begin{equation}
K_{SUSY} = \frac{\sigma_{Effective}}{\sigma_{Tree}}\ \ ,
\label{eq:ksusy}
\end{equation}
which under the narrow-width approximation of
eq.~(\ref{eq:sigmaintoBRs})
can be computed as a ratio of partial decay widths:
\begin{eqnarray}
K_{SUSY} &=& \frac{K_{partial}}{K_{full}} \nonumber\\ 
K_{partial} &=& \frac{\Gamma^{Eff}(\squark\rightarrow  
{q}^{'}\chi_{r})}{\Gamma^{Tree}(\squark\rightarrow  
{q}^{'}\chi_{r})}\times
\frac{\Gamma^{Eff}(\squark^{\ast}\rightarrow {\bar{q}}^{''}\chi_{s})}
{\Gamma^{Tree}(\squark^{\ast}\rightarrow {\bar{q}}^{''}\chi_{s})}\nonumber\\  
K_{full} &=& \left(\frac{\Gamma^{Eff}_{full}}{\Gamma^{Tree}_{full}}\right)^{2}
\label{eq:ksusygroup}
\end{eqnarray}
where $\Gamma_{full}$ is the full decay width and
 \textit{Eff} and \textit{Tree} stand for 
the effective approximation and the tree-level computation. 
For the parameter space points of the present study, all bosonic
$\stopp_1$ decay channels are kinematically closed, and therefore the
full decay width consists only of the chargino/neutralino
(\textit{inos}) channels. Bosonic channels have some impact on the
$\stopp_2$ full decay width. 
$K_{SUSY}$ is composed of two factors: 
$K_{partial}$, regarding the \textit{partial} decay widths of the produced 
squarks decaying into a quark and the selected 
final particles ($\neut_{1,2}$ or $\cpmin_1$  analyzed in section~\ref{resultsGamma})
and $K_{full}$ the radiative correction to the total decay width. 
For the parameter space explored in the present work we have found
that the radiative corrections to the total decay width are always
negative ($K_{full}<1$).
Of course, using the narrow-width approximation~(\ref{eq:sigmaintoBRs})
one can not compute angular distributions or correlations, something
that it is possible to perform with the MadGraph implementation.

\begin{figure}[t!]
\centering
{\resizebox{16cm}{!}
{\begin{tabular}{cc}
\includegraphics[width=8cm]{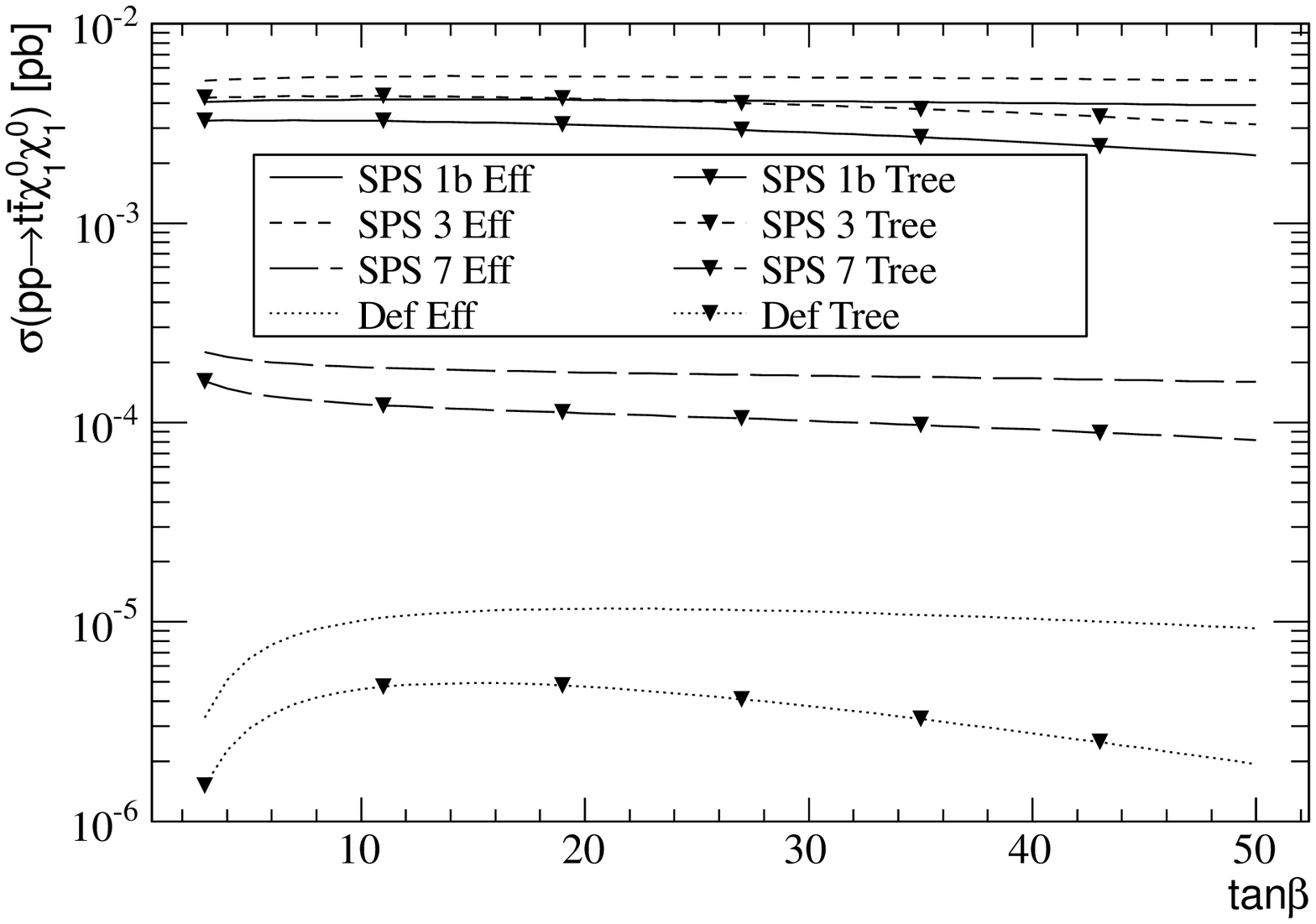}&
\includegraphics[width=8cm]{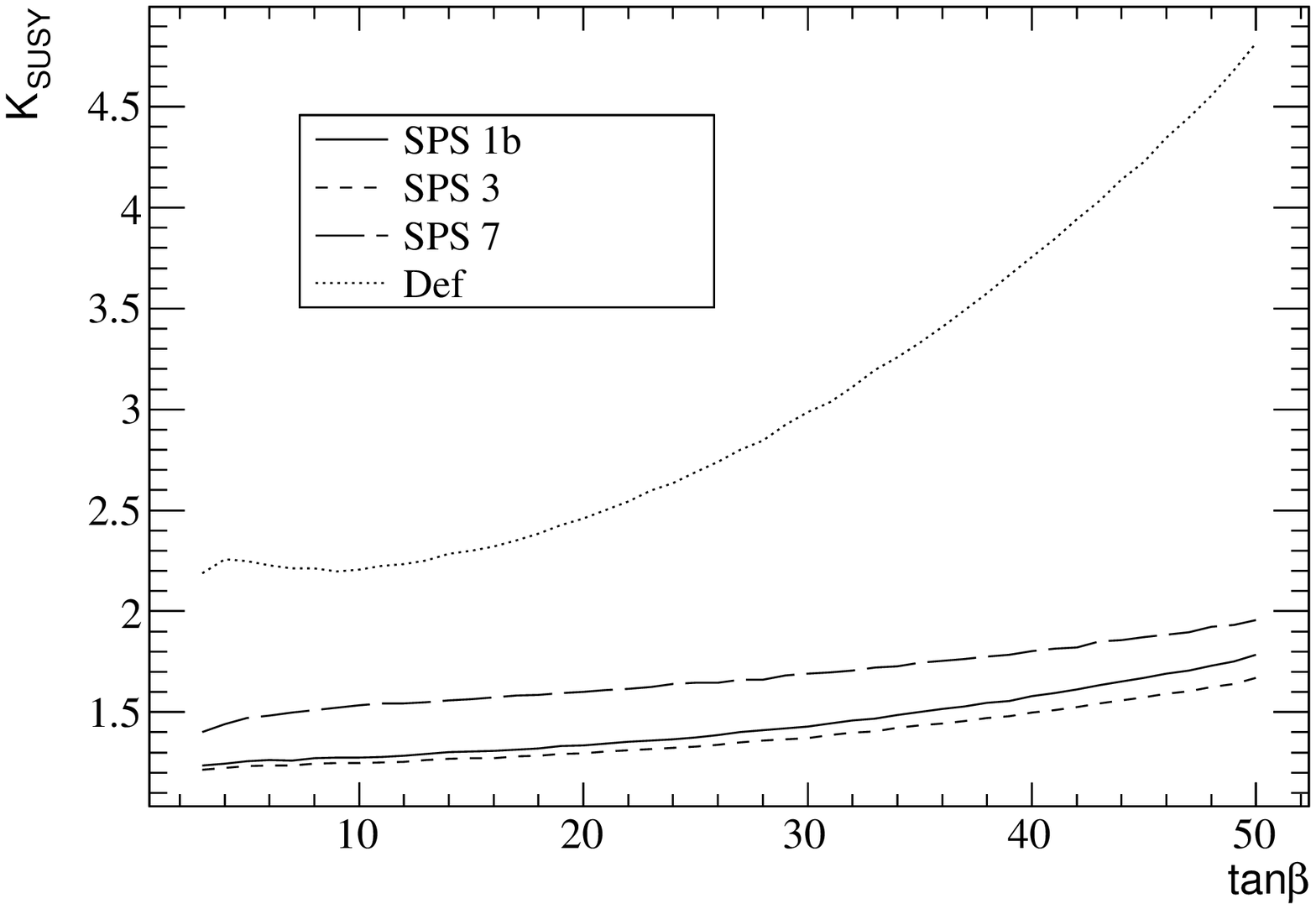}\\
\includegraphics[width=8cm]{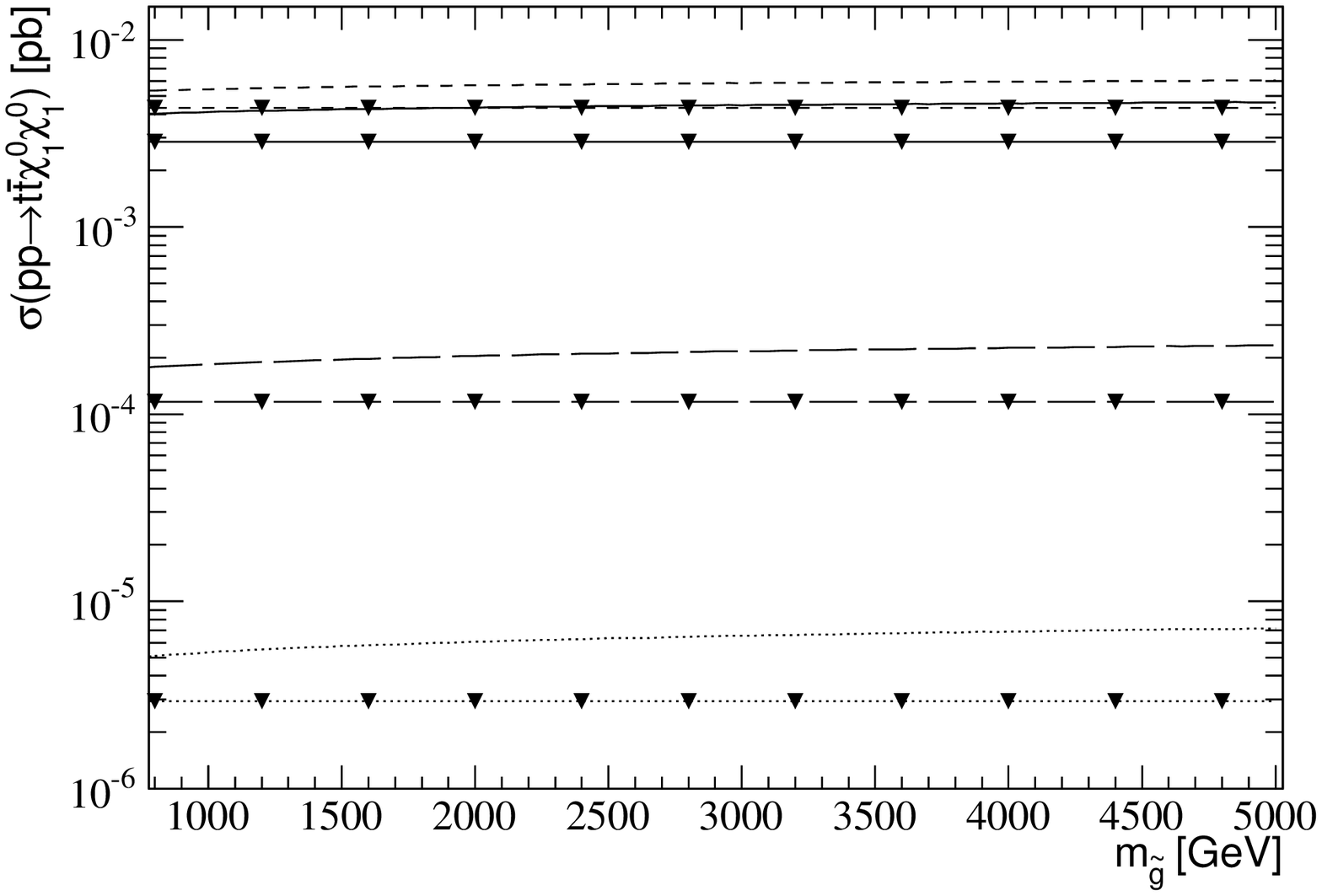}&
\includegraphics[width=8cm]{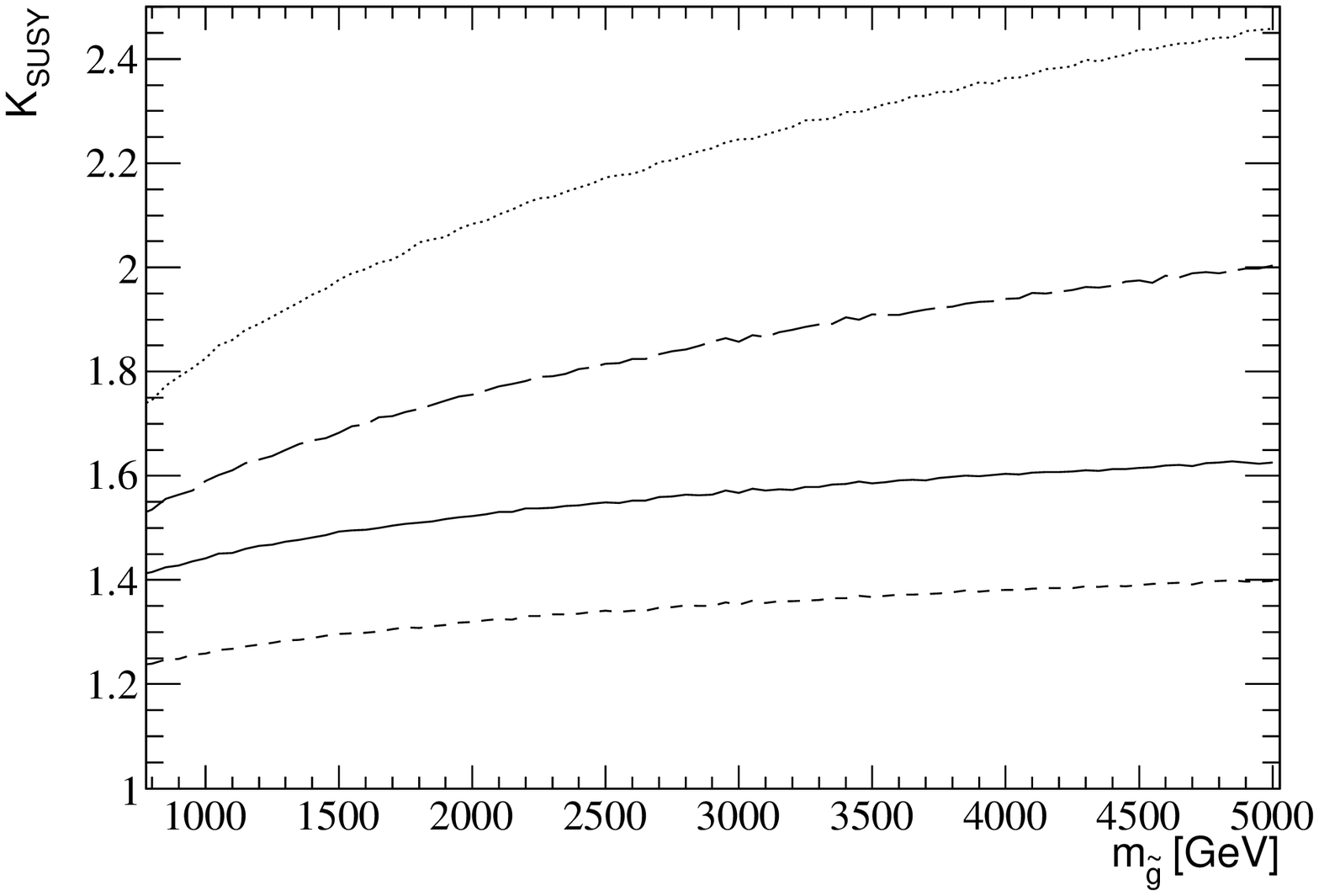}\\
(a)&(b)\\
\end{tabular}}}
\caption{\textbf{(a)} Cross-section of 
$pp\rightarrow [\stopp_1]\rightarrow 
t\neut_{1}\bar{t}\neut_{1}$  and
\textbf{(b)} $K_{SUSY}$,  
as a function of $\tan\beta$ and $\mg$ for the input parameters
as defined in SPS1b, 3, 7 and \textit{Def}~(\ref{eq:SUSYparams}).}
\label{stop1n1n1}
\end{figure}

\begin{figure}[t!]
\centering
{\resizebox{16cm}{!}
{\begin{tabular}{cc}
\includegraphics[width=8cm]{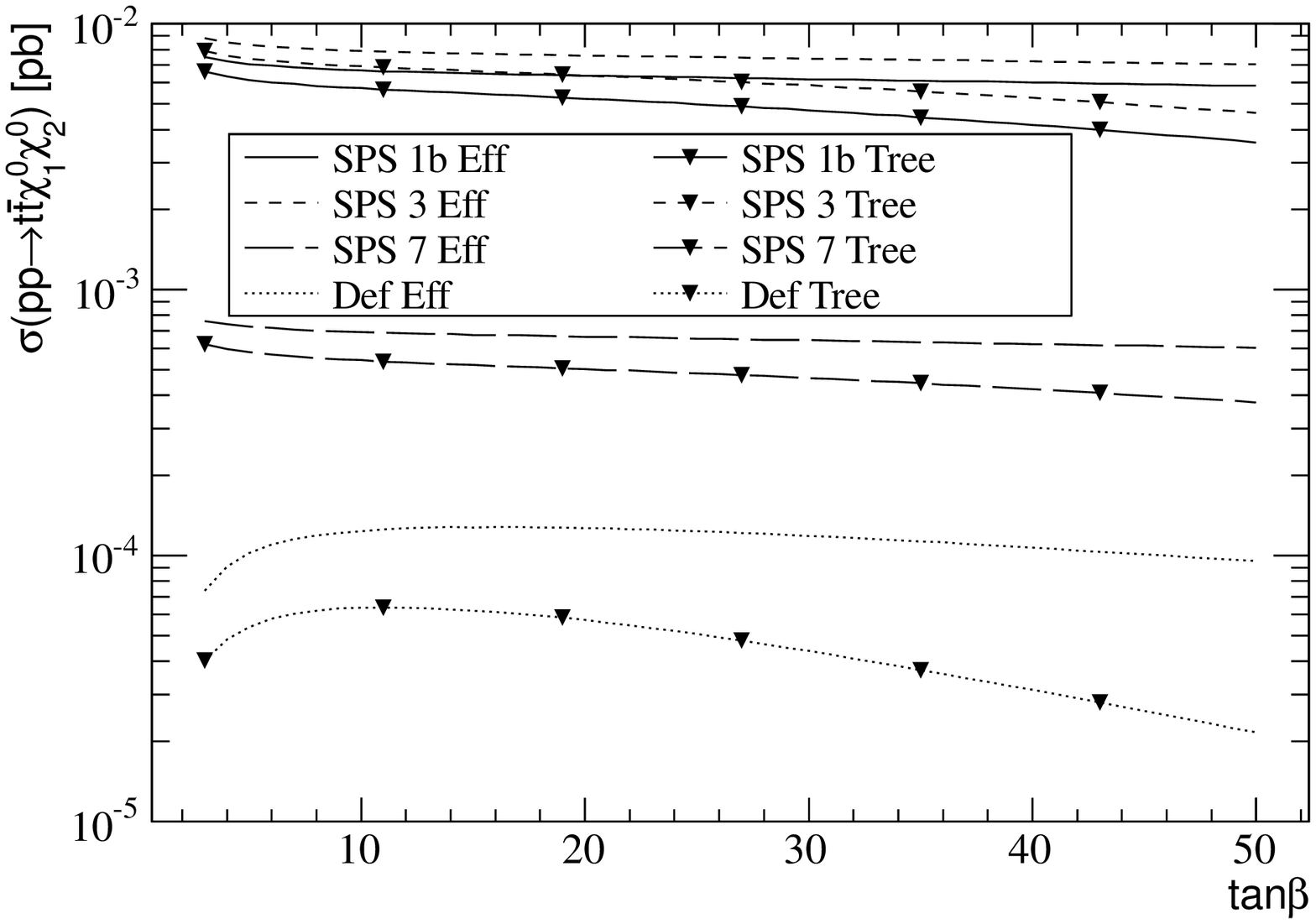}&
\includegraphics[width=8cm]{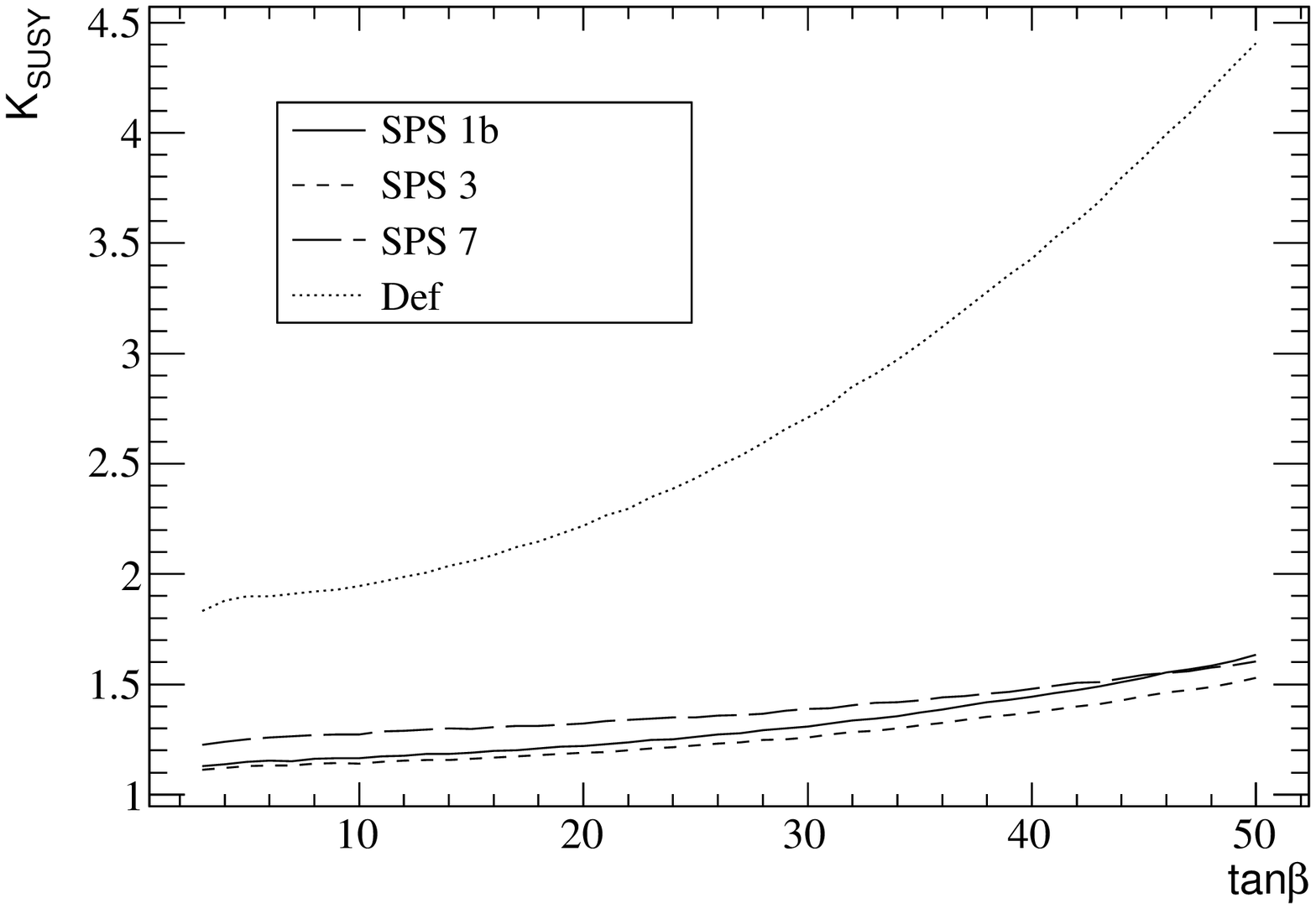}\\
\includegraphics[width=8cm]{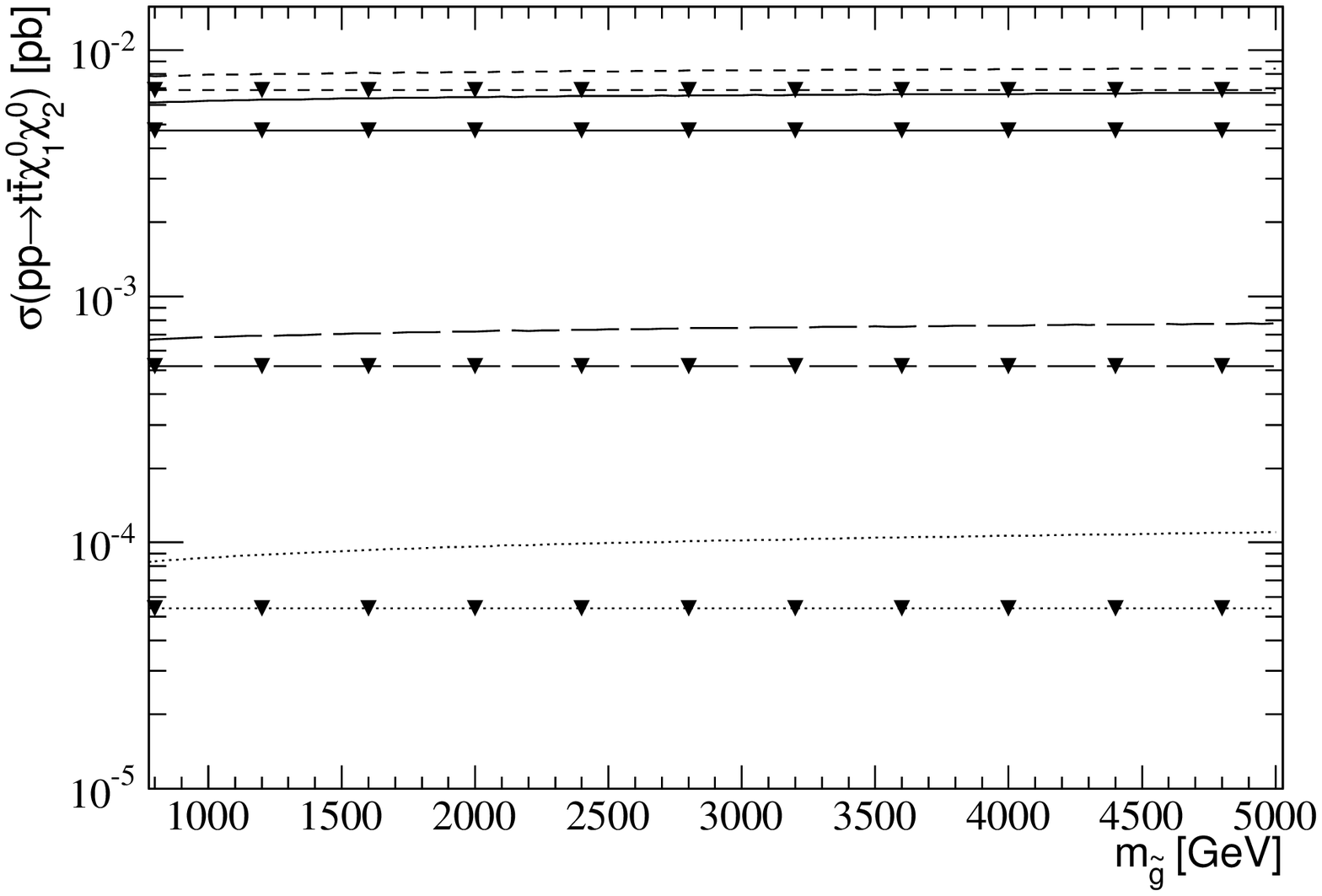}&
\includegraphics[width=8cm]{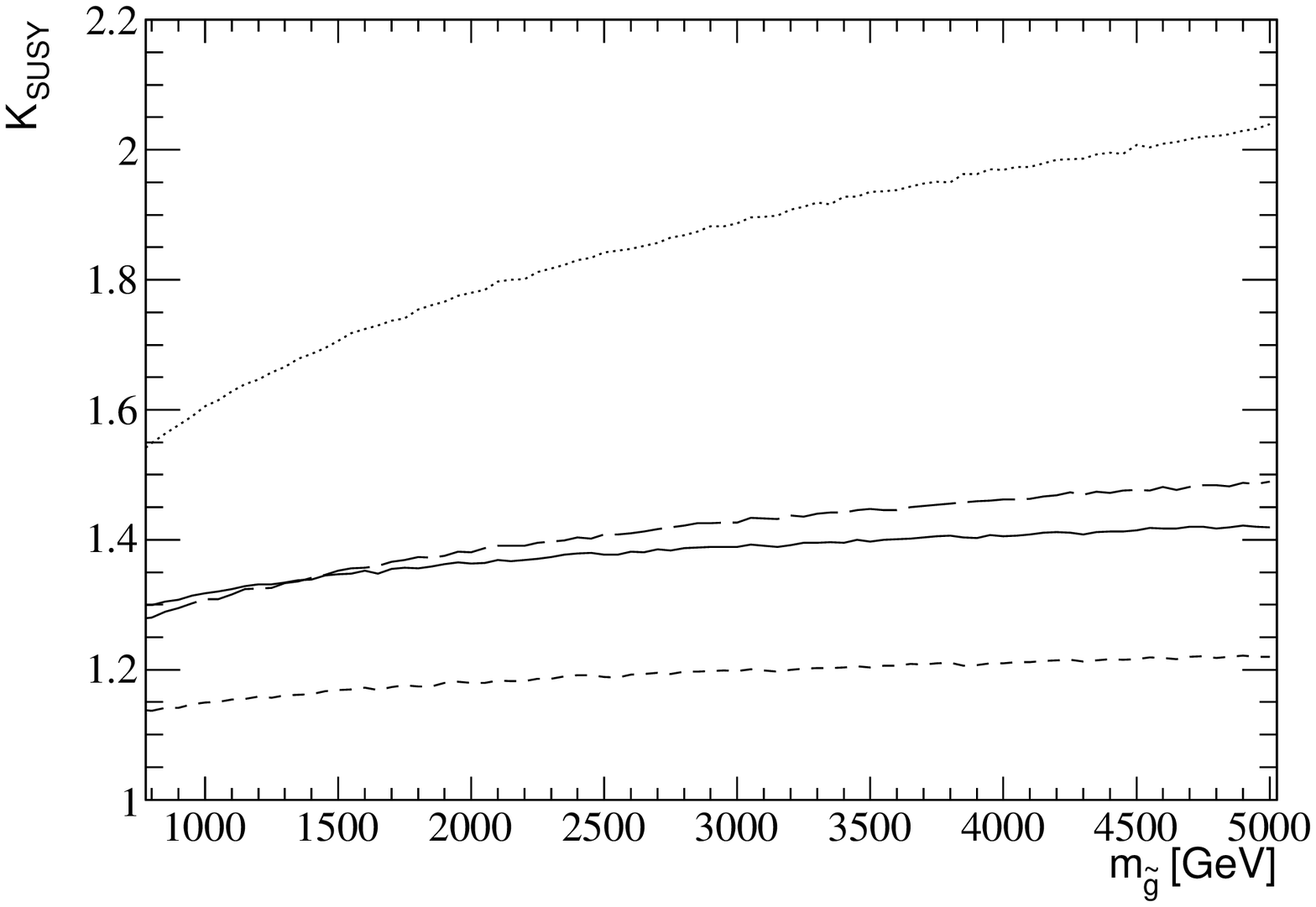}\\
(a)&(b)\\
\end{tabular}}}
\caption{\textbf{(a)} Cross-section of 
$pp\rightarrow[\stopp_1]
\rightarrow t\neut_2\bar{t}\neut_{1}$ and
\textbf{(b)} $K_{SUSY}$,  
as a function of $\tan\beta$ and $\mg$ for the input parameters
as defined in SPS1b, 3, 7 and  \textit{Def}~(\ref{eq:SUSYparams}).}
\label{stop1n1n2}
\end{figure}

\begin{figure}[t!]
\centering
{\resizebox{16cm}{!}
{\begin{tabular}{cc}
\includegraphics[width=8cm]{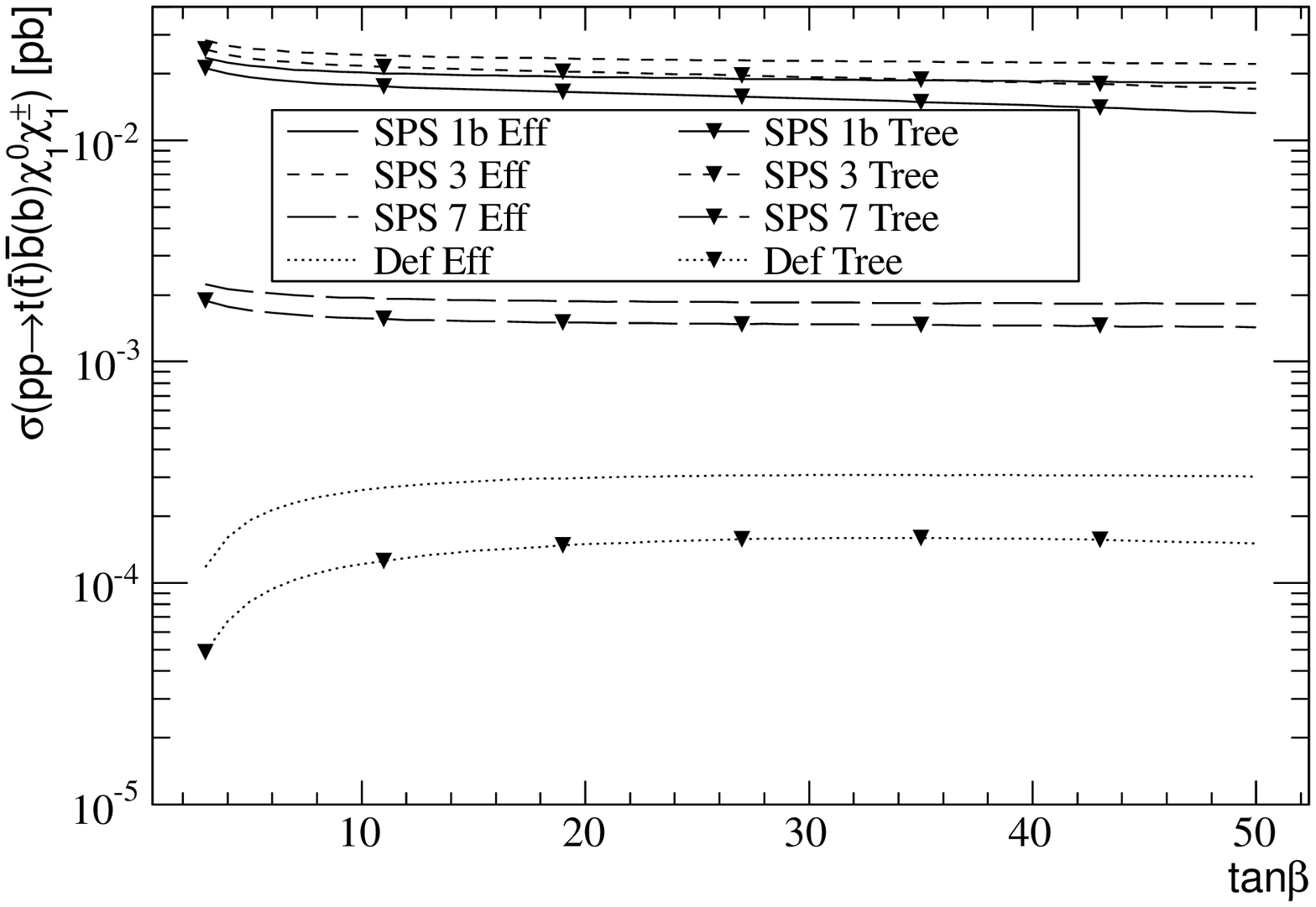}&
\includegraphics[width=8cm]{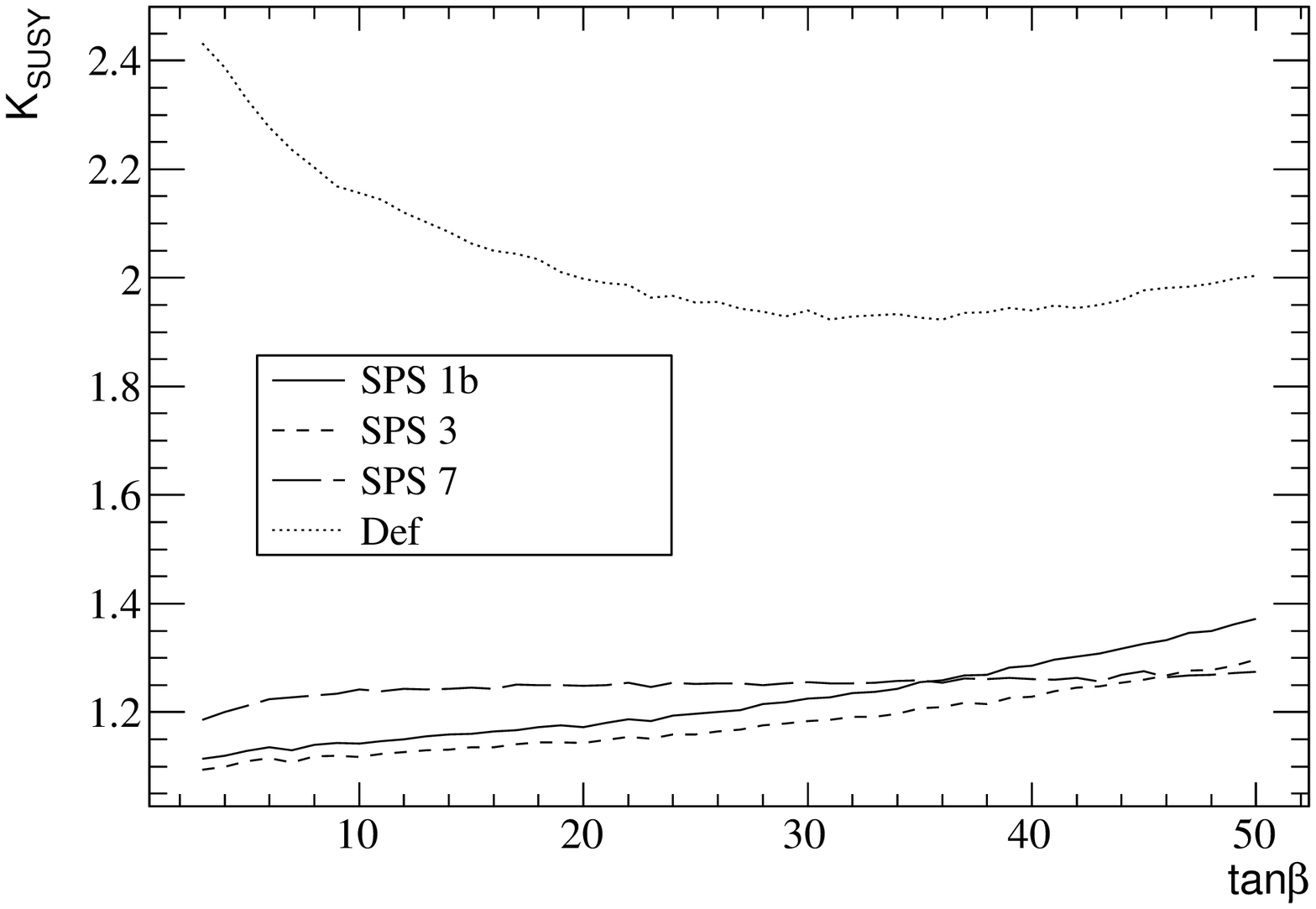}\\
\includegraphics[width=8cm]{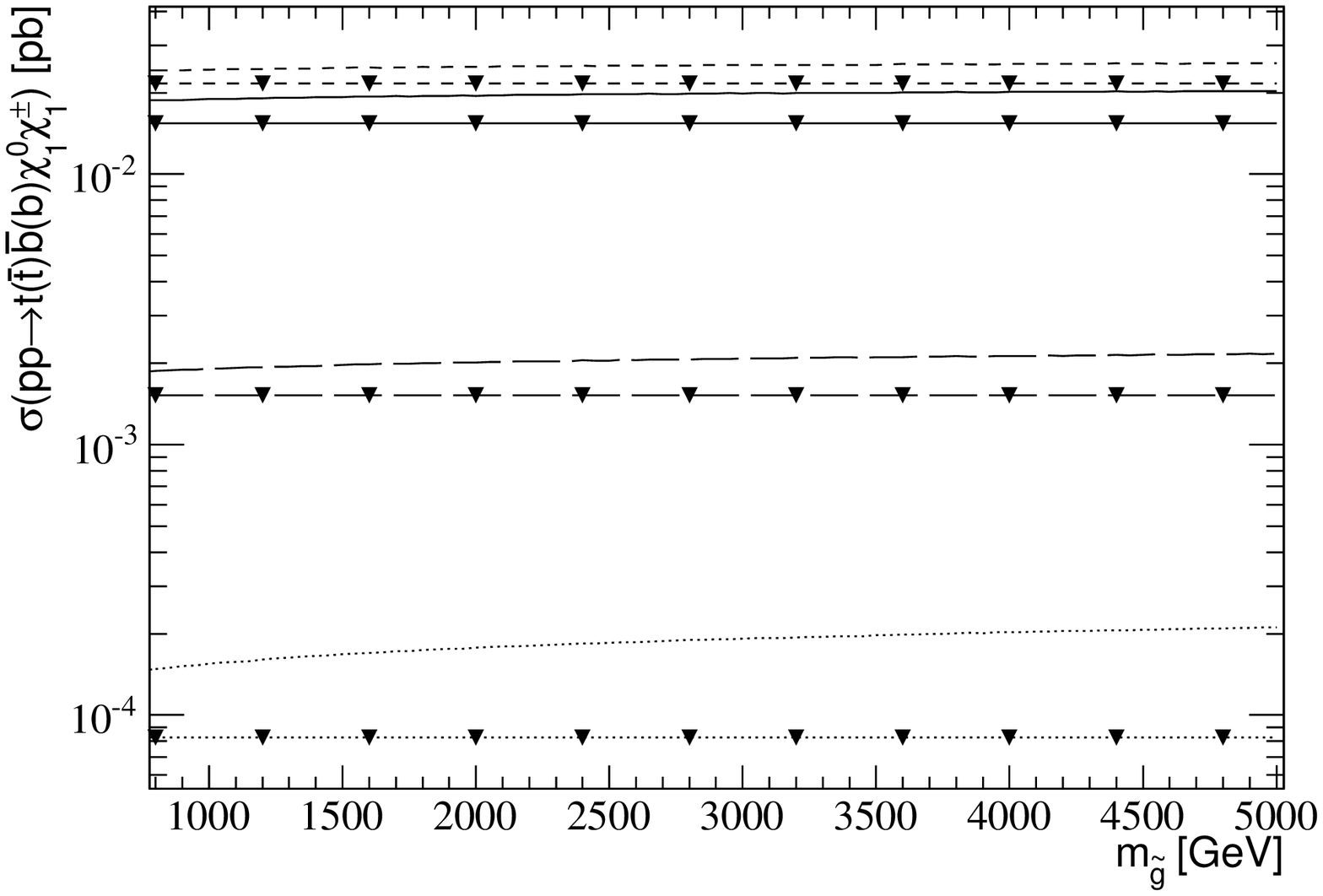}&
\includegraphics[width=8cm]{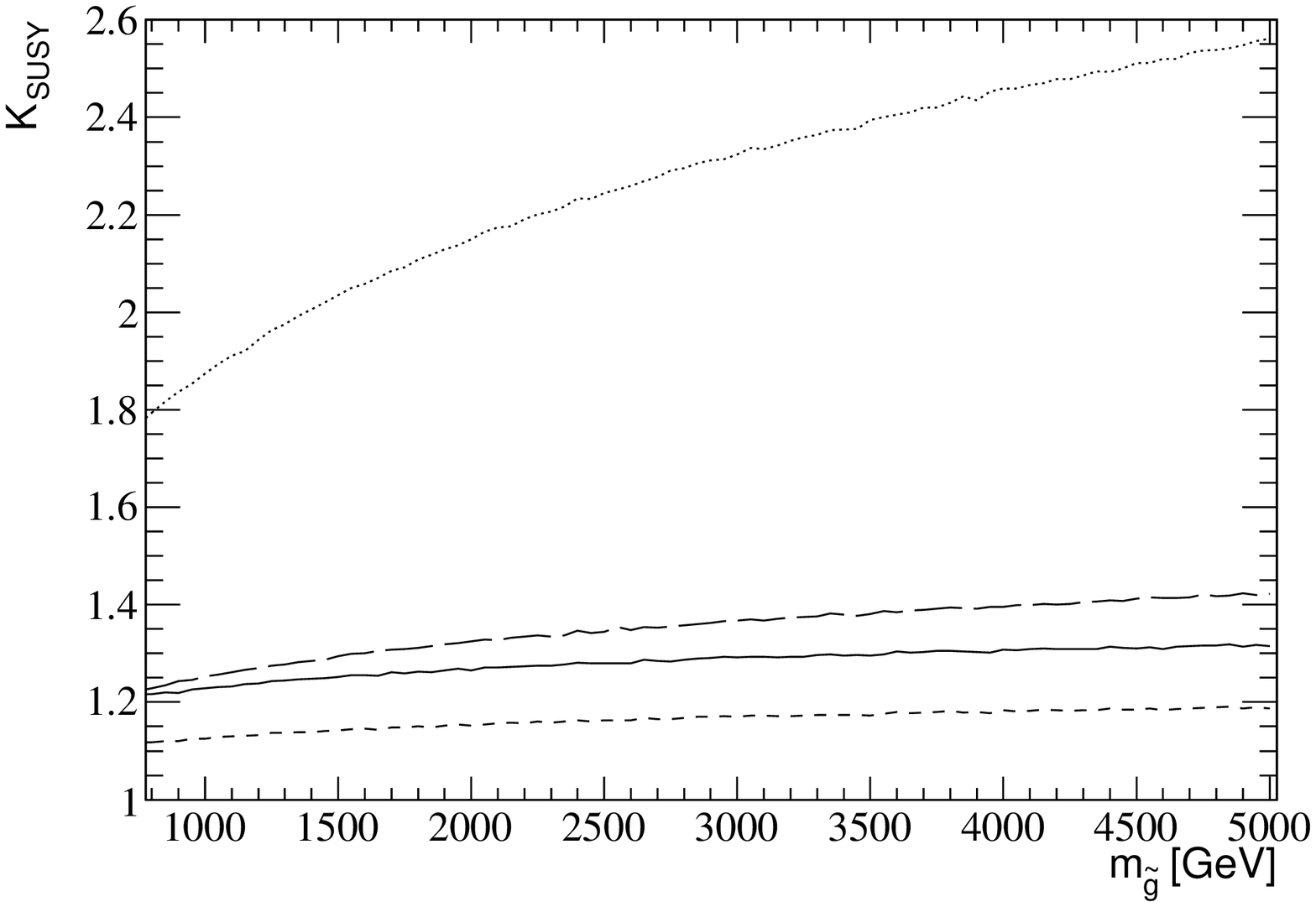}\\
(a)&(b)\\
\end{tabular}}}
\caption{\textbf{(a)} Cross-section  
$\sigma(pp\rightarrow[\stopp_1]
\rightarrow b\cplus_{1}\bar{t}\neut_{1})
+\sigma(pp\to [\stopp_1] \to \bar{b}\cmin_1 t\neut_1)$ and
\textbf{(b)} $K_{SUSY}$,  
as a function of $\tan\beta$ and $\mg$ for the input parameters
as defined in SPS1b, 3, 7 and \textit{Def}~(\ref{eq:SUSYparams}).}
\label{stop1n1x1}
\end{figure}

\begin{figure}[t!]
\centering
{\resizebox{16cm}{!}
{\begin{tabular}{cc}
\includegraphics[width=8cm]{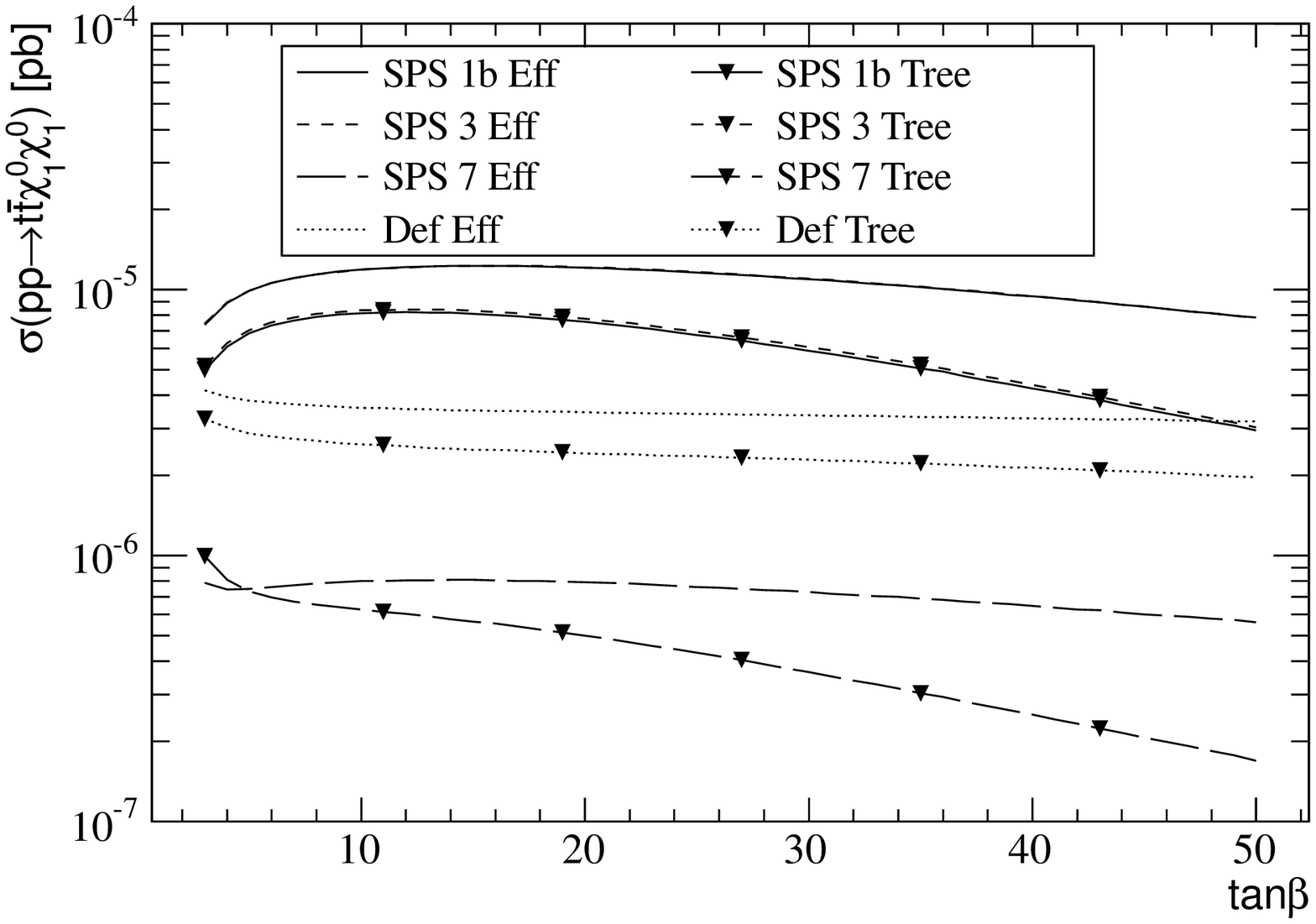}&
\includegraphics[width=8cm]{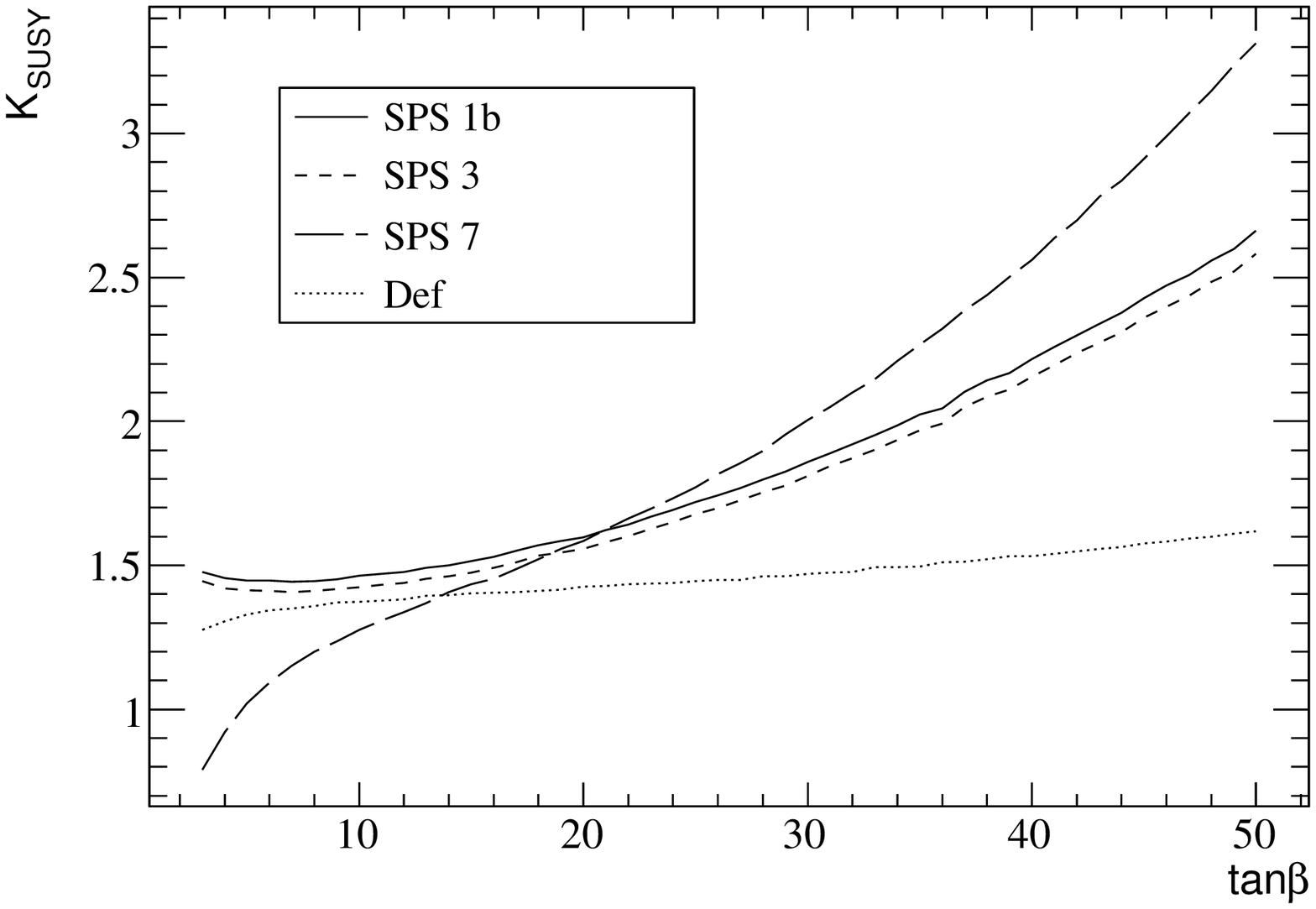}\\
\includegraphics[width=8cm]{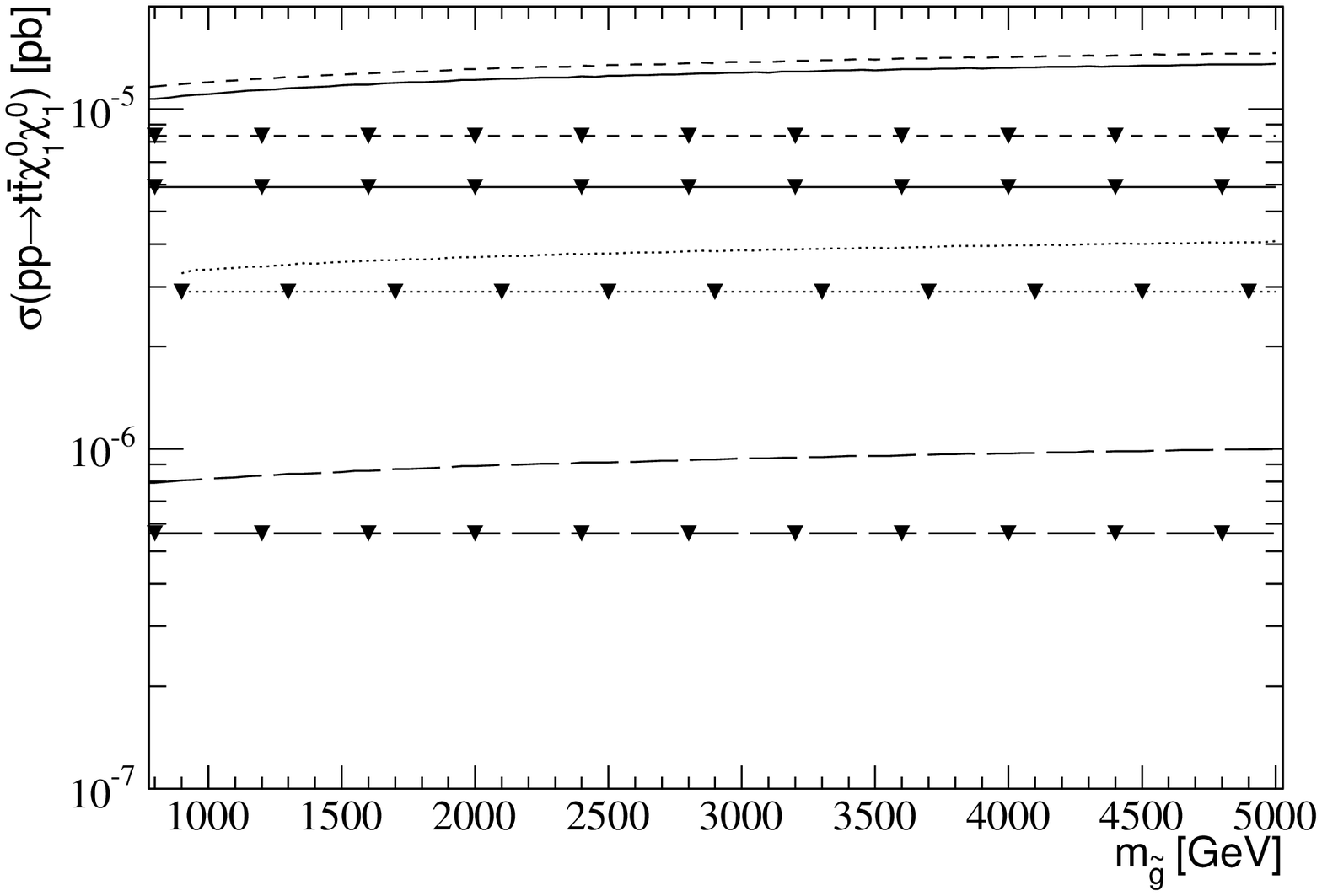}&
\includegraphics[width=8cm]{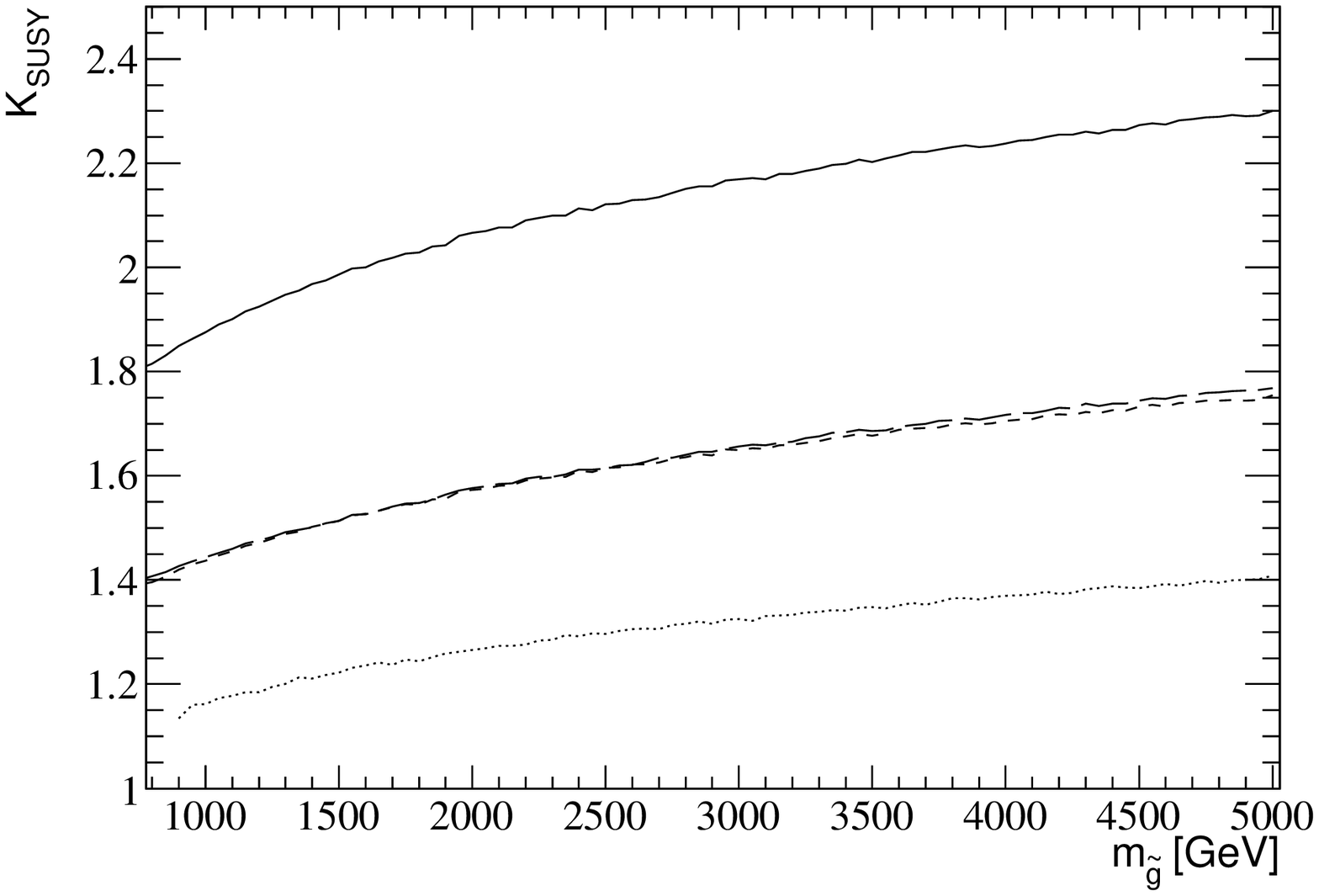}\\
(a)&(b)\\
\end{tabular}}}
\caption{\textbf{(a)} Cross-section of $pp\rightarrow[\stopp_2]
\rightarrow t\neut_{1}\bar{t}\neut_{1}$ and
\textbf{(b)} $K_{SUSY}$,  
as a function of $\tan\beta$ and $\mg$ for the input parameters
as defined in SPS1b, 3, 7 and  \textit{Def}~(\ref{eq:SUSYparams}).}
\label{stop2n1n1}
\end{figure}

\begin{figure}[t!]
\centering
{\resizebox{16cm}{!}
{\begin{tabular}{cc}
\includegraphics[width=8cm]{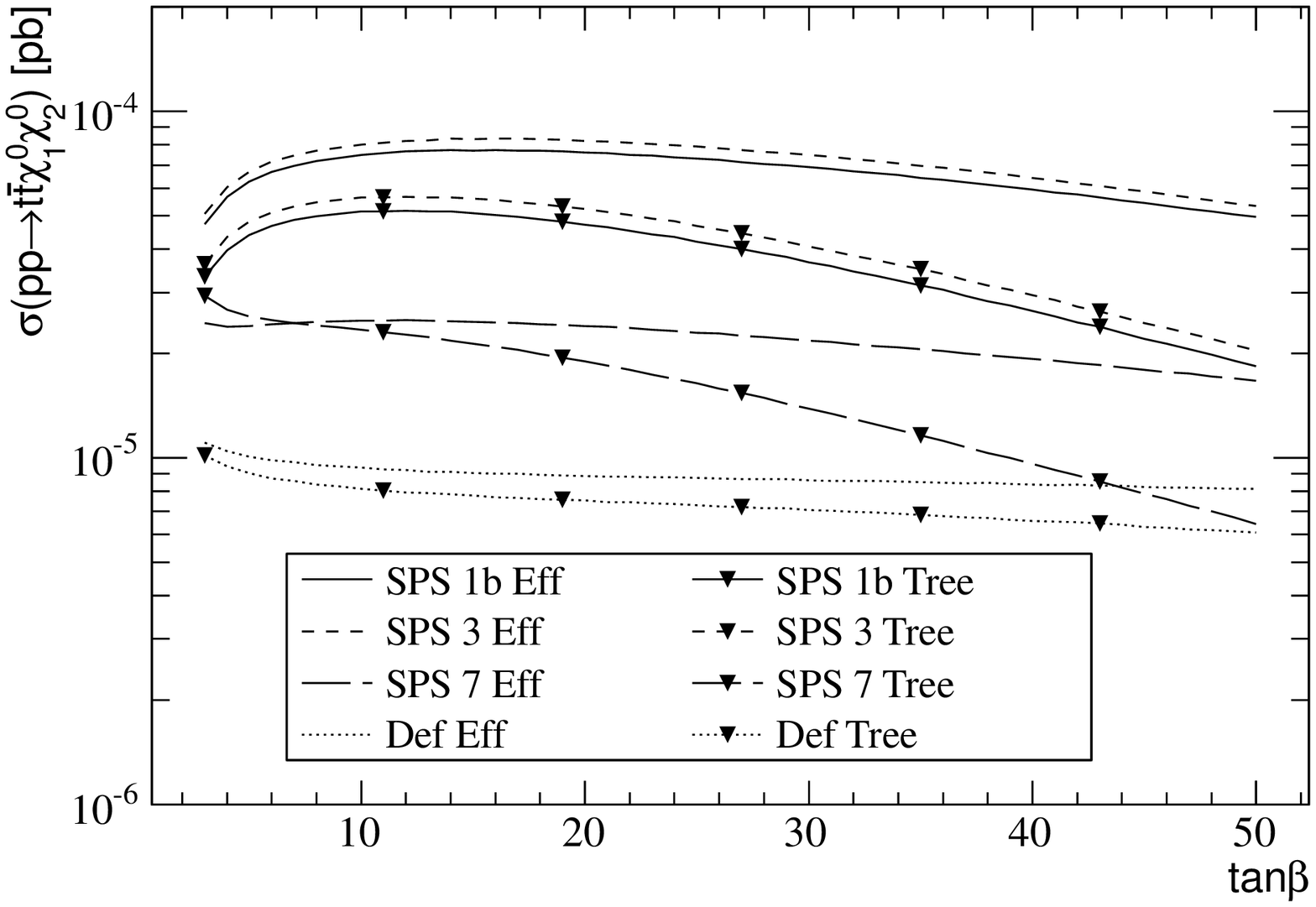}&
\includegraphics[width=8cm]{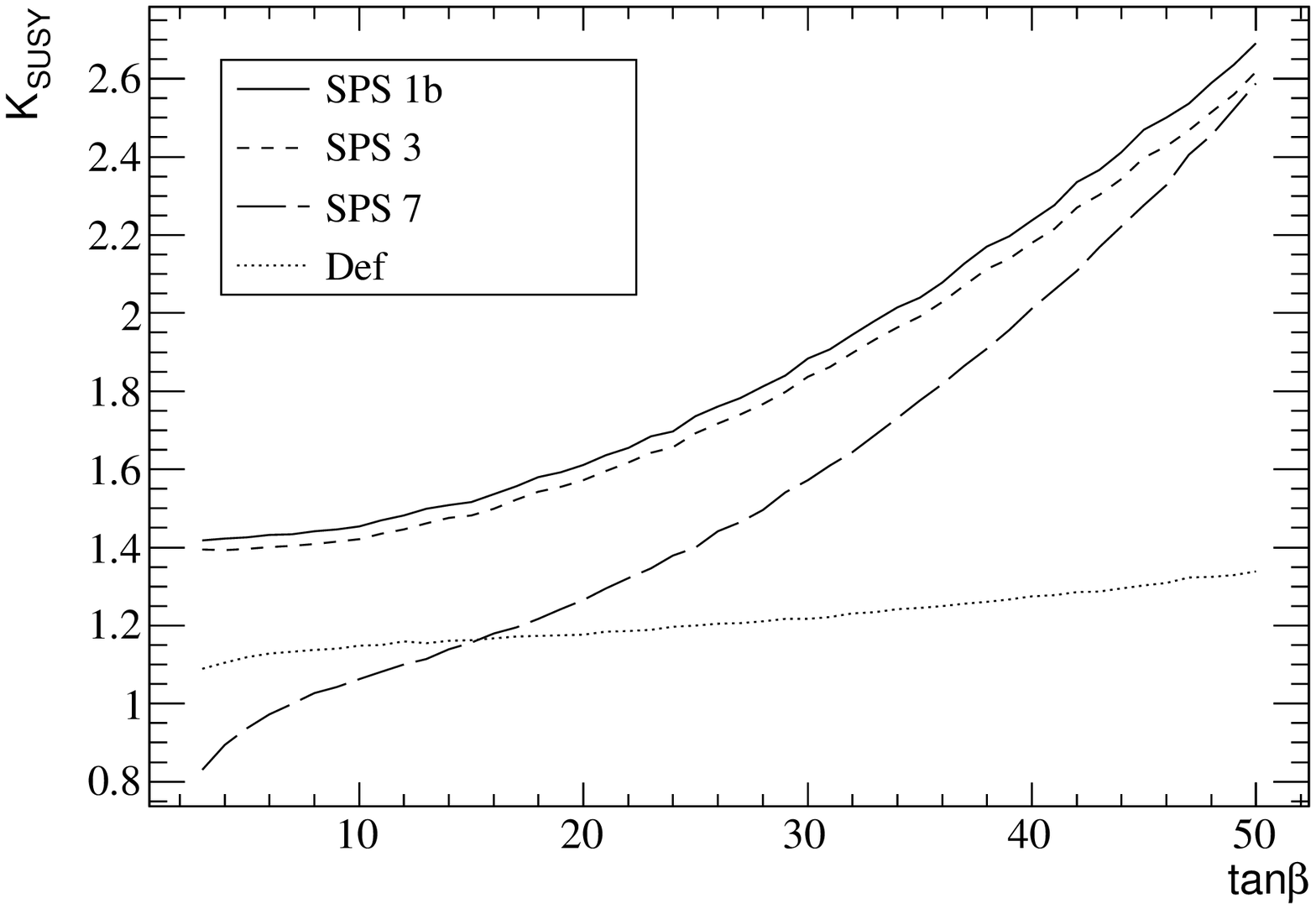}\\
\includegraphics[width=8cm]{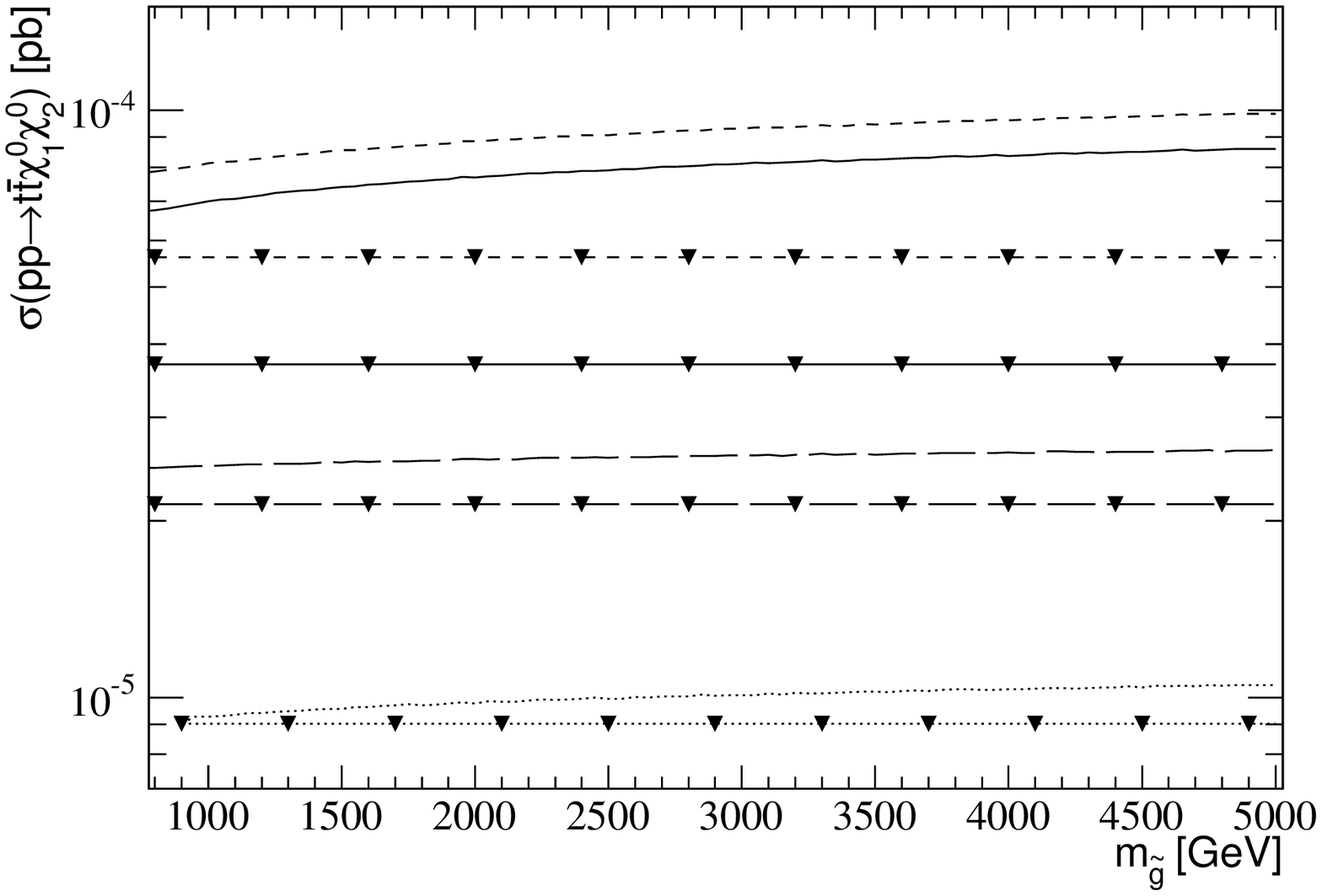}&
\includegraphics[width=8cm]{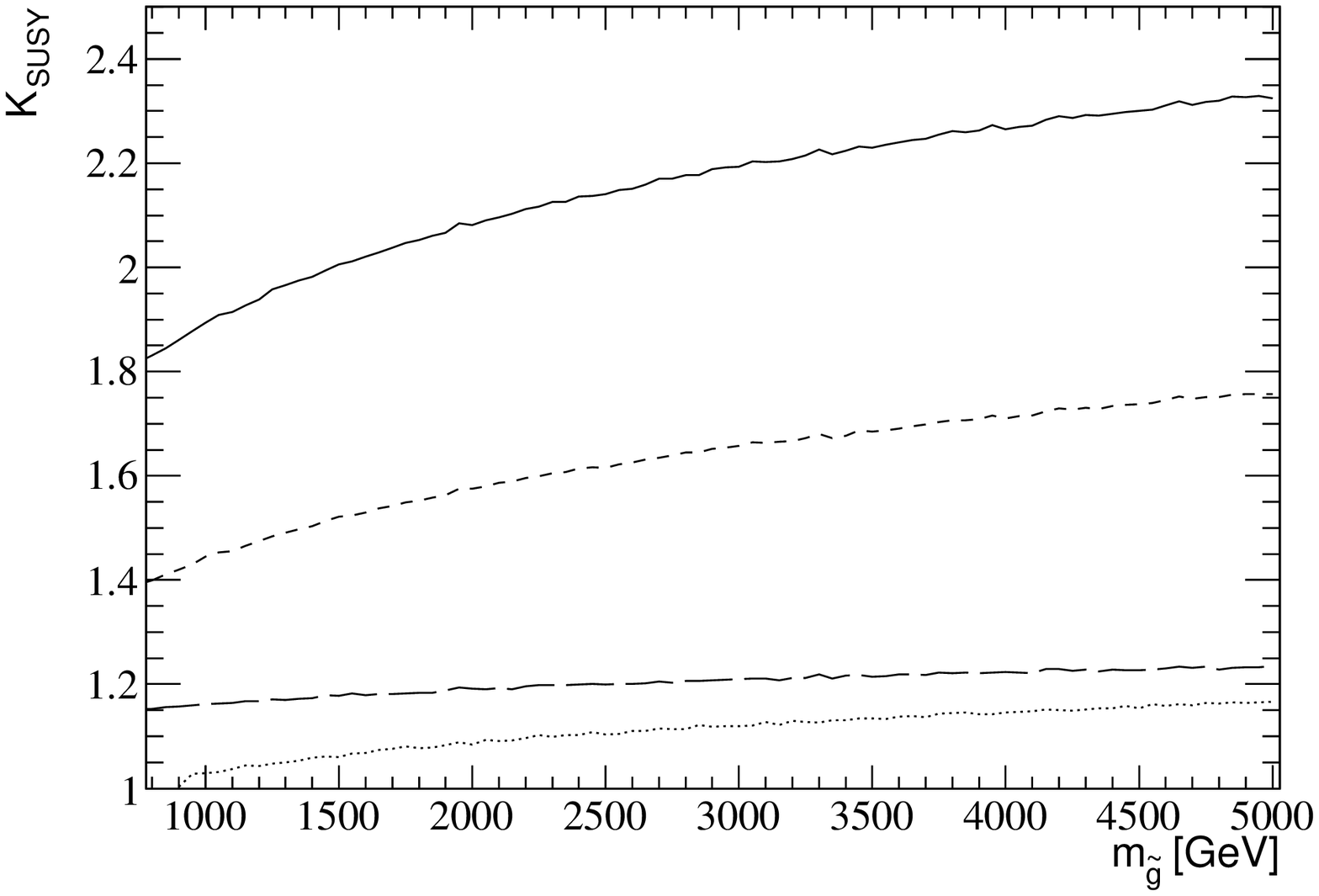}\\
(a)&(b)\\
\end{tabular}}}
\caption{\textbf{(a)} Cross-section of $pp\rightarrow [\stopp_2]
\rightarrow t\neut_2\bar{t}\neut_{1}$ and
\textbf{(b)} $K_{SUSY}$,  
as a function of $\tan\beta$ and $\mg$ for the input parameters
as defined in SPS1b, 3, 7 and \textit{Def}~(\ref{eq:SUSYparams}).}
\label{stop2n1n2}
\end{figure}

\begin{figure}[t!]
\centering
{\resizebox{16cm}{!}
{\begin{tabular}{cc}
\includegraphics[width=8cm]{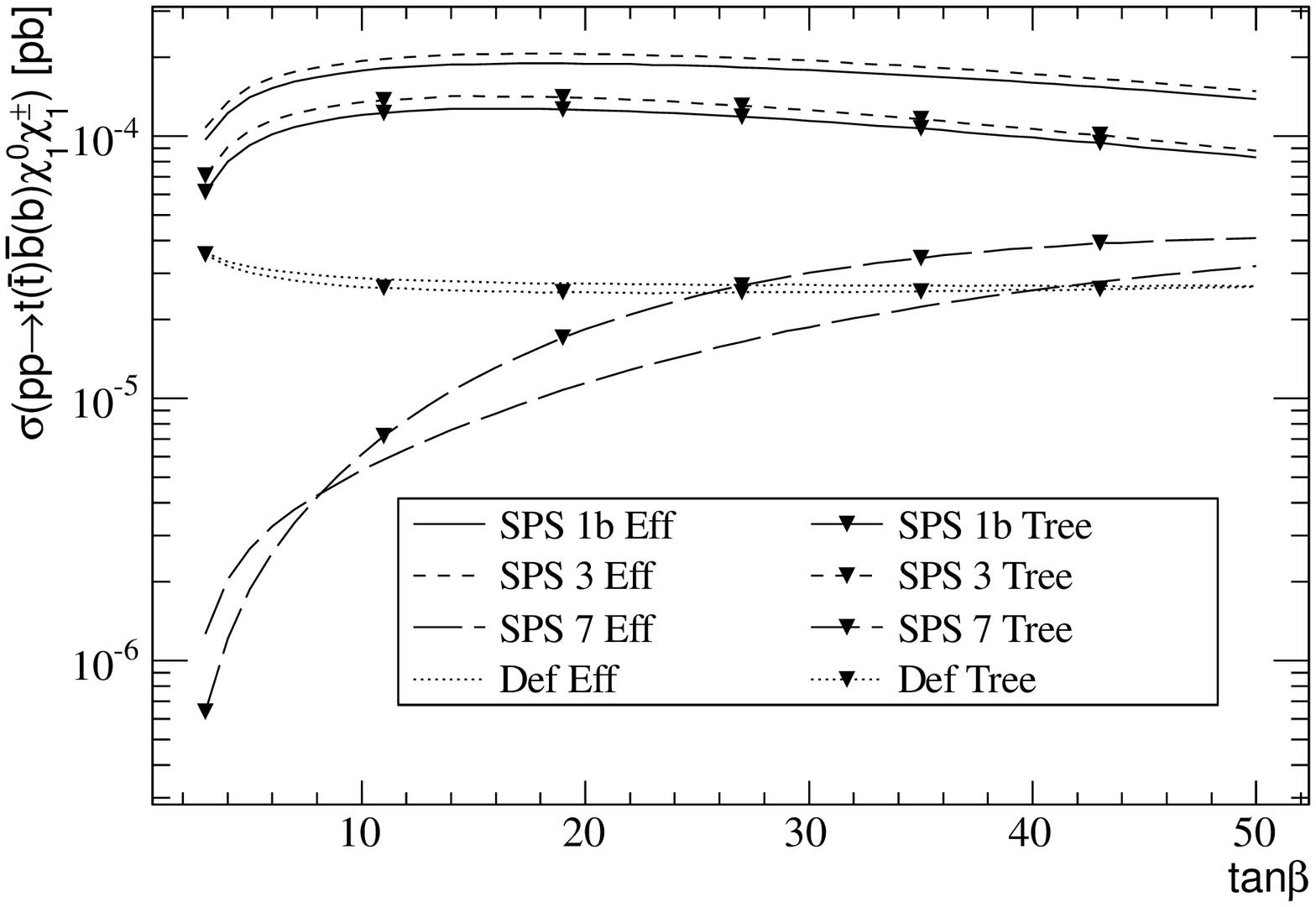}&
\includegraphics[width=8cm]{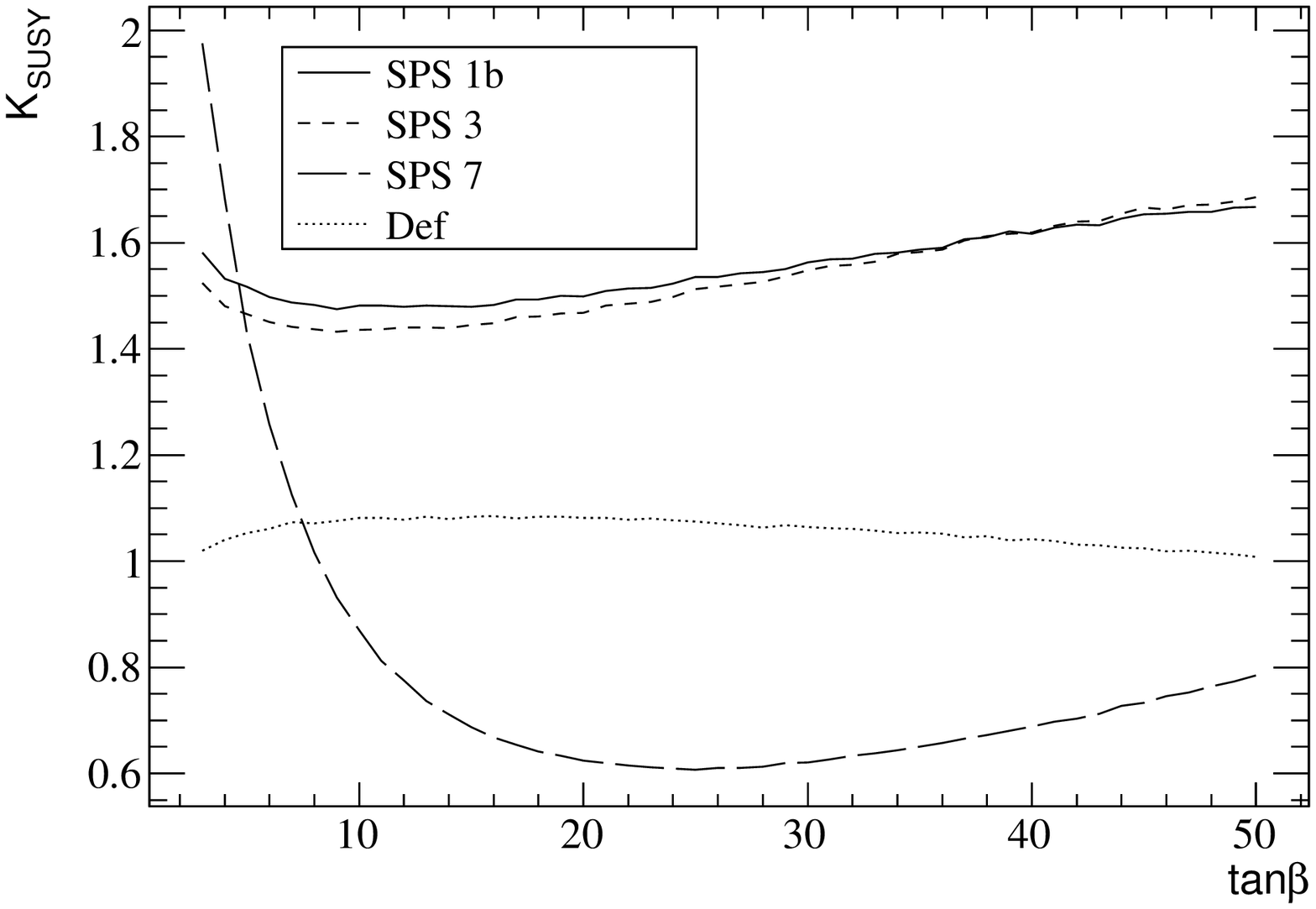}\\
\includegraphics[width=8cm]{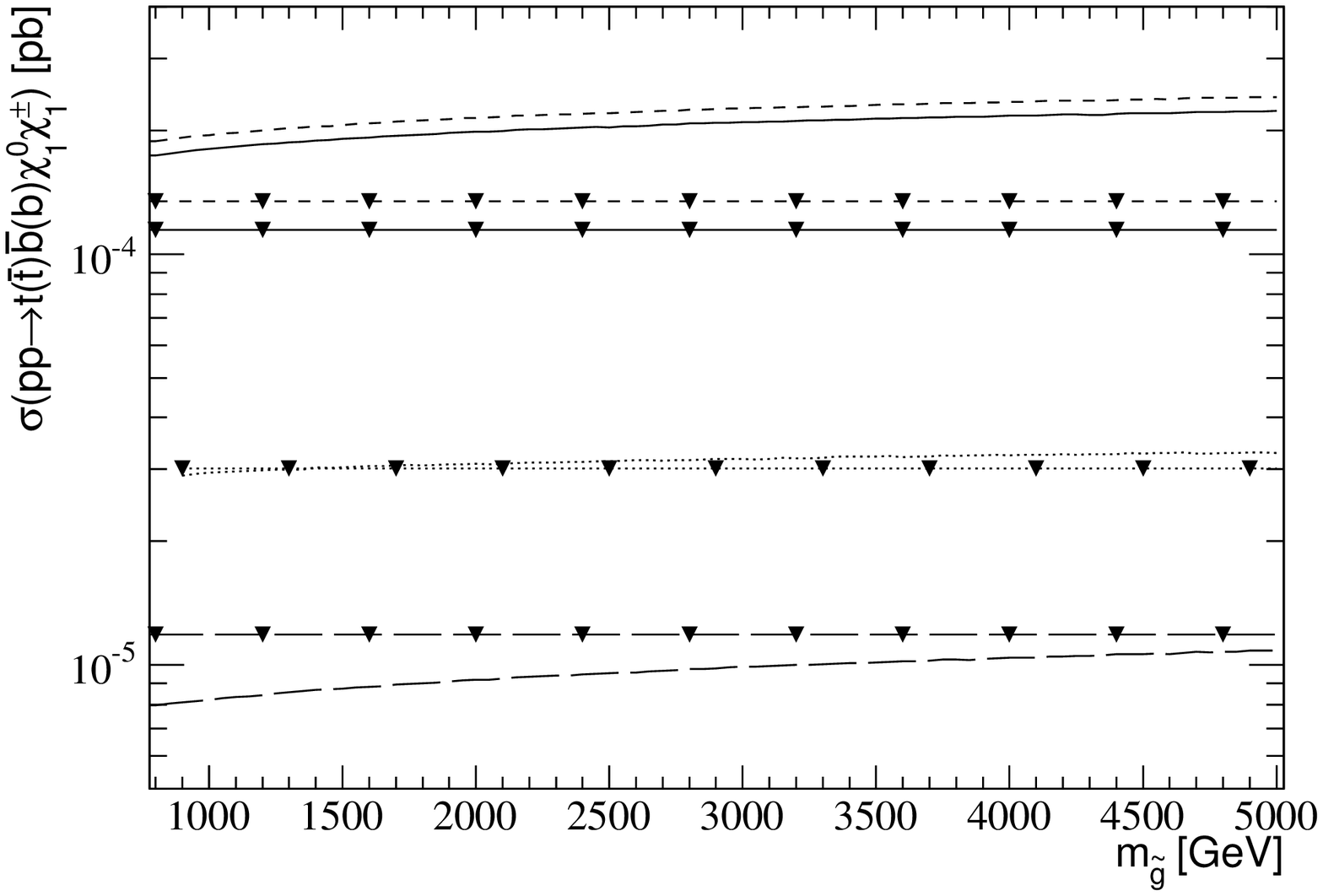}&
\includegraphics[width=8cm]{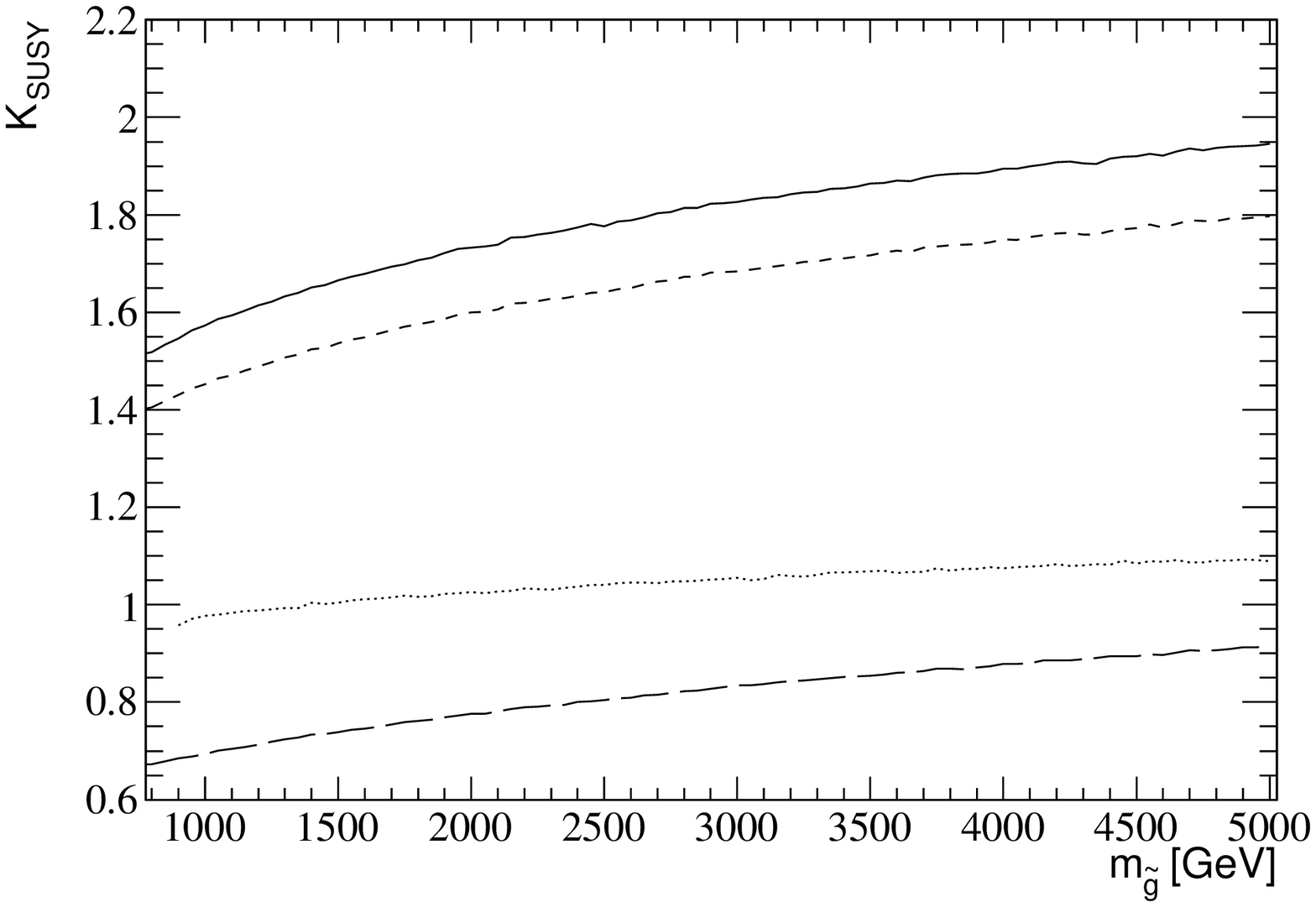}\\
(a)&(b)\\
\end{tabular}}}
\caption{\textbf{(a)} Cross-section  
$\sigma(pp\rightarrow [\stopp_2] \to b\cplus_1\bar{t}\neut_1)+
\sigma(pp\to [\stopp_2] \to \bar{b}\cmin_1t \neut_1)$ and 
\textbf{(b)} $K_{SUSY}$,  
as a function of $\tan\beta$ and $\mg$ for the input parameters
as defined in SPS1b, 3, 7 and  \textit{Def}~(\ref{eq:SUSYparams}).}
\label{stop2n1x1}
\end{figure}

The results for the cross-sections are summarized in 
Figures~\ref{stop1n1n1}-\ref{stop2n1x1}, where
we show the tree-level and effective approximation cross-section
computations, as well as the $K_{SUSY}$~(\ref{eq:ksusy}) factor, 
as a function of $\tb$ and $\mg$.

Figures \ref{stop1n1n1}-\ref{stop1n1x1} present the cross-section 
corresponding to the lightest top-squark $\stopp_{1}$.
Fig.~\ref{stop1n1n1}a shows the results for 
$\sigma(pp\rightarrow [\stopp_1]
\rightarrow (t\neut_1)
(\bar{t}\neut_1))$ 
as a function of $\mg$ and $\tb$  for the different
scenarios of the SUSY input parameters analyzed in this work. The ratio
$K_{SUSY}$ is shown in Fig.~\ref{stop1n1n1}b. The radiative corrections
are positive in all scenarios, this is due to the combination of two
factors:
 first, the
radiative corrections to the partial decay width $\Gamma(\stopp_1\to
b\neut_1)$ are positive in the scenarios SPS1b, SPS3 and \textit{Def}
(Figs.~\ref{partialsps1b37-mgl},~\ref{partialsps1b37-tb} and
Table~\ref{tablepartial}), providing a $K_{partial}>1$, but, more
importantly, the radiative corrections to the total decay width are
negative, providing a $K_{full}<1$, and therefore a
$K_{SUSY}>1$ (see eq.~(\ref{eq:ksusygroup})), as can be seen in
Table~\ref{tablepartial}. For example, in SPS1b the leading decay
channels (at the tree-level) are $b\cplus_1$ and $b\cplus_2$, but they
have negative radiative corrections of $-13.7\%$ and $-39.9\%$
respectively, in the end, when taking into account all channels one obtains
$\Gamma_{full}^{Tree}=4.49\GeV$, $\Gamma_{full}^{Eff}=3.78\GeV$,
providing $K_{full}=(0.84)^2\simeq0.71\simeq 1/1.41$, so a
$\delta_{full}^{Eff}=-26\%$ radiative correction to the total decay
width translates to a $41\%$ enhancement of the production cross-section.
The increase of the radiative corrections $K_{SUSY}$ with $\mg$ is given
by the $\log(\mg)$-terms~(\ref{eq:logterms}) of the effective
computation. This figure shows the importance of the radiative
corrections, which can enhance the cross-section by a large factor
(between $1.2$ and $5$ in Fig.~\ref{stop1n1n1}), and also of the newly
included $\log(\mg)$-terms, the log terms produce an increase of the
$K_{SUSY}$-factor (and hence the cross-section) of $1.6/1.4=1.14$,
$1.4/1.2=1.17$, $2/1.5=1.33$, $2.4/1.7=1.41$ in the range
$\mg=1000-5000\GeV$ for the scenarios SPS1b, SPS3, SPS7, \textit{Def} respectively.
The largest cross-sections are obtained in the 
SPS1b and SPS3 scenarios, which have a similar behaviour of the corrections. 
At $\tb=30$, the cross-section values in the effective approximation 
are $4.08\times 10^{-3}$ pb (SPS1b) and $5.38\times 10^{-3}$ pb (SPS3),
a factor $1.43$ and $1.37$ larger than the tree-level prediction
respectively. 
The SPS7 scenario
provides an intermediate value of the cross-section in our
analysis. For $\tb=15$ and $\mg=926\GeV$ (as fixed in this scenario) 
the effective prediction for the cross-section is $1.82\times 10^{-4}$
pb, a factor $1.56$ larger than the tree-level prediction (see
Fig.~\ref{stop1n1n1}b). 
The tree-level cross-sections decrease as a function of $\tb$ in the
region $\tb>10$, this behaviour is softened by the positive radiative
corrections. 
We recall that the squark masses (and hence phase-space factors) also
change with $\tb$, i.e. $\msto$ changes from $772\GeV$ to $780\GeV$.
The smaller values of the cross-section are obtained in the \textit{Def}
scenario~(\ref{eq:SUSYparams}), e.g.  $\sigma^{Eff}=6.54 \times10^{-6}$ pb for
the effective computation prediction at $\tb=5$, a factor $2.25$ larger
than the tree-level prediction. The largest radiative corrections are
obtained at large $\tb$, with $K_{SUSY}=4.81$. We recall that in this scenario
$\mg=3\TeV$, enhancing the radiative corrections.

The results for $\sigma(pp\rightarrow[\stopp_1]
\rightarrow t\neut_2\bar{t}\neut_{1})$ are presented
in Fig.~\ref{stop1n1n2}. 
For SPS1b/SPS3 the cross-sections are slightly larger than the
$(t\neut_1)(t\neut_1)$ channel (Fig.~\ref{stop1n1n1}), whereas they are
much larger for SPS7/\textit{Def}. We can trace back this behaviour to the
relative value of the partial decay widths
(Figs.~\ref{partialsps1b37-mgl}, \ref{partialsps1b37-tb},
Table~\ref{tablepartial}) and combinatorial factors. 
First of all, there is a factor $2$ combinatorial enhancement factor
because now the particles in the final state are different. Then in
SPS1b/SPS3 there is a slight suppression because 
$BR(\stopp_1\to t\neut_2) < BR(\stopp_1\to t\neut_1)$, whereas in
SPS7/\textit{Def} there is a large enhancement because $BR(\stopp_1\to t\neut_2) \gg BR(\stopp_1\to t\neut_1)$.
The radiative corrections
(Fig.~\ref{stop1n1n2}b) are smaller than in the $(t\neut_1)(t\neut_1)$
channel (Fig.~\ref{stop1n1n1}b), this is due to the fact that the
radiative corrections are smaller (more negative) for the partial decay
width $\Gamma(\stopp_1\to t\neut_2)$ than for $\Gamma(\stopp_1\to
t\neut_1)$ (Table~\ref{tablepartial}). This channel in the \textit{Def}
scenario has a cross-section one order of magnitude larger than the
$(t\neut_1)(t\neut_1)$ channel.

To finish with the $\stopp_1$ production analysis,
Fig.~\ref{stop1n1x1} shows the cross-sections and radiative
corrections for the lightest neutralino ($\neut_1$) and chargino
($\cpmin_1$) production mediated by $\stopp_1$. 
Since the experimental
analysis does not perform charge identification, both
charge-conjugate channels
$\sigma(pp\rightarrow [\stopp_1]\to t\neut_{1}\bar{b}\cmin_{1})$ and
$\sigma(pp\rightarrow [\stopp_1]\to \bar{t}\neut_{1}b\cplus_{1})$
contribute to the same experimental signal, therefore 
Fig.~\ref{stop1n1x1}a shows the sum of both
cross-sections\footnote{Bottom-squark mediated channels also contribute
  to this final state, we do not take into account their contribution
  because they have different kinematical resonances, and the
  interferences with $\stopp$ channels are small.}.
This channel provides the largest cross-section for $\stopp_1$-pair
production in all studied scenarios. 
For example, at $\tb=30$, the cross-section values in 
the effective approximation are $1.89\times 10^{-2}$ pb (SPS1b),
$2.29\times 10^{-2}$ pb (SPS3), $1.85\times 10^{-3}$ pb (SPS7) and 
$3.08\times 10^{-4}$ pb (\textit{Def}). The radiative corrections
(Fig.~\ref{stop1n1x1}b) are slightly smaller in this channel for the SPS
scenarios, providing $K_{SUSY}$ factors in the range $K_{SUSY}\simeq
1.1-1.4$ ($10\%-40\%$ increase of the cross-section). The \textit{Def}
scenario has the largest radiative corrections for the reference value
$\tb=5$~(\ref{eq:SUSYparams}), but they have a different behaviour than
the previous channels (Figs.~\ref{stop1n1n1}, \ref{stop1n1n2}) at large
$\tb$: they decrease instead of growing at large $\tb$. 
In this scenario the leading $\stopp_1$ decay channels are
$\neut_{3,4}$ and $\cplus_2$, all of them having negative corrections
(Table~\ref{tablepartial}) which grow with $\tb$. This provides a
growing contribution to $K_{SUSY}$ through
$1/K_{full}$~(\ref{eq:ksusygroup}). The other contribution,
$K_{partial}$, has two distinct factors: the radiative corrections to
$\Gamma(\stopp_1\to t\neut_1)$ have a nearly flat behaviour with $\tb$
at a value around $18\%-22\%$, but the radiative corrections to
$\Gamma(\stopp_1\to b\cplus_1)$ have a strong decreasing behaviour
from $\sim 29\%$ ($\tb=5$) to $\sim -52\%$ ($\tb=50$). This decrease
in $K_{partial}$ partially compensates for the increase due to
$1/K_{full}$ resulting in a more flat behaviour with $\tb$ as compared
with the $t\neut_\alpha \bar{t}\neut_\beta$ channels.

The results for the $\stopp_{2}$ mediated cross-sections are shown in 
figures~\ref{stop2n1n1}-\ref{stop2n1x1}. The values of
these cross-sections are smaller than for the $\stopp_{1}$ mediated cross-sections. 
Fig.~\ref{stop2n1n1} presents the cross-section 
$\sigma(pp\rightarrow[\stopp_2]
\rightarrow t\neut_{1}\bar{t}\neut_{1})$ and 
the corresponding radiative corrections. 
Radiative corrections are positive (except for a small corner at low
$\tb$ for SPS7), enhancing the cross-section, which
stays in the $10^{-7}-10^{-5}$ pb range.
The largest results for the cross-section are obtained in the 
SPS1b and SPS3 scenarios, which have similar values, and they 
overlap in the plots as function of $\tb$. 
At $\tb=30$, the corresponding values in the effective approximation 
are $1.094\times 10^{-5}$ pb (SPS1b) 
and $1.098\times 10^{-5}$ pb (SPS3). 
The different results between these points 
(when plotted as a function of $\mg$)
arise from the different
nominal values of $\tb$ (see appendix~\ref{sec:SPS}). 
The same similarities occur in the other channels
  (Figs.~\ref{stop2n1n2}, \ref{stop2n1x1}), but not as pronounced as the
  present one.
In scenario
\textit{Def} we obtain
an intermediate value of the cross-section ($3.37\times 10^{-6}$ pb), and
the lowest value is obtained in the SPS7 scenario 
($7.29\times 10^{-7}$ pb).
Note the quite distinct behaviour of the radiative corrections
(Fig.~\ref{stop2n1n1}b) as compared with the $\stopp_1$ channels
(Fig.~\ref{stop1n1n1}b): in the present channel the \textit{Def}
scenario has smaller radiative corrections, and a flatter evolution with
$\tb$. Quite opposite, the SPS scenarios have larger radiative
corrections and a steeper slope as a function of $\tb$. 
The radiative corrections tend to soften the slopes of the
cross-sections as a function of $\tb$. We note also a small region of
negative radiative corrections for SPS7 at low $\tb<7$.

Figure~\ref{stop2n1n2} shows the results for 
$\sigma(pp\rightarrow [\stopp_2]
\rightarrow t\neut_2\bar{t}\neut_{1})$. 
The cross-sections are larger than in the $\neut_1\neut_1$ channel in
all scenarios, being in the range of $10^{-6}-10^{-5}$ pb. 
The largest values are again obtained for the
SPS1b and SPS3 scenarios. In these two scenarios, the partial decay 
widths of $\stopp_{2}$ decaying into $\neut_{1,2}$ have always positive
radiative corrections in the whole explored region 
of gluino masses and $\tb$ (Figs.~\ref{partialsps1b37-mgl}b,
\ref{partialsps1b37-tb}b), 
providing an enhancement to  
$K_{SUSY}$, additional to the negative correction to the total decay
width ($ K_{full}<1$). The cross-sections for SPS7 and
\textit{Def} have a mild evolution with $\tb$, and the radiative
corrections are smaller than in the $\neut_1\neut_1$ channel. 

Finally, we present the results for the chargino-neutralino channel in 
Fig.~\ref{stop2n1x1}. Again, the plots in Fig.~\ref{stop2n1x1}
contain the sum of the two charge-conjugate modes:
$\sigma(pp\rightarrow [\stopp_2] \to b\cplus_1\bar{t}\neut_1)+
\sigma(pp\to [\stopp_2] \to \bar{b}\cmin_1t \neut_1)$.
This channel provides the largest cross-section for $\stopp_2$ 
production in the SPS1b, SPS3 and \textit{Def} scenarios. In these
scenarios the radiative corrections are smaller than previous channels,
and they have a quite flat evolution with $\tb$ and $\mg$. In contrast,
in the SPS7 scenario we find a steep evolution with $\tb$ of both: the
cross-section and the radiative corrections. This evolution is inherited
from the partial decay width $\Gamma(\stopp_2\to b\cplus_1)$
(Fig.~\ref{partialsps1b37-tb}). The radiative corrections are
negative in a wide region of parameters, producing a decrease of the
cross-section.

Let us note that the main $\stopp_2$ decay channels correspond to
$\neut_{3,4}$ in the neutralino sector and $\cplus_2$ in the chargino
sector (Table~\ref{tablepartial}), together these channels are
responsible for more than $60\%$ of the $\stopp_2$ decay branching
ratio in all studied scenarios. At the same time, these leading channels
have large negative radiative corrections ($|\delta^{Eff}|>20\%$),
providing a large $1/K_{full}$~(\ref{eq:ksusygroup}). For the SPS1b, SPS3,
\textit{Def} scenarios the radiative corrections to $\Gamma(\stopp_2\to
b\cplus_1)$ are quite moderate and negative
(Figs.~\ref{partialsps1b37-mgl}, \ref{partialsps1b37-tb}, Table~\ref{tablepartial}) producing
a small reduction of $K_{SUSY}$ through
$K_{partial}$. But for SPS7 the radiative
corrections are larger, and have a very strong $\tb$ dependence
$\delta^{Eff}\simeq(80\%, 40\%,-75\%)$ for $\tb=(2,5,50)$
(Fig.~\ref{partialsps1b37-tb}),
for $\tb\gtrsim 10$ the negative corrections to $\Gamma(\stopp_2\to
b\cplus_1)$ (negative contribution to $K_{SUSY}$) overcompensate the
negative corrections to the total decay width $\Gamma_{full}$
(positive contribution to $K_{SUSY}$), providing an overall negative
contribution to the cross-section radiative corrections.

To complete the discussion in this section, Table~\ref{tablexsection}
shows the radiative correction factor $K_{SUSY}$~(\ref{eq:ksusy}) for
all the production channels cross-sections and all the scenarios
presented in this work. 
Results are shown for three different values of
$\tb$, to make the comparison more meaningful. For the same reason the
\textit{Def} scenario is shown also for a light gluino mass
$\mg=1\TeV$. In general, the radiative corrections are positive,
increasing the production cross-section. As analyzed above, this is
mainly (but not only) due to the fact that the radiative corrections
decrease the top-squark total decay width.
For $\mg=1\TeV$ the positive corrections provide a $K_{SUSY}$
  factor between $1.05$ and $3.91$, depending on the channel and scenario.
Negative radiative
corrections are only obtained for 
$\sigma(pp\rightarrow [\stopp_{2}]
\rightarrow(b\cplus_1)(\bar{t}\neut_1))$ in the
SPS7 scenario
and the \textit{Def} scenario at $\mg=1\TeV$, due to large negative corrections to $\Gamma(\stopp_2\to
b\cplus_1)$ (Fig.~\ref{partialsps1b37-tb} and
Table~\ref{tablepartial}), they provide a $K_{SUSY}$ between
$0.62$ and $0.99$ depending on the scenario. 
The largest radiative
corrections are obtained at large $\tb$. 
We recall that the corrections grow with the gluino mass 
(e.g. $K_{SUSY}=4.81$ for $\tb=50$, $\mg=3\TeV$ in the
  $(\neut_1\neut_1)$ channel and \textit{Def} scenario).

\begin{table}
\small
\begin{tabular}{|c|c|c|c|c|c|c|c|c|c|}
  \hline
  $K_{SUSY}$
& \multicolumn{3}{|c|}{SPS7} &\multicolumn{6}{|c|}{\textit{Def}}\\
  & \multicolumn{3}{|c|}{$\mg=926.0\GeV$} &\multicolumn{3}{|c|}{$\mg=1\TeV$}&\multicolumn{3}{|c|}{$\mg=3\TeV$}\\
  \hline
  \multicolumn{1}{|r|}{$\tb$}  
  &10&30&50&10&30&50&10&30&50\\
  \hline
  $pp\rightarrow[\stopp_{1}]\rightarrow(t\neut_1)(\bar{t}\neut_1)$&
  1.53&1.69&1.95& 1.83&2.46 &3.91 &2.20&2.98&4.81\\
  $pp\rightarrow[\stopp_{1}]\rightarrow(t\neut_2)(\bar{t}\neut_1)$&
  1.27&1.38&1.60& 1.66& 2.32& 3.71&1.94&2.70&4.40\\
  $pp\rightarrow[\stopp_{1}]\rightarrow(b\cplus_1)(\bar{t}\neut_1)$&
  1.24&1.25&1.27&1.77 &1.62 & 1.62&2.15&1.93&2.00\\
  
  \hline
        $pp\rightarrow[\stopp_{2}]\rightarrow(t\neut_1)(\bar{t}\neut_1)$& 
        1.28&2.00&3.31& 1.20& 1.28&1.40 &1.37&1.47&1.62\\
  
  $pp\rightarrow[\stopp_{2}]\rightarrow(t\neut_2)(\bar{t}\neut_1)$& 
  1.06&1.57&2.59& 1.05& 1.11& 1.22&1.15&1.22&1.34\\
  
    $pp\rightarrow[\stopp_{2}]\rightarrow(b\cplus_1)(\bar{t}\neut_1)$& 
    0.87&0.62&0.78& 0.99&0.98 & 0.93&1.08&1.06&1.01\\
  
     \hline
     \multicolumn{10}{|c|}{}\\
   \hline
  &\multicolumn{3}{|c|}{SPS1b} & \multicolumn{3}{|c|}{SPS3} \\
  & \multicolumn{3}{|c|}{$\mg=916.1\GeV$} &\multicolumn{3}{|c|}{$\mg=914.3\GeV$}\\
  \cline{1-7}
  \multicolumn{1}{|r|}{$\tb$}  
  &10&30&50&10&30&50\\
  \cline{1-7}
  $pp\rightarrow[\stopp_{1}]\rightarrow(t\neut_1)(\bar{t}\neut_1)$& 
  1.27&1.42&1.78&1.24&1.37&1.66\\
  
  $pp\rightarrow[\stopp_{1}]\rightarrow(t\neut_2)(\bar{t}\neut_1)$& 
  1.16&1.30&1.63&1.13&1.25&1.52\\
  
    $pp\rightarrow[\stopp_{1}]\rightarrow(b\cplus_1)(\bar{t}\neut_1)$& 
    1.14&1.22&1.37&1.11&1.18&1.29\\
    
    \cline{1-7}
    
      $pp\rightarrow[\stopp_{2}]\rightarrow(t\neut_1)(\bar{t}\neut_1)$& 
      1.46&1.86&2.66&1.42&1.81&2.58\\
  
  $pp\rightarrow[\stopp_{2}]\rightarrow(t\neut_2)(\bar{t}\neut_1)$& 
  1.45&1.88&2.69&1.42&1.84&2.62\\
  
    $pp\rightarrow[\stopp_{2}]\rightarrow(b\cplus_1)(\bar{t}\neut_1)$& 
    1.48&1.56&1.66&1.44&1.55&1.69\\
    
        \cline{1-7}
      \end{tabular}
  \caption{Effects of the radiative corrections to production cross-sections in the
    effective approximation, $K_{SUSY}$ eq.~(\ref{eq:ksusy}), for SPS1b, SPS3, SPS7 and
    \textit{Def} SUSY parameters choice at different $\tan\beta$.}
\label{tablexsection}
\end{table}

\section{Conclusions}
\label{sec:conclusions}

We have implemented and tested an effective description of squark
interactions with charginos and neutralinos in the
MSSM\cite{Guasch:2008fs} into
MadGraph\cite{Stelzer:1994ta,Alwall:2007st,Alwall:2011uj}. 
A careful check of our implementation has been done by comparing 
the computation of the partial decay width of squarks into charginos and
neutralinos with FeynArts/FormCalc/LoopTools-based
programs~\cite{Hahn:2000kx,Hahn:2001rv,Hahn:1998yk}. 
We find perfect agreement.
We have reproduced previous results in the
literature~\cite{Guasch:2008fs}. 
This implementation allows to perform any Monte-Carlo computation taking
into account the leading SUSY radiative corrections to
squark-chargino/neutralino couplings.

Using this implementation, we have computed the partial
decay widths of top-squarks into charginos and neutralinos in the effective
coupling approximation, for sets of SUSY input
parameters as defined in the Snowmass Points and Slopes SPS1b, SPS3 and
SPS7, which correspond to scenarios in which the effective
approximation can be applied, eq.~(\ref{eq:gluinocondition}). 
We have also analyzed a particular scenario which we denote as
  \textit{Def}, eq.~(\ref{eq:SUSYparams}).
We have checked that 
the effective approximation provides a good description of the radiative
corrections if the gluino mass is larger than the squark mass ($\mg\gtrsim
800\GeV$ for the chosen scenarios).

For the first time we have computed SUSY particle production
cross-sections at the $14\TeV$ LHC using the effective description of
squark interactions. 
We have analyzed the cross-sections performing a comprehensive
scan of the SUSY parameter space around the scenarios SPS1b, SPS3, SPS7
and \textit{Def}.
We have focussed on reactions involving top-squarks, giving rise to a
final state involving the lightest neutralinos and chargino
($\neut_{1,2}$, $\cplus_1$). These are the channels used by the CMS and
ATLAS collaborations to perform squark searches. The
radiative corrections are positive in most of the explored parameter
space, producing an increase of the SUSY production cross-section. For a
gluino mass of $\mg\simeq1\TeV$ they provide up to a factor $4$
enhancement of the cross-section (Table~\ref{tablexsection}). This
enhancement factors are mostly (but not only) driven by the negative
radiative corrections to the top-squark total decay width
$K_{full}$~(\ref{eq:ksusygroup}). The corrections grow with the gluino
mass ($\mg$). This leads to the lucky situation that, if the gluino is
heavy (and hence, has a small production rate at the LHC) the radiative
corrections to the squark-chargino/neutralino couplings will be
large, and easier to study at the LHC. On the other hand, if 
the gluino is light the radiative corrections to the
squark-chargino/neutralino couplings will be small, but then the gluino
production rate at the LHC will be quite large, and will be easy to
study! all in all, it is a win-win situation (provided SUSY exists at
all). We leave the proof or rejection of SUSY to our experimental
colleages, and give them a new tool to explore with better precision the
SUSY parameter space at the LHC or the future ILC.

\section*{Acknowledgements}
J.G. and R.S.F. have been supported by MICINN (Spain)
(FPA2010-20807-C02-02); 
J.G. also by DURSI (2009-SGR-168) and by DGIID-DGA (FMI45/10); 
S.P. and A.A. by  grant (FPA2009-09638); S.P. also by a
Ram\'on y Cajal contract from MICINN (PDRYC-2006-000930), 
DGIID-DGA (2011-E24/2) and DURSI (2009-SGR-502); 
A.A. by an {\'A}nimo-Ch{\'e}vere project from Erasmus Mundus Program of
the European Commission and by a SANTANDER Scholarship Program for 
Latinoamerican students. The 
Spanish Consolider-Ingenio 2010 Program CPAN (CSD2007-00042) has
supported this work. 
J.G. and R.S.F. wish to thank the hospitality of the
Universidad de Zaragoza. A.A.  wishes to thank the hospitality of the
Universitat de Barcelona.

\appendix
\section{Snowmass points and slopes parameters}
\label{sec:SPS}
Snowmass Points and Slopes parameters from
  Ref.\cite{Allanach:2002nj}. Mass parameters in $\GeV$.\\
{\scriptsize
\begin{tabular}{|c|c|c|c|c|c|c|c|c|c|c|c|c|c|c|c|}
\hline
SPS&$\mg$&$\mu$&$M_{A}$&$\tan\beta$&$M_{1}$&$M_{2}$&$A_{t}$&$A_{b}$&$M_{\sfr_{L}}$&$M_{\sbottom_{R}}$&$M_{\stopp_{R}}$&$M_{\squark_{L}}$&$M_{\tilde{d}_{R}}$&$M_{\tilde{u}_{R}}$\\
\hline
1b&916.1&495.6&525.5&30&162.8&310.9&-729.3&-987.4&762.5&780.3&670.7&
836.2&803.9&807.5\\
3&914.3&508.6&572.4&10&162.8&311.4&-733.5&-1042.2&760.7&785.6&661.2&
818.3&788.9&792.6\\
7&926.0&300.0&377.9&15&168.6&326.8&-319.4&-350.5&836.3&826.9&780.1&
861.3&828.6&831.3\\
\hline
\end{tabular}
}

\providecommand{\href}[2]{#2}

\end{document}